\documentclass[graybox,natbib]{svmult}

\usepackage{mathptmx}       
\usepackage{helvet}         
\usepackage{courier}        
\usepackage{type1cm}        
\usepackage{makeidx}         
\usepackage{graphicx}        
\usepackage{multicol}        
\usepackage[bottom]{footmisc}

\makeindex             

\begin{document}

\title*{Resolved Stellar Populations as Tracers of Outskirts}
\author{Denija Crnojevi\'c}
\institute{Denija Crnojevi\'c, Texas Tech University,
  Department of Physics and Astronomy, Box 41051, Lubbock, TX 79409, USA, \email{denija.crnojevic@ttu.edu}}
%
%
\maketitle


\abstract{Galaxy haloes contain fundamental clues about the galaxy 
formation and evolution process: hierarchical cosmological models
predict haloes to be ubiquitous, and to be (at least in part) the
product of past merger and/or accretion events. The advent of
wide-field surveys in the last two decades has revolutionized our view of our
own Galaxy and its closest ``sister'', Andromeda, revealing copious
tidal streams from past and ongoing accretion episodes, as well as
doubling the number of their known faint satellites. The focus shall
now be shifted to galaxy haloes beyond the Local Group: resolving
individual stars over significant areas of galaxy haloes will enable
estimates of their ages, metallicities and gradients. 
The valuable information collected for galaxies with a
range of masses, morphologies and within diverse environments will 
ultimately test and quantitatively inform theoretical models of galaxy formation, and shed
light onto the many challenges faced by simulations on galactic
scales.
}


\section{The Importance of Haloes\index{hierarchical assembly, stellar haloes}} \label{intro}

Our understanding of galaxy formation and evolution has dramatically
evolved in the past fifty years.
The first and simplest idea for the formation scenario of our own
Milky Way (MW) Galaxy was put forward by
\cite{eggen62}, who proposed the bulk of a stellar halo to be formed
in a rapid collapse of gas in the protogalaxy. This scenario, often referred to as
``monolithic'' collapse, is a dissipative process and 
takes place on dynamical timescales of the order of
$\sim10^8$\,yr. This process gives birth to a metal-poor stellar
component in the halo outer regions, while the inner regions ends up
being more metal-rich due to the reprocessing of the gas as it collapses
deeper into the protogalaxy potential well.
This idea was later challenged by an alternative explanation, based on
the observation that globular clusters (GCs) at different 
Galactocentric distances have a wide range of metallicities.
In this scenario, the halo is formed on longer timescales ($\sim10^9$\,yr) and, instead of
being a self-contained system, it comes together as the product of
several protogalactic fragments (\citealt{searle78}). These fragments 
can be pre-enriched before they are accreted. While both scenarios are
capable of explaining many observed quantities of the Galactic halo,
they cannot individually give a comprehensive picture
(\citealt{norris91, chiba00}), which has led
to the development of hybrid ``two-phase'' models. In the latter,
the inner Galaxy regions are formed in a first phase as a result of a
monolithic-like process, while the outer halo regions are built up
over the Galaxy's lifetime through dissipationless accretion events
(\citealt{freeman02}).

In the past couple of decades, the most widely accepted paradigm of
the hierarchical Lambda-Cold Dark Matter ($\rm \Lambda$CDM)
structure formation model has prevailed, favouring the predominance of
merger and accretion events in the build-up of galactic haloes 
(\citealt{white91, bullock05, springel06, johnston08}).
These models predict the ubiquitous presence of haloes, which are
characterized by old and metal-poor populations and often shows signs
of recent interactions, in contrast with the smooth haloes predicted by
dissipative models (\citealt{bullock05, abadi06, font06}).
The interaction events provide a mine of information on the assembly of haloes: 
dynamical timescales become relatively long (up to several Gyr) in the
outer regions of a galaxy, and thus accretion/merger events that
occurred a long time ago are often still visible as coherent
structures like disrupting galaxies or streams, which readily testify the past
assembly history of their
host. The assembly itself depends on a variety of factors, such as 
number, mass, stellar content and structural properties of the
accreted satellites, as well as orbital properties, timing and energy of the
accretion event. Even when the progenitor is completely dissolved in the host's
halo (which is particularly true in the inner halo regions where
dynamical timescales are relatively short),
its stripped stellar content still retains a characteristic coherence
in velocity space as well as in metallicity content, thus giving important clues 
about the progenitor's properties. Observing the stellar ``fossils''
that populate galaxy haloes thus offers a unique
opportunity to reconstruct the modes, timing, and statistics 
of the halo formation process.

Besides being taletellers of their host system's merger history,
the shape and size of haloes also hold vital clues to the process of
galaxy formation. In particular, they can teach us about the
primordial power spectrum of density fluctuations at the smallest
scales; about the reionization process, that shall lead to faint and
concentrated haloes for an early suppression of star formation in
low-mass dark matter (DM) subhaloes; or about the triaxiality of DM haloes, which are
predicted to be more flattened for dissipationless formation scenarios
(\citealt{abadi06}).
Despite only accounting for a mere $\sim1\%$ of a 
galaxy's total mass (e.g., \citealt{morrison93}),
extended haloes are clearly extremely valuable to test and refine
theoretical predictions on the halo assembly process. 
Due to their extreme faintness, however, haloes have not been as fully exploited
as they should have been as key tests of galaxy formation models:
they are not easily detected above
the sky level, i.e., surface brightness values of $\mu_V\sim25$\,mag\,arcsec$^{-2}$,
posing a serious observing challenge to their investigation.
Cosmological simulations predict the majority of past and ongoing accretion events
to have surface brightness values well below this value 
(e.g., \citealt{bullock05}). According to some models, reaching a
surface brightness of $\mu_V\sim29$\,mag\,arcsec$^{-2}$ should allow the
detection of at least one stream per observed galaxy
(\citealt{johnston08, cooper10}).
How is it then possible to extract the 
information locked in the faint outskirts of galaxies?

\subsection{Resolved Stellar Populations\index{Resolved stellar
    populations, Low surface brightness}} \label{rsp}

The best method to study faint haloes and their substructure in nearby
galaxies is to resolve
individual stars. Even when sparse and faint, resolved stars can be
individually counted, and a stellar number density can easily be
converted into a surface brightness value. When the 
Galactic extinction presents a high degree of spatial inhomogeneity
(possibly mimicking faint irregular substructures),
and the sky level is higher than the integrated light signal coming 
from extremely faint sources,
resolved populations provide a very powerful means to trace them. This
method is not free from complications: there will always be
contamination coming both from foreground Galactic stars as well as
from background unresolved galaxies. This can be accounted for
statistically, by observing ``field'' regions away from the main
target and quantifying the contaminants, while a direct confirmation
of a star's membership requires spectroscopy. At the same time, resolving
individual stars poses constraints on the inherent
nature and on the distance of the putative targets: for systems
where the stellar density is so high that stars fall on top of each
other on the sky, the ``crowding'' prevents the resolution of
individual objects. This can of course occur also in the case of a
relatively sparse galaxy which has a large line-of-sight distance,
so that the stars are packed in a small region of the sky. Distance
is also the principal enemy of depth: the larger the distance, the brighter the 
detection limit, i.e., the absolute magnitude/surface brightness that we can reach for a
fixed apparent magnitude. Nonetheless, resolved stellar populations
are able to deliver powerful information for galaxies located within
$\sim10$\,Mpc, i.e., within the so-called Local Volume.

The discovery of the Sagittarius dwarf galaxy by \citet{ibata94} 
from the identification of a comoving group of stars opened the door
to the era of halo studies and their substructure: a galaxy resembling the properties
of classical dwarf spheroidals was clearly in the process
of being disrupted by its giant host, our own MW. This evidence was
the first to support theoretical predictions for the hierarchical assembly models and the
existence of observable accretion events. Soon thereafter, stellar 
density maps allowed the discovery of a prominent low surface 
brightness stream around the MW's closest
giant spiral Andromeda (M31), the so-called Giant Stellar Stream (\citealt{ibata01}).
This feature, invisible to the naked eye, is a clear example
of the elusive nature of haloes and their substructure: the surface brightness of
the Giant Stellar Stream is $\mu_V\sim30$\,mag\,arcsec$^{-2}$, which is
prohibitive for integrated light images.

As challenging as it is, the mere detection of haloes and their
substructures is not enough to provide quantitative constraints on
models of galaxy evolution. From the stars' photometry and thus position in the
colour-magnitude diagram (CMD), i.e., the observational counterpart of
the Hertzsprung-Russel diagram, it is possible to characterize the
properties of the considered stellar system. First and foremost, 
in contrast to integrated light, accurate distance measurements can be
obtained from CMD features that act as standard candles, e.g., the
luminosity of the tip of the red giant branch (TRGB) or of the
horizontal branch (HB).
Another key advantage of resolved populations is the possibility to 
constrain ages and metallicities more tightly than with integrated
light alone. The CMD is used to quantify the star
formation rate as a function of lookback time, and thus derive
the star formation history (SFH) of a composite stellar population
 (e.g., \citealt{gallart05}, and references therein).
Spectroscopy of individual stars is the ultimate method to constrain
their metallicity content and kinematical properties, such as radial
velocity and proper motion, which allows for the full six-dimensional
phase space to be investigated. The latter cannot, for the moment, be
achieved beyond the LG limits, and still only occasionally for M31.

Besides giving precious insights into galaxy haloes and their accretion
histories, resolved stellar populations can help us characterizing the  
``surviving'' low-mass galaxies that have not been accreted to
date and reside in the outskirts of giant hosts.

\subsection{The Low-mass End of the Galaxy Luminosity Function\index{Dwarf galaxies}} \label{satellites}

The low-mass end of the galaxy luminosity function (LF) is of no less
interest than haloes themselves. Besides the MW and M31, the Local Group (LG)
contains tens of smaller galaxies which can be studied in detail due
to their proximity (see \citealt{tolstoy09} for a review).
While the $\rm \Lambda$CDM cosmological
model has provided a convincing match to the large-scale structures
observed in the high-redshift Universe, it falls short at
the smallest, galactic scales, indicating an incomplete understanding
of the physics involved in the evolution of galaxies: for example, 
the ``missing-satellite problem'' has been highlighted for the first time
by \citet{moore99} and \citet{klypin99}. Briefly, the number of DM subhaloes predicted in simulations exceeds the observed number
of MW satellites by almost two orders of magnitude. The shape of the
DM profile in the innermost regions of dwarf galaxies is also a matter
of debate (the ``cusp-core'' problem; \citealt{walker11}).
In addition, the more massive among the MW satellites are less dense
than what is expected from simulations, which is puzzling because they
should be affected by fewer observational biases than their smaller,
sparser siblings (the ``too-big-to-fail'' problem; \citealt{boylan11}). 
In addition, the fact that many of the MW and M31
satellites are distributed along planes does not have a straightforward
explanation in $\rm \Lambda$CDM models (e.g., \citealt{pawlowski14}).
 
From the theoretical point of view, the inclusion of baryonic physics
in DM-only simulations is key to reconcile predictions with
observations of the smallest galaxies. In particular, effects such as
supernova feedback, stellar winds, cosmic reionisation, and tidal/ram
pressure stripping all concur to reduce star formation
efficiency in the least massive DM haloes. Tremendous progress is being made on this
front, taking into account realistic physics as well as increasing the
resolution of simulations (e.g., \citealt{stinson09, brooks13, 
sawala16, wetzel16}). At the same time, new
observational discoveries keep offering intriguing challenges at the
smallest galactic scales, as further described in Sect.~\ref{mw_sats}
and \ref{m31_sats}.

\section{Local Group\index{Local Group}} \label{sec:lg}

The galaxies closest to us give us the most detailed information
because of the large number of stars that can be resolved. Here, I
will summarize what we have learnt in the past two decades about our own Galaxy
(even though an extensive picture of the MW outskirts goes beyond the
scope of this contribution and can be found in Figueras, this volume), about its
closest spiral neighbour M31 and about their lower-mass satellites.

\subsection{Milky Way} \label{mw}

The MW is traditionally divided into discrete components, i.e., the
bulge, the disks (thin and thick) and the halo. The spheroidal portion
of the MW is given by the central bulge, which consists mainly of
metal-rich populations, and an extended diffuse
component, which has a lower mean metallicity. Overall, stars and GCs in the
halo have ages $\sim11-13$\,Gyr (\citealt{carollo07}).
The halo can be further deconstructed into an inner halo and an outer
halo, even though the distinction could 
partly arise from observational biases (\citealt{schoenrich14}).
The inner and outer haloes also seem to have
different chemical composition ([Fe/H]$\sim-1.6$ and [Fe/H]$\sim-2.2$,
respectively; \citealt{ryan91, carollo07}).
According to simulations, the two halo components should 
also have formed on different timescales:
the inner halo ($<20$\,kpc) is partly constituted by early-formed in-situ stars,
partly due to a violent relaxation process, and partly assembled from
early, massive merging events that provide metal-rich populations
(\citealt{abadi06, font11, tissera13, pillepich14, cooper15}); the outer halo is
assembled more recently, with its mass beyond
$\sim30$\,kpc being mainly accreted in the past $\sim8$\,Gyr
(\citealt{bullock05, cooper10}). These predictions are, however, still not sufficient
at a quantitative level, and unconstrained as to the exact
ratio of accreted stars versus in-situ populations.
At the same time, observations of the MW halo
with better statistics and precision are needed to inform them.

Our position within the MW puts us at a clear disadvantage for global studies
of its outskirts: the distant and sparse halo stars are
observed from within the substantial disk component, which produces
contamination both in terms of extinction and numerous disk stars
along the line of sight, which completely ``obscure'' the sky at low
Galactic latitudes. Nonetheless, thanks to the advent of wide-field imagers,
the past two decades have revolutionized the large scale view of our
Galaxy. Several stellar tracers can be used to dig into the MW halo at
different distance ranges: old main sequence turnoff (MSTO) stars are
identified mostly out to $\sim20$\,kpc, brighter RGB
stars out to $\sim40-50$\,kpc, while RR~Lyrae and blue horizontal branch
(BHB) stars can be detected out to 100\,kpc. Spatial clustering of
these stellar components indicate non-mixed substructure, which is
often confirmed to be kinematically coherent.

\subsubsection{The Emergence of Streams\index{Milky Way streams, SDSS, PanSTARRS}} \label{mw_streams}

\begin{figure}[h]
\includegraphics[scale=.7]{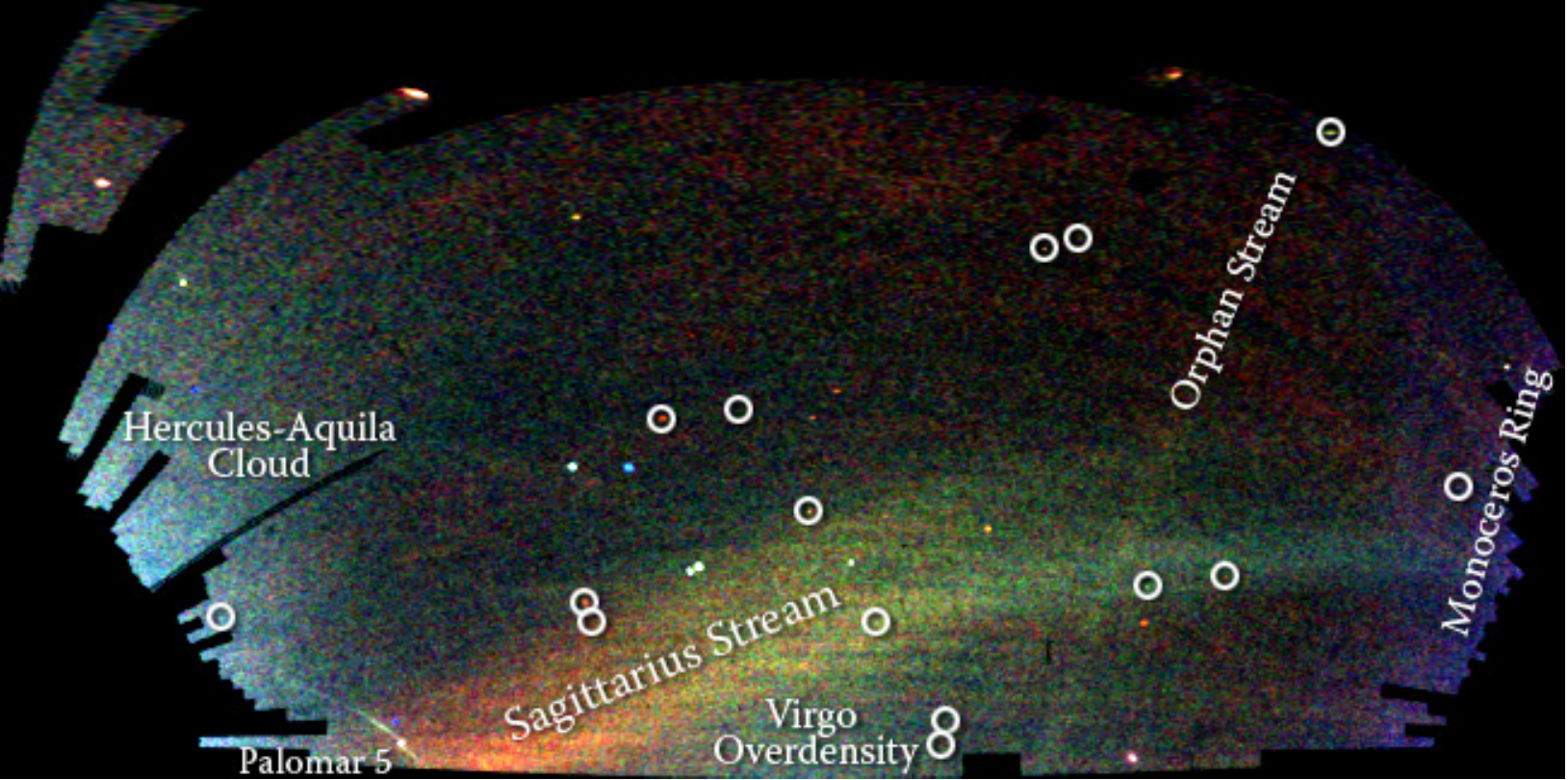}
\caption{Spatial density of SDSS stars around the Galactic cap, binned
  in $0.5\times0.5$\,deg$^2$; the colour scale is such that
  blue indicates the nearest stars while red is for the furthest
  ones. Labelled are the main halo substructures, which are in some
  cases streams associated with a GC or a dwarf galaxy;
  the circles show some newly
  discovered dwarf satellites of the MW. Plot adapted from
  \citet{belokurov06a} (http://www.ast.cam.ac.uk/$\sim$vasily/sdss/field\_of\_streams/dr6/)
}
\label{fig:fos}   
\end{figure}

\begin{figure}[h]
\includegraphics[width=12cm]{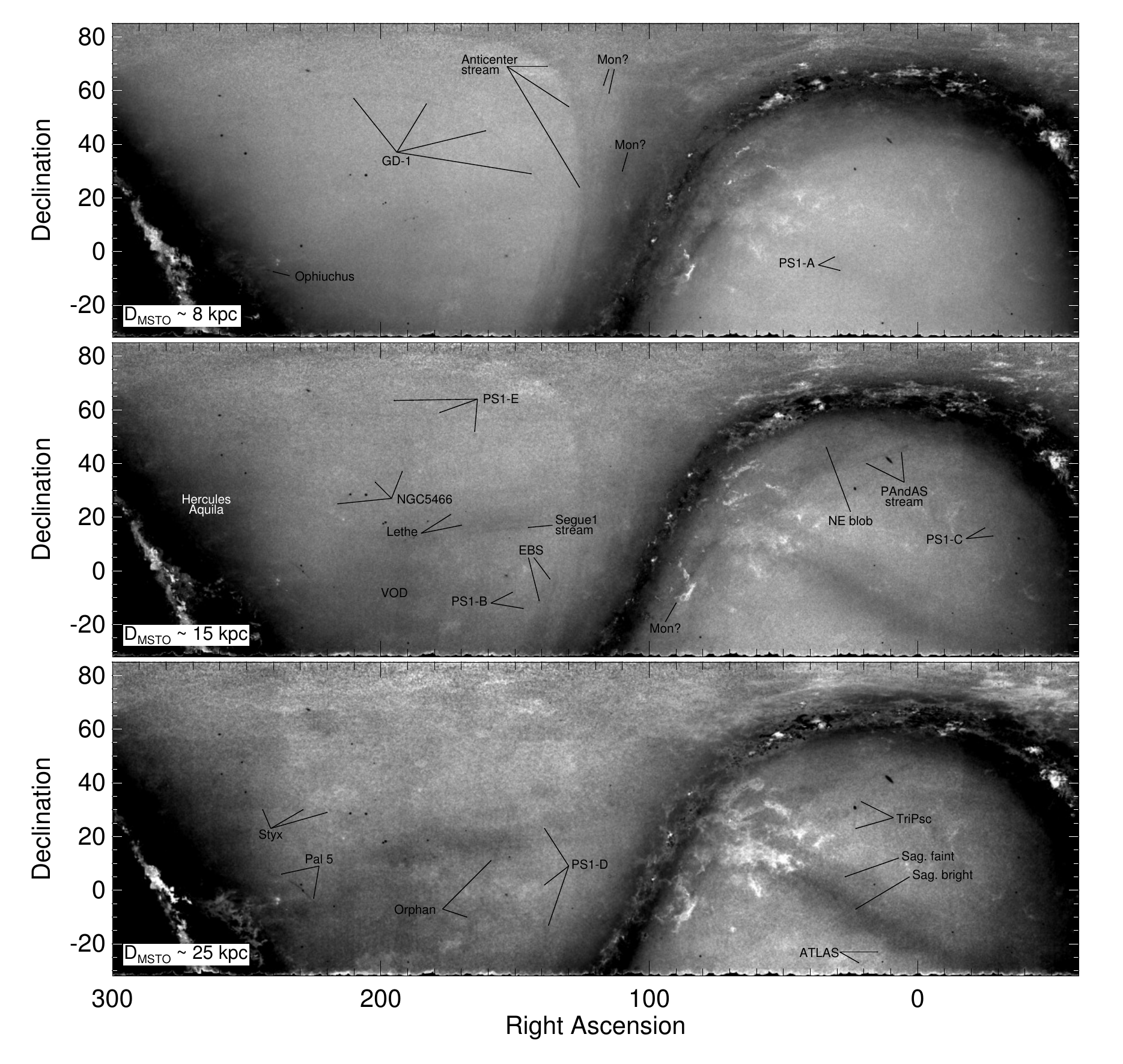}
\caption{Stellar density maps of the whole PanSTARRS
  footprint, obtained by selecting MSTO stars at
  a range of heliocentric distances (as indicated in each panel). The map is on a logarithmic scale, with darker areas
  indicating higher stellar densities. The many substructures are
  highlighted in each panel. Reproduced from \cite{bernard16}, their Fig.~1,
with permission of MNRAS}
\label{fig:panstarrs}   
\end{figure}

After the cornerstone discovery of the disrupting Sagittarius dwarf, 
it became clear that substructure is not only present in the
MW halo, but it also might constitute a big portion of it. To put it
in S. Majewski's words (\citealt{majewski99a}),
\begin{quotation} 
``There is good reason to believe that within a decade we will have a
firm handle on the contribution of satellite mergers in the formation
of the halo, as we move observationally from serendipitous discoveries
of circumstantial evidence to more systematic surveys for the fossils
left behind by the accretion process'' .
\end{quotation}
In the following decade, several stream-like features have indeed emerged
from a variety of multi-band photometric and spectroscopic 
surveys indeed, and the Sloan Digital Sky Survey (SDSS) 
proved to be an especially prolific mine for such discoveries
around the northern Galactic cap.
The Sagittarius stream has been traced further, including in the
Galactic anti-centre direction
(e.g., \citealt{mateo98, majewski03}), 
and independent substructures have been uncovered 
(\citealt{ivezic00, yanny00, newberg02, grillmair06, juric08}),
most notably the Monoceros ring, the Virgo overdensity, the
Orphan stream and the Hercules-Aquila cloud. Some of these have 
later been confirmed to be coherent with radial velocities
(\citealt{duffau06}). Note that most of these substructures are
discovered at Galactocentric distances $>15$\,kpc, while the
inner halo is smooth due to its shorter dynamical timescales.

During the past decade, one of the most stunning vizualisations of the ongoing accretion
events in the MW halo was provided by the Field of Streams (\citealt{belokurov06a}),
reproduced in Fig.~\ref{fig:fos}. The stunning stellar density map is 
derived from SDSS data of stars around the old MSTO at the distance of Sagittarius, 
with a range of magnitudes to account for a range in distances. 
This map not only shows the Sagittarius stream and its distance
gradient, but also a plethora of less massive streams, as well as an
abundance of previously unknown dwarf satellites
(see Sect.~\ref{mw_sats}).
The Field of Streams has been now complemented with results from
the latest state-of-the-art surveys, most notably the all-sky Panoramic 
Survey Telescope and Rapid Response System (PanSTARRS), which covers
an area significantly larger than that of SDSS. In
Fig.~\ref{fig:panstarrs} the first stellar density maps from
PanSTARRS are shown, obtained in a similar way as the Field of Streams
(\citealt{bernard16}). The map highlights the fact that the deeper and
wider we look at the Galaxy
halo, the more substructures can be uncovered and used to constrain
its past accretion history and the underlying DM halo properties.
From this kind of maps, for example, the halo stellar mass that 
lies in substructures can be estimated, amounting to $\sim2-3 \times
10^8\,M_\odot$ (see \citealt{belokurov13}).
Using SDSS, \cite{bell08} also highlight the predominant role of accretion in the
formation of the MW's halo based on MSTO star counts, adding to up to
$\sim40\%$ of the total halo stellar mass (note that, however,
different tracers could indicate much smaller values; e.g.,
\citealt{deason11}).

Many of the known halo streams arise from tidally disrupting
GCs, of which Palomar 5 is one of 
the most obvious examples (\citealt{odenkirchen01}). This demonstrates the
possible role of GCs, besides dwarf satellites, in building up the
halo stellar population, and additionally implies that
some of the halo GCs may be stripped remnants of nucleated accreted satellites
(see \citealt{freeman02}, and references therein).
In order to discern between a dwarf or a cluster origin of halo stars, we need
to perform chemical ``tagging'', i.e., obtain spectroscopic abundances
for tens of elements for these stars (e.g., \citealt{martell10}): stars born within the
same molecular cloud will retain the same chemical composition and
allow us to trace the properties of their birthplace. A number of
ambitious ongoing and upcoming spectroscopic surveys (SEGUE, APOGEE,
Gaia-ESO, GALAH, WEAVE, 4MOST) 
is paving the path for this promising research line, even though
theoretical models still struggle to provide robust
predictions for the fraction of GC stars lost to the MW halo
(e.g., \citealt{schiavon16}, and references therein).

\subsubsection{The Smooth Halo Component\index{Milky Way halo}} \label{mw_halo}

Once the substructure in the halo is detected, it is important that it
is ``cut out'' in order to gain insights into the smooth, in-situ stellar
component (note that, however, the latter will inevitably suffer from residual
contamination from accreted material that is now fully dissolved).
The stellar profile of the Galactic halo is, in fact, not smooth at all: several
studies have found a break at a radius $\sim25$\,kpc, with a marked
steepening beyond this value (\citealt{watkins09, sesar13}), 
in qualitative agreement with halo formation models.
Some of the explanations put forward suggest that a density break in
the halo stellar profile is the likely consequence of a massive accretion
event, corresponding to the apocentre of the involved stars
(\citealt{deason13}).

The kinematics of halo stars, of GCs and of satellite
galaxies, as well as the spatial distribution of streams and tidal
features in satellites, can be further used as mass tracers for the DM halo. 
The total MW mass is to date still poorly constrained, given the
difficulty of evaluating it with a broad range of different
tracers. The general consensus is for a virial mass value of
$\sim1.3\pm0.3 \times 10^{12}\,M_\odot$, even though values discrepant up to
a factor of two have recently been suggested (see \citealt{bland16}
for a compilation of estimates).
Besides providing estimates for the total MW mass, studies of SDSS 
kinematical data, of the Sagittarius stream and of GCs tidal
streams have provided discording
conclusions on the shape of the MW DM halo: nearly spherical 
from the modelling of streams or strongly oblate
from SDSS kinematics at Galactocentric distances $<20$\,kpc,
while nearly spherical and oblate based on stream geometry
or prolate from kinematical arguments for distances as
large as $\sim100$\,kpc (see \citealt{bland16} for details). These
constraints need a substantial improvement in the future to be able to
inform cosmological models:
the latter predict spherical/oblate shapes once baryons are included
in DM-only flattened haloes (see \citealt{read14}).

\subsubsection{Dwarf Satellites\index{Milky Way ultra-faint satellites}} \label{mw_sats}

As mentioned above, the SDSS has revolutionized our notions of dwarf
satellites of the MW. Bright enough to be easily recognized on
photographic plates, a dozen ``classical'' MW dwarf
satellites has been known for many decades before the advent of
wide-field surveys (\citealt{mateo98, grebel00}). Starting with
the SDSS, an entirely new class of
objects has started to emerge with properties intermediate between the
classical dwarfs and GCs (see \citealt{willman10}, and references therein).
The so-called ultra-faint satellites have
magnitudes higher than $M_V\sim-8$ and surface brightness values so
low that the only way to find them is to look for spatial overdensities of
resolved main sequence/BHB stars. Their discovery ten years ago doubled the
number of known MW satellites and revealed the most DM-dominated
galaxies in the Universe, with mass-to-light ratios of up to
several times $10^3\,M_\odot/L_\odot$ (\citealt{simon07}). 

More recently, the interest in the low end of the galaxy LF
has been revitalized once again with deep,
wide-field surveys performed with CTIO/DECam, VST/Omegacam, and
PanSTARRS: these have led to the discovery of more than 20 southern
dwarfs in less than two years 
(\citealt{bechtol15, koposov15, kim15a, drlica16, torrealba16},
and references therein). Some of these discoveries 
represent extremes in the properties of MW
satellites, with surface brightness values as low as $\sim30$\,mag\,arcsec$^{-2}$, total luminosities of only a few hundred $L_\odot$ and
surprisingly low stellar density regimes. One of the perhaps most
intriguing properties of the newly discovered dwarfs is that many of
them appear to be clustered around the Large Magellanic Cloud (LMC): this
might be the smoking gun for the possible infall of a group of dwarfs
onto the MW, which is predicted by simulations (\citealt{donghia08,
  sales16}). Low-mass galaxies are expected to have satellites on
their own and to provide a large fraction of a giant galaxy's dwarf
companions (e.g., \citealt{wetzel15}). The properties of the possible LMC
satellites will give us a glimpse onto the conditions of galaxy
formation and evolution in an environment much different from the LG
as we know it today.

These faintest galaxies, or their accreted and fully dispersed
counterparts, are also excellent
testbeds to look for the very most metal-poor stars and to investigate
the star formation process in the early stages of the Universe
(e.g., \citealt{frebel15}). 
The study of the lowest mass galaxies holds the promise to 
challenge our knowledge of galaxy physics even further and pushes us
to explore unexpected and exciting new limits.

\subsection{M31 (Andromeda)\index{M31 wide-field surveys, PAndAS, SPLASH}} \label{sec:m31}

Our nearest giant neighbour has received growing attention in the past
decade. Having a remarkable resemblance with the MW and a comparable
mass (e.g., \citealt{veljanoski14}), it is a natural
ground of comparison for the study of spiral haloes. 
In terms of a global perspective, the M31 halo is arguably 
known better than that of the MW: our external point of view allows us to
have a panoramic picture of the galaxy and its surrounding regions. The
other side of the medal is that, at a distance of $\sim780$\,kpc, we
can only resolve the brightest evolved stars in M31, and we are mostly
limited to a two-dimensional view of its populations. Its proximity also implies a
large angular size on the sky, underlining the need for wide
field-of-view imagers to cover its entire area.

At the distance of M31, ground-based observations are able to resolve
at best the uppermost $\sim3-4$ magnitudes below the TRGB,
which is found at a magnitude $i\sim21$. The RGB is an excellent
tracer for old ($>1$\,Gyr) populations, but suffers from a degeneracy in age and
metallicity: younger, metal-rich stars overlap in magnitude and
colour with older, metal-poor stars (\citealt{koch06}). Despite
this, the RGB colour is often used as a photometric indicator for metallicity,
once a fixed old age is assumed (\citealt{vandenberg06, crnojevic10}). 
This assumption is justified as long as a prominent young and intermediate-age population
seems to be absent (i.e., as judged from the lack of luminous main sequence and
asymptotic giant branch, AGB, stars), and it shows very good agreement with
spectroscopic metallicity values where both methods have been applied.

The very first resolved studies of M31's halo introduced 
the puzzling evidence that the M31 halo stellar populations along
the minor axis have a higher metallicity than that of the MW at similar galactocentric
distances (e.g., \citealt{mould86}). This was further confirmed by 
several studies targeting projected distances from 5 to 30\,kpc and
returning an average value of [Fe/H]$\sim-0.8$:
in particular, \cite{durrell01} study a halo region at a galactocentric
distance of $\sim20$\,kpc and underline the difference between the
properties of M31 and of the MW, suggesting that our own Galaxy might
not represent the prototype of a typical spiral. In fact, it has
later been suggested that the MW is instead fairly atypical based on
its luminosity, structural parameters and the metallicity of its halo
stars when compared to spirals of similar mass (\citealt{hammer07}). This
result was interpreted as the consequence of an abnormally quiet
accretion history for the MW, which apparently lacked a major
merger in its recent past. 

The wide-area studies of M31's outskirts were pioneered $\sim15$
years ago with an Isaac Newton Telescope survey mapping $\sim40$\,deg$^2$ around M31,
reaching significantly beyond its disk out to galactocentric distances
of $\sim55$\,kpc (\citealt{ibata01, ferguson02}). As mentioned
before, the southern Giant Stream was first uncovered with this
survey, and the halo and its substructures could be studied with a
dramatically increased detail. A metal-poor halo component
([Fe/H]$\sim-1.5$) was finally uncovered for regions beyond 30\,kpc 
and out to 160\,kpc (\citealt{irwin05, kalirai06, chapman06}),
similar to what had been observed for the MW both in terms of metallicity and
for its stellar density profile. These studies do not detect
a significant gradient in metallicity across the covered
radial range. Nonetheless, the properties of
the inner halo remained a matter of debate: 
while \cite{chapman06} found a metal-poor halo population within 30\,kpc
above the disc, \cite{kalirai06} analysed a kinematically selected
sample of stars within 20\,kpc along the minor axis and derived a
significantly higher value of [Fe/H]$\sim-0.5$. At the same time, 
\cite{brown06} used deep, pencil beam {\it Hubble Space Telescope} ({\it HST})
pointings in M31's inner halo to
conclude that a significant fraction of its stellar populations
have an intermediate age with an overall high metallicity. These
results were later interpreted by \cite{ibata07} in light of their
wider-field dataset: the samples from \cite{kalirai06} and
\cite{brown06} are simply part of regions dominated by an
extended disc component and with a high contamination from various
accretion events, respectively. This underlines, once again, the
importance of wide-field observations to reach a global
understanding of halo properties.

\begin{figure}[h]
\centering
\includegraphics[width=7.5cm]{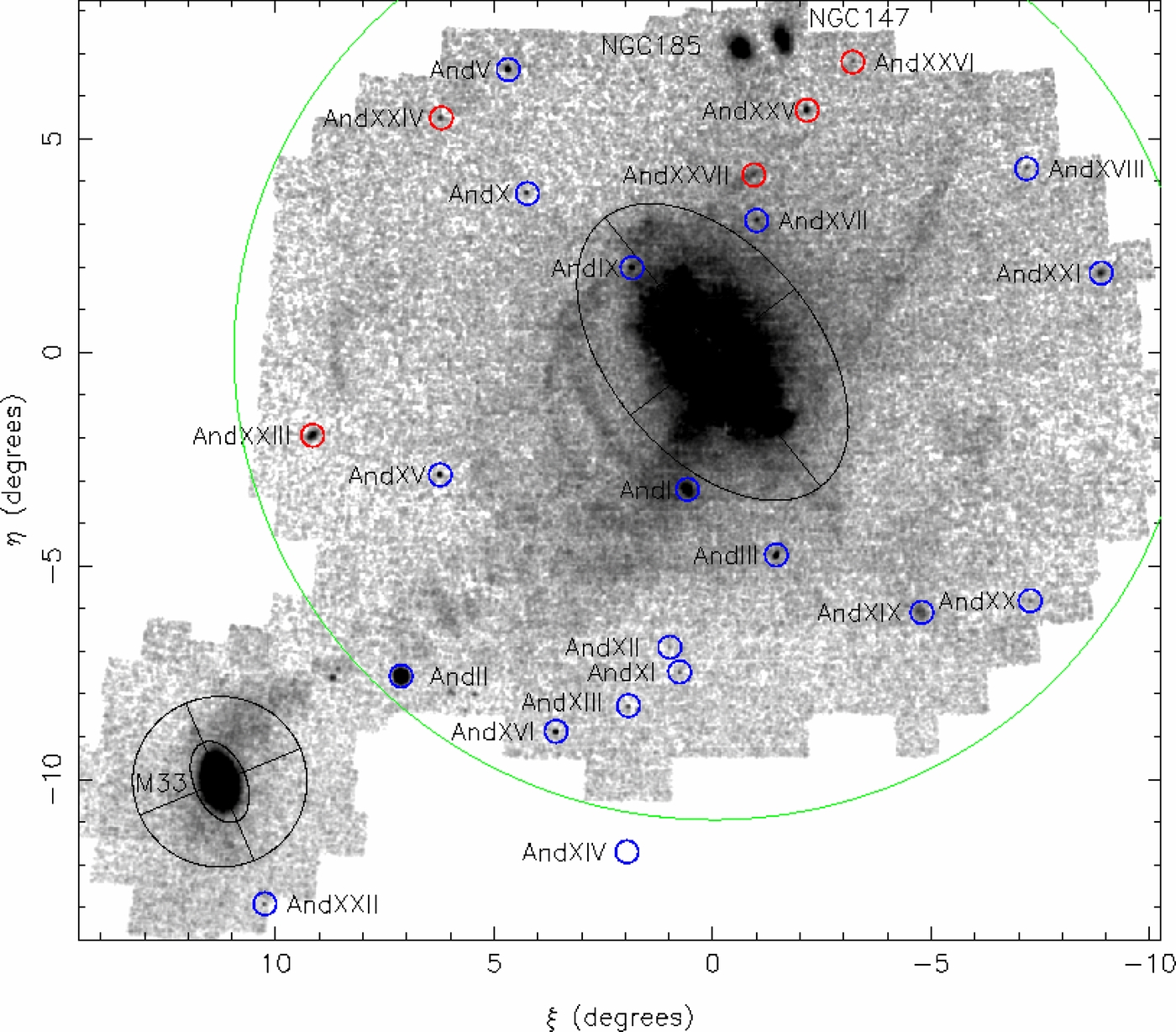}
\includegraphics[width=6.8cm]{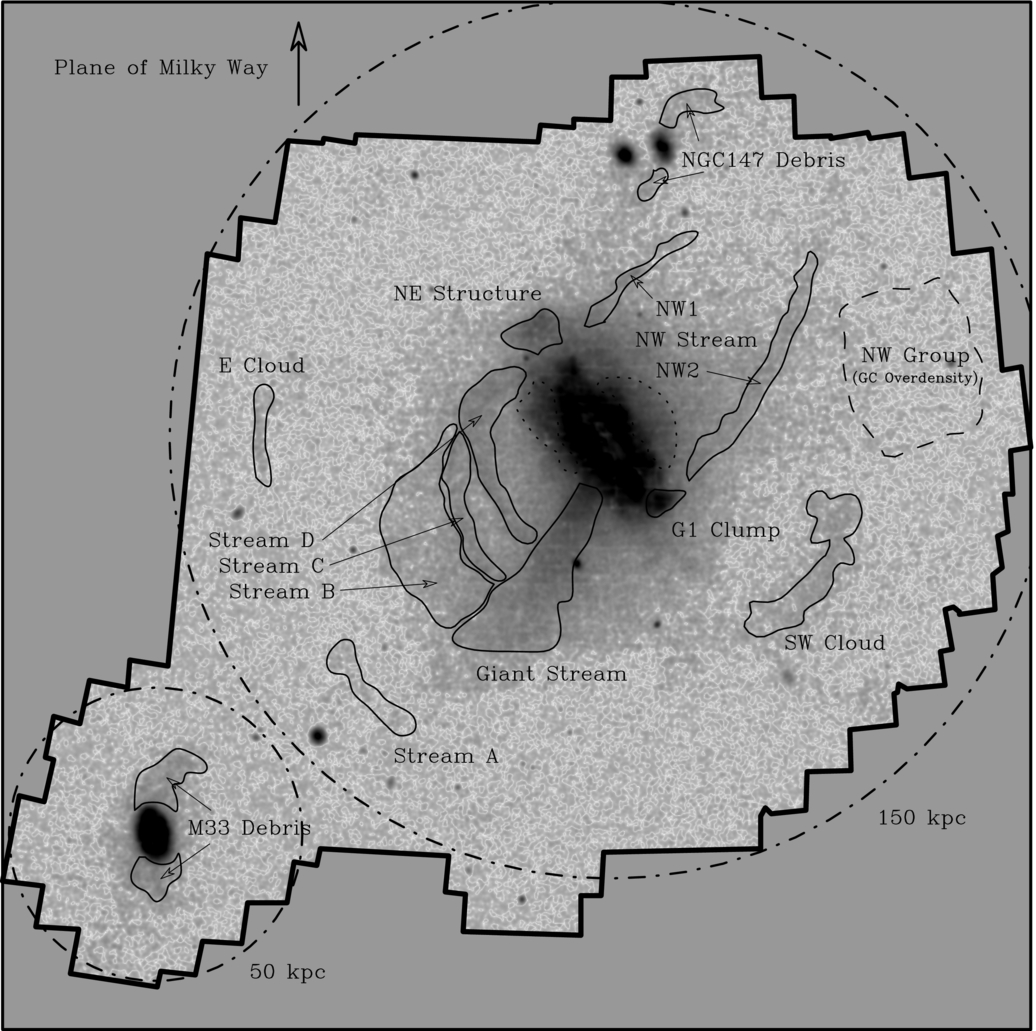}
\caption{Stellar density maps of metal-poor RGB populations at the distance
  of M31, as derived from the PAndAS survey. The large circles lie at
  projected radii of 150\,kpc and 50\,kpc from M31 and M33, respectively. 
{\it Upper panel}: The Andromeda satellites are visible as clear overdensities
  and are marked with circles. The vast majority of them were
  uncovered by the PAndAS survey. Reproduced by permission of the AAS
  from \cite{richardson11}, their Fig.~1. {\it Lower panel}: The main
  substructures around M31 are highlighted, showcasing a broad range
  of morphologies and likely progenitor type. Tidal debris is also
  present in the vicinities of the low-mass
  satellites M33 and NGC~147, indicating an ongoing interaction with M31. Reproduced
  by permission of the AAS from \cite{lewis13}, their Fig.~1}
\label{pandas1}  
\end{figure}

\begin{figure}[h]
\includegraphics[width=12cm]{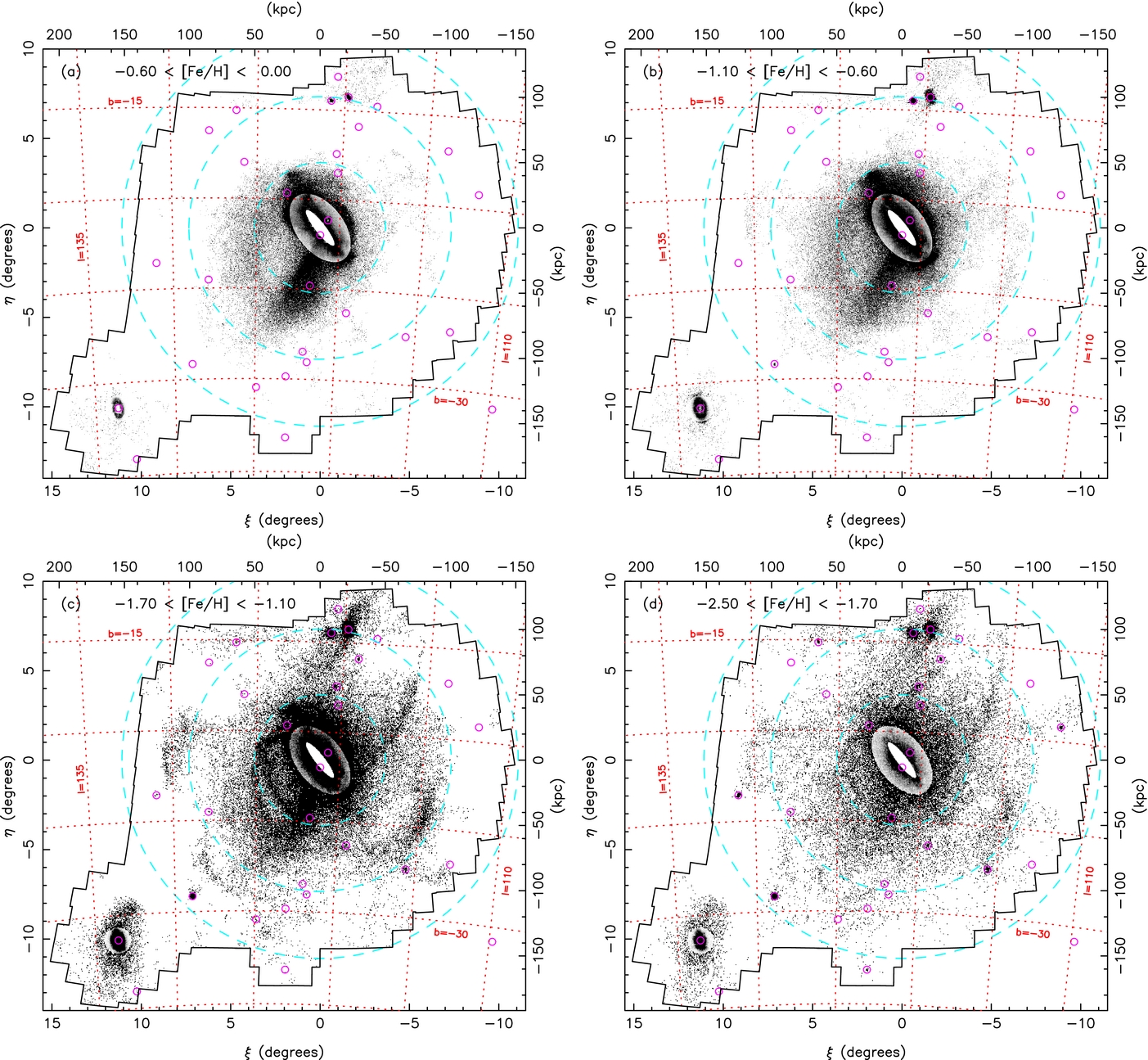}
\caption{Stellar density map of M31 (akin to Fig.~\ref{pandas1}), this
  time subdivided into photometric metallicity bins (as indicated in
  each subpanel). The {\it upper} panels show high metallicity cuts, where
  the Giant Stream and Stream C are the most prominent features; note
  that the shape of the Giant Stream changes as a function of
  metallicity. The {\it lower} panels show lower metallicity cuts: the
  {\it lower left} panel is dominated by substructure at large radii, while
  the most metal-poor panel ({\it lower right}) is smoother and believed to
  mostly contain in-situ populations. Reproduced by permission of the
  AAS from \cite{ibata14}, their Fig.~9.}
\label{pandas2}   
\end{figure}

The M31 INT survey was further extended out to $150$\,kpc (200\,kpc in
the direction of the low-mass spiral M33) with the 
Canada-France-Hawaii Telescope/Me\-ga\-cam and dubbed 
Pan-Andromeda Archaeological Survey (PAndAS; \citealt{ibata07,
  mcconnachie09}). This survey contiguously covered an impressive
380\,deg$^2$ around M31, reaching 4\,mag below the TRGB.
The PAndAS RGB stellar density map (see Fig.~\ref{pandas1})
is a striking example of an active accretion history, with a copious
amount of tidal substructure at both small and large galactocentric radii.
PAndAS also constituted a mine for the discovery of a number of very
faint satellites and GCs (see below; \citealt{richardson11, huxor14,
  martin16}). Fig.~\ref{pandas2} further shows the RGB stellar map
broken into bins of photometric metallicity.
The parallel Spectroscopic and Photometric Landscape of Andromeda's Stellar
Halo (SPLASH) survey (\citealt{guha06, kalirai06}) provides a comparison
dataset with both photometric and spectroscopic information, the
latter obtained with Keck/DEIMOS. The SPLASH pointings are
significantly smaller than the PAndAS ones but strategically cover M31
halo regions out to $\sim225$\,kpc.
Deeper, pencil-beam photometric follow-up studies have further made use of the {\it HST} to
target some of the substructures uncovered in M31's outskirts,
resolving stars down to the oldest MSTO
(e.g., \citealt{brown06, bernard15}).
These observations reveal a high complexity in the stellar
populations in M31, hinting at a high degree of mixing in its outskirts.
Overall, M31 has evidently had a much richer recent accretion history
than the MW (see also \citealt{ferguson16}).

\subsubsection{Streams and Substructures\index{M31 streams}} \label{m31_streams}

As seen from the maps in Figs.~\ref{pandas1} and \ref{pandas2}, while
the inner halo has a flattened shape and contains prominent,
relatively metal-rich substructures (e.g., the Giant Stream), the
outer halo ($>50$\,kpc) hosts significantly less extended, narrow, metal-poor 
tidal debris.

The features in the innermost regions of M31 can be connected to its
disk populations (e.g., the north-east structure or the G1 clump):
kinematic studies show that a rotational component is present in
fields as far out as 70\,kpc, and they retain a fairly high
metallicity (\citealt{dorman13}). This reinforces the possible
interpretation as a vast structure, which can be explained as
disk stars torn off or dynamically heated due to satellite accretion
events. Deep {\it HST} pointings of these features indeed reveal relatively young
populations, likely produced from pre-enriched gas in a continuous
fashion, comparable to the outer disk (\citealt{ferguson05, brown06, bernard15}).

The most prominent feature in M31's outer halo, the Giant Stream, was
initially thought to originate from the disruption of either M32 or 
NGC~205, the two dwarf ellipticals located at only $\sim25-40$\,kpc from
M31's centre (\citealt{ibata01, ferguson02}). While both these dwarfs
shows signs of tidal distortion, it was soon clear that none of them
could produce the vast structure extending $\sim100$\,kpc into M31's
halo. Great effort has been spent into mapping 
this substructure both photometrically and spectroscopically, in order
to trace its orbit and define its nature: a
gradient in its line-of-sight distance was first highlighted by
\cite{mcconnachie03}, who found the outer stream regions to be located
behind M31, the innermost regions at about the distance of M31, and
an additional stream component on the opposite (northern) side of M31
to be actually in front of M31.
The stream presents a metallicity gradient,
with the core regions being more metal-rich and the envelope more
metal-poor (see also Fig.~\ref{pandas2}), as well as a very narrow
velocity dispersion, with the addition of a
puzzling second kinematic component (\citealt{gilbert09});
possible interpretations for the latter may be a wrap or bifurcation
in the stream, as well as a component from M31's populations.

A number of increasingly sophisticated theoretical studies have tried to reproduce
the appearance of the Giant Stream and picture its progenitor, which
is undetected to date. The general consensus seems to be that a
relatively massive ($\sim10^9\,M_\odot$) satellite, possibly with a
rotating disk, impacted M31 from behind with a pericentric passage
around $1-2$\,Gyr ago (most recently, \citealt{fardal13, sadoun14}). In
particular, simulations can
reproduce the current extension and shape of the stream and predict
the progenitor to be located to the north-east of M31, just beyond its
disk (\citealt{fardal13}). This study also concludes that some of the 
substructures linked to M31's inner regions are likely to have arisen 
from the same accretion event, i.e., the north-east
structure and the G1 clump (Fig.~\ref{pandas1}): these shelf features would
trace the second and third passage around M31, which is also supported by
their radial velocities.
CMDs of the Giant Stream populations are in agreement with these
predictions: its stellar populations have mixed properties, consistent
with both disk and stream-like halo features (\citealt{ferguson05,
  richardson08}). Detailed reconstruction of its SFH indicate that most
star formation occurred at early ages, and was possibly quenched at the
time of infall in M31's potential (around 6\,Gyr ago)
(\citealt{bernard15}). Again, these studies deduce a likely origin of
these populations as a dwarf elliptical or a spiral bulge.

Besides the Giant Stream, the only other tidal feature with a
relatively high metallicity is Stream C (see Fig.~\ref{pandas1} and
\ref{pandas2}), which appears in the metal-poor RGB maps as well. The
origin of this feature is obscure, even though it is tempting to
speculate that it could be part of the Giant Stream event.
The lower left panel of Fig.~\ref{pandas2}, showing metal-poor
populations, encompasses all of the narrow streams and arcs beyond
100\,kpc, which extend for up to several tens of kpc in length. All
these substructures are extremely faint ($\mu_V\sim31.5$\,mag\,arcsec$^{-2}$), and their origin is mostly unknown because of the
difficulty in following up such faint and sparse populations.
As part of the {\it HST} imaging of these features, \cite{bernard15} find
that their populations are mainly formed at early ages and undergo a more
rapid chemical evolution with respect to the disk populations.
Despite the metal-poor nature of these features,
the hypothesis of a single accretion event producing most of the tidal
features observed in the outer halo is not that unlikely, given the
metallicity gradient present in the Giant Stream itself.

An efficient alternative to investigate the nature of these streams 
is to study the halo GC population: the wide-field surveys of M31 have
allowed to uncover a rich population of GCs beyond a radius of
$\sim25$\,kpc (e.g., \citealt{huxor14}, and references therein),
significantly more numerous than that of the MW halo. \cite{mackey10}
first highlighted a high spatial correlation between the streams in
M31's halo and the GC population, which would be extremely unlikely in
a uniform distribution. Following the hypothesis that the
disrupting satellites might be providing a high fraction of M31's halo 
GCs, \cite{veljanoski14} obtained spectroscopic follow-up: they
were able to confirm that streams and GCs often have correlated
velocities and remarkably cold kinematics. This exciting result gives
hope for studies of more distant galaxies, where halo populations
cannot be resolved and GCs could be readily used to trace possible
substructure.

\subsubsection{Smooth Halo\index{M31 halo}} \label{m31_halo}

One of the first spatially extended datasets to investigate the halo
of M31 in detail is described in \cite{tanaka10}: their Subaru/SuprimeCam
photometry along the minor axis in both directions are deeper, even
though less extended, than PAndAS. The stellar density profile derived
in this study extends out to 100\,kpc and shows a consistent power law
for both directions. The authors also suggest that, given the
inhomogeneities in the stellar populations, the M31 halo is likely
not fully mixed.

In the most metal-poor (lower right) panel of Fig.~\ref{pandas2}, 
the substructures in the outer halo fade away, displaying a smoother
component that can be identified with the in-situ M31 halo. Once the
substructures are decoupled based on the lack of obvious spatial
correlation and with an additional photometric metallicity cut,
\cite{ibata14} derive a stellar density profile out to 150\,kpc. Again,
the profile follows a power-law, which turns out to be steeper when
increasingly more metal-rich populations are considered.
\cite{ibata14} also conclude that only $5\%$ of M31's total halo
luminosity lies in its smooth halo, and the halo mass is as high as
$\sim10^{10}\,M_\odot$, significantly larger than what estimated for
the MW.

The SPLASH survey extends further out than PAndAS, and benefits from
kinematical information that is crucial to decontaminate the studied
stellar samples from foreground stars and decreases the scatter in the
radial profiles. Based on this dataset, 
\cite{gilbert12} find that the halo profile does not reveal 
any break out to 175\,kpc. This is somewhat surprising given the
prediction from simulations that accreted M31-sized stellar haloes
should exhibit a break beyond a radius of $\sim100$\,kpc
(\citealt{bullock05, cooper10}). Beyond a radius of 90\,kpc, significant
field-to-field variations are identified in their data, which suggests
that the outer halo regions are mainly comprised of stars from
accreted satellites, in agreement with previous studies. At the
outermost radii probed by SPLASH ($\sim230$\,kpc), there is a tentative
detection of M31 stars, but this is hard to confirm given the high
contamination fraction. Finally, the \cite{gilbert12} stellar halo profile
suggests a prolate DM distribution, which is also consistent with being
spherical, in agreement with \cite{ibata14}.

Both \cite{ibata14} and \cite{gilbert14} investigate the existence of
a metallicity gradient in the smooth halo of M31: they found a
steady decrease in metallicity of about 1\,dex from the very inner
regions out to 100\,kpc. This might indicate the past accretion of (at
least) one relatively massive satellite. At the same time, a large
field-to-field metallicity variation could mean that the outer halo
has been mainly built up by the accretion of several smaller progenitors.

\subsubsection{Andromeda Satellites\index{M31 satellites}} \label{m31_sats}

Similarly to the boom of satellite discoveries around the MW, the vast
majority of dwarfs in M31's extended halo has been uncovered by the SDSS,
PAndAS, and PanSTARRS surveys in the past decade 
(see \citealt{martin16}, and references therein).
The M31 satellites follow the same relations between
luminosity, radius and metallicity defined by MW satellites, with the
exception of systems that are likely undergoing tidal disruption
(\citealt{collins14}).
Once more, the characterization of the lowest-mass galaxies raises new,
unexpected questions: from the analysis of accurate distances and kinematics,
\cite{ibata13} conclude that half of the M31 satellites lie in a
vast ($\sim200$\,kpc) and thin ($\sim12$\,kpc) corotating plane, and
share the same dynamical orbital properties. The extreme thinness of
the plane is very hard to reconcile with $\rm \Lambda$CDM predictions, 
where such structures should not survive for a Hubble time. While
several theoretical interpretations have been offered
(e.g., \citealt{fernando16}), none is conclusive, and this reinforces the allure of
mystery surrounding low-mass satellites.

\subsection{Low-mass Galaxies In and Around the Local Group\index{Nearby dwarf galaxies}} \label{dwarfs_halo}

Besides the detailed studies of the two LG spirals, increasing
attention is being paid to lower-mass galaxies and their outskirts.
Given the self-similar nature of DM, low-mass galaxies should naively
be expected to possess haloes and satellites of their own; however,  
our difficulty in constraining star formation efficiency and physical
processes affecting galaxy evolution at these scales blurs these
expectations. In the last couple of years, the increasing resolution
of cosmological simulations has allowed to make quantitative
predictions about the halo and substructures in sub-MW-mass galaxies,
and about the number of satellites around them
(\citealt{wheeler15, dooley16}). Observations are thus much
needed to test these predictions.

Since the late 90s, numerous studies of star-forming dwarfs within or
just beyond the LG have claimed the detection of an RGB component extending beyond the
blue, young stars (see \citealt{stinson09}, and references therein),
hinting at a generic mode of galaxy formation independent on galaxy size.
Such envelopes, however, were not characterized in detail, and in fact
could not be identified uniquely as the product of hierarchical
merging without, e.g., accurate age and metallicity estimates.

The presence of extended haloes in the most luminous satellites of the
MW and M31, i.e., the irregular LMC and the low-mass spiral M33,
respectively, has not been confirmed to
date despite the availability of exquisite datasets.
\cite{gallart04} demonstrate how, out to a galactocentric distance of
7\,kpc, the stellar density profile of the LMC disk does not show a clear
break, in contrast to previous tentative claims. Clearly, the question
is complicated by the fact that the LMC is undergoing tidal
disruption, and stripped stellar material could easily be
misinterpreted as a halo component. Nonetheless, \cite{mcmonigal14}
suggest to have found a sparse LMC halo population from a wide-field
dataset around the nearby dwarf galaxy Carina, at galactocentric distances as large
as $20$\,deg. The question might be settled in the near future with the
help of wide-field surveys such as the Survey of MAgellanic Stellar History (\citealt{martin15}). With
regard to possible low-mass satellites, there is now tantalizing indication that the LMC might
have fallen onto the MW with its own satellite system, as mentioned in
Sect.~\ref{mw_sats}.
As part of the PAndAS survey, deep imaging of M33 has revealed
prominent substructure in its outer disk reminiscent of a tidal
disturbance, and a faint, diffuse substructure possibly identified as a halo
component (\citealt{cockcroft13}). This result was, however, carefully reconsidered by
\cite{mcmonigal16}, who claim that a definitive sign of a halo structure
cannot be confirmed, and if present it must have a surface brightness
below $\mu_V\sim35$\,mag\,arcsec$^{-2}$.

Besides the investigation of haloes and satellites, deep and wide-field views
of low-mass galaxies are crucial to, e.g., assess the presence of tidal
disturbances, which in turn are key to estimate mass values and
constrain DM profiles (e.g., \citealt{sand12}). As demonstrated by
\cite{crnojevic14a}, a striking similarity in the global properties
(luminosity, average metallicity, size) of two
low-mass galaxies, such as the M31 satellites NCG~185 and
NGC~147, can be quite misleading: once deep imaging was obtained around
these galaxies (within PAndAS), NCG~147 revealed extended, symmetric
tidal tails, returning a much larger extent and luminosity for this
dwarf than what was previously thought. This
dataset further showed a flat metallicity gradient for NGC~147, in
contrast with the marked gradient found in NGC~185. All these pieces of evidence
point at an ongoing interaction of NGC~147 with M31.
Large-scale studies of LG dwarfs also provide
useful insights into their evolutionary history: by studying CMDs
reaching below the MSTO, \cite{hidalgo13} trace significant age gradients that
advocate an outside-in mode of star formation for dwarf galaxies.

Clearly, systematic deep searches are needed to detect and
characterize the outskirts of low-mass satellites. With this goal in
mind, wide-field surveys of nearby ($<3$\,Mpc) dwarfs have started to be
pursued. The first of these efforts targets NGC~3109, a sub-LMC-mass
dwarf located just beyond the boundaries of the LG: several candidate
satellites of NGC~3109 are identified from a CTIO/DECam survey targeting regions out to
its virial radius (\citealt{sand15}). One of them, confirmed to be at the
distance of NGC~3109, is relatively bright ($M_V\sim-10$),
and is already in excess of the predicted number by \citet{dooley16}
for this system. Other ongoing surveys are similarly looking for
halo substructures and satellites in several relatively isolated dwarfs,
e.g., the SOlitary LOcal dwarfs survey (\citealt{higgs16}) and
the Magellanic Analog Dwarf Companions And Stellar Halos
survey (\citealt{carlin16}), by using wide-field imagers on large
telescopes such as CFHT/MegaCam, Magellan/Megacam, CTIO/DECam and
Subaru/HyperSuprimeCam. These datasets will constitute a mine of
information to constrain the role of baryonic processes at the
smallest galactic scales. 

\section{Beyond the Local Group\index{Local Volume galaxies, Halo-to-halo scatter}} \label{sec:beyondlg}

The ground-breaking photometric and kinematic surveys carried out in
the past two decades have significantly advanced our knowledge of 
haloes and their substructures within LG galaxies. 
Nonetheless, the MW and M31 may not be representative
of generic MW-sized haloes, given the stochasticity of the 
hierarchical assembly process: several marked differences in the
stellar populations of their haloes underline the need for observations
of a statistically significant sample of galaxy haloes with different
morphologies, with surveys targeting large portions of their haloes.

Cosmological simulations of  MW-mass analogues show a wide variation in
the properties of their haloes.
As already mentioned, the relative contribution of in-situ
star formation and disrupted satellites remains unclear: depending on
the models (e.g., full hydrodynamical simulations, $N$-body models with
particle tagging), they can vary from a negligible number of
accretion events for a MW-sized halo, to making up for most of a
stellar halo content (e.g., \citealt{lu14, tissera14}).
Even within the same set of simulations,
the number, mass ratio and morphology of accretion and merger
events span a wide range of possible values 
(\citealt{bullock05, johnston08, garrison14}). The chemical content of
extended haloes can provide useful insights into their assembly
history: mergers or accretion events of similar-mass satellites 
will generally tend to produce mild to flat
gradients; in-situ populations will feature increasingly
metal-poor populations as a function of increasing galactocentric
radius, similarly to the accretion of one or two massive companions
(e.g., \citealt{cooper10, font11}).
More extended merger histories are also expected to return younger and
relatively metal-rich populations with respect to those coming from a
shorter assembly, and to produce more massive stellar haloes, with the
final result that the mean halo metallicities of MW-mass spirals can
range by up to 1\,dex (e.g., \citealt{renda05}).

Comprehensive observational constraints are key to guide future
simulations of galaxy haloes: the past decade has seen a 
dramatic increase in the observational census of resolved galaxy haloes
beyond the LG, thanks to deep imaging obtained with space facilities,
as well as to the advent of wide-field imagers on large ground-based
telescopes.

While the increasing target distance means that it is easier to survey
larger portions of their haloes, the drawback is that the depth of the
images decreases dramatically, and thus we are only able to detect the
brightest surface brightness features in the haloes, i.e., the
uppermost $\sim2-3$\,mag below the TRGB in terms of resolved stars
(see Fig.~6 in \citealt{radburn11} for a schematic visualization of 
the different stellar evolutionary phases recognizable in such shallow CMDs).
A number of studies has surveyed relatively nearby and more distant 
galaxy haloes in integrated light despite the serious challenges posed
by sky subtraction at such faint magnitudes, masking of bright stars, 
flat-fielding and scattered light effects, point spread function modelling,
and/or spatially variable Galactic extinction.

A few early studies have been able to uncover a halo
component and tidal debris in the target galaxies
(e.g., \citealt{malin83, morrison94,
  sackett94}), without, however, settling the questions about their
existence, nature or ubiquity.
Different approaches have been adopted to detect haloes and their
substructures, i.e., targeting either individual galaxies 
(e.g., \citealt{zheng99, pohlen04, jablonka10,
  janowiecki10, martinez10, adams12, atkinson13}) 
or stacking the images of thousands of objects 
(e.g., \citealt{zibetti04, vandokkum05, tal09}).
A precise quantification of the occurrence of faint substructure 
in the outskirts of nearby galaxies seems as uncertain as it can be, 
ranging from a few percent to $\sim70\%$
(see, e.g., \citealt{atkinson13}, and references therein).
This is perhaps unsurprising given the heterogeneity of 
methods used, target galaxy samples, and surface brightness
limits in such studies.
Besides the identification of such features, the characterization of
unresolved halo stellar populations constitutes an even harder challenge:
integrated colours and spectra can at most reach a few effective radii,
thus missing the outer haloes. Even for the available datasets, the
degeneracies between age, metallicity and extinction are generally
challenging to break (e.g., \citealt{dejong07}); in addition, tidal
features can rarely tell us about the mass ratio of a merger event or
its orbit (with the exception of tails).
Here, we do not intend to discuss the detection of haloes and the variety of 
fractions and morphologies for tidal features observed in 
integrated light studies; Knapen \& Trujillo (this volume)
treat this topic in detail, 
while this contribution focusses on resolved populations.

Obtaining resolved photometry beyond the LG is a daunting task as
well, due to the very faint luminosities involved---the brightest
RGB stars for galaxies at \mbox{$\sim4-10$\,Mpc} have magnitudes of
$I\sim24-28.5$, and thus this approach is so far really
limited to the Local Volume.
Early attempts to perform photometry of individual stars in the outskirts
of nearby galaxies have been made using large photographic plates and
the first CCDs (e.g., \citealt{humphreys86, davidge89, georgiev92}). 
The brightest populations (i.e., the youngest)
could often be reconciled with being members of the parent galaxy, but 
the critical information on the faint, old stars was still out of reach.
With the advent of wide-format CCDs in the mid 90s, photometry finally
became robust enough to open up new perspectives on the resolved stellar 
content of our closest neighbours.
 
The first studies of this kind date back to twenty years ago and
mainly focus on the inner regions of the 
target galaxies, most commonly their disks or inner haloes, with the goal of 
studying their recent star formation and of deriving TRGB
distances (see, e.g., \citealt{soria96} for CenA, \citealt{sakai99}
for M81 and M82). 
\cite{elson97}, in particular, resolved individual stars in the halo
of the S0 galaxy NGC~3115 with {\it HST}. By analysing the uppermost 1.5\,mag
of the RGB at a galactocentric distance of 30\,kpc, they derived a
distance of $\sim11$\,Mpc, and additionally
discovered for the first time a bimodality in the photometric
metallicity distribution function of this early-type galaxy.
\cite{tikhonov03} studied for the first time the resolved 
content of the nearest ($\sim3.5$\,Mpc) S0 
galaxy NGC~404 with combined ground-based and {\it HST} imaging. 
Their furthermost {\it HST} pointings ($\sim20$\,kpc in projection) 
contain RGB stars that are clearly older than the main disk 
population, with similar colour (metallicity). The authors 
conclude that the disk of NGC~404 extends out to this 
galactocentric distance, but they do not mention a halo component.

Beyond these early studies of individual galaxies, the need for
systematic investigations of resolved stellar haloes was soon recognized. Next we
describe the design and results of some systematic surveys targeting
samples of galaxies in the Local Volume.

\subsection{Systematic Studies\index{Systematic halo studies, GHOSTS}} \label{systematic}

A decade ago, \cite{mouhcine05a, mouhcine05b, mouhcine05c} 
started an effort to systematically observe the haloes of eight 
nearby ($<7$\,Mpc) spiral galaxies with the resolution of {\it HST}. In particular,
they utilized WFPC2 to target fields off of the 
galaxies' disks (2 to 13\,kpc in projection along the minor axis) 
with the goal of investigating their stellar populations, and obtaining 
accurate distance estimates as well as photometric metallicity
distribution functions, to gain insights into
the halo formation process.
\cite{mouhcine05c} find the haloes to predominantly contain
old populations, with no younger components and little
to no intermediate-age populations.
Interestingly, \cite{mouhcine05b} find a correlation
between luminosity and metallicity for the target galaxies,
where the metallicity is derived from the mean colour of 
the resolved RGB. Both the spiral galaxies from their sample
(NGC~253, NGC~4244, NGC~4945, NGC~4258, NGC~55, NGC~247,
NGC~300, and NGC~3031 or M81) and the two ellipticals
(NGC~3115 and NGC~5128 or Centaurus~A, included in their 
comparison from previous literature data)
fall on the same relation, indicating that haloes might
have a common origin regardless of the galaxy morphological
type. Interestingly enough, the MW halo turns out to be 
substantially more metal-poor than those of the other galaxies
of comparable luminosity, based on kinematically
selected pressure-supported halo stars within $\sim10$\,kpc
above the disk (see also Sect.~\ref{sec:m31}). This relation is
consistent with
a scenario where halo field stars form in the potential
well of the parent galaxy in a gradual way from pre-enriched
gas. Moreover, the relatively high metallicities of the
target haloes seem to suggest that they likely originate from
the disruption of intermediate-mass galaxies, rather than
smaller metal-poor dwarf galaxies (\citealt{mouhcine05c}).

Interestingly, the dataset presented and studied in
\cite{mouhcine05a, mouhcine05b, mouhcine05c} is further analyzed by
\cite{mouhcine06} to find that each spiral of the sample
presents a bimodal metallicity distribution. In particular,
both a metal-poor and a metal-rich component are present in the 
outskirts of the target galaxies, and both components correlate
with the host's luminosity. This is taken as a hint that these
populations are born in subgalactic fragments that were already
embedded in the dark haloes of the host galaxy; the metal-poor
component additionally has a broader dispersion than that of
the metal-rich population. These properties show similarities
with GC subpopulations in the haloes of early-type 
galaxies (e.g., \citealt{peng06}). \cite{mouhcine06} argues that the metal-poor
component may arise from the accretion of low-mass satellites,
while the metal-rich one could be linked to the formation of
the bulge or the disk.

The shortcoming of this ambitious study is, however, twofold: first,
the limited field of view (FoV) of {\it HST} hampers global conclusions on the galaxies' haloes,
and the stellar populations at even larger radii may have
different properties than those in the observed fields;
second, perhaps most importantly, it is not obvious what structure
of the galaxy is really targeted, i.e., the halo, the outer 
bulge or disk, or a mixture of these.

Along the same lines of these studies,
\cite{radburn11} present an even more ambitious {\it HST} survey of
14 nearby disk galaxies within 17\,Mpc, with a range of
luminosities, inclinations and morphological types.
The Galaxy Halos, Outer disks, Substructure, Thick disks, and Star
clusters (GHOSTS) survey aims at investigating radial light profiles, axis
ratios, metallicity distribution functions (MDFs), SFHs, possible
tidal streams and GC populations, all to be considered as a function 
of galaxy type and position within the galaxies.
The 76 ACS pointings of the survey are located along both major and minor axes
for most of the targets, and reach $\sim2-3$\,mag below the TRGB,
down to surface brightness values of $V\sim30$\,mag\,arcsec$^{-2}$.
This dataset thus represents a very valuable resource for testing
hierarchical halo formation models.
\cite{monachesi16} investigate six of the galaxies in this sample 
(NGC~253, NGC~891, M81, NGC~4565, NGC~4945, and NGC~7814)
and conclude that all of them contain a halo component out to 50\,kpc, and
two of them out to 70\,kpc along their minor axis. The colour (i.e.,
photometric metallicity) distribution of RGB stars in the target haloes
is analysed and reveals a non-homogeneity which likely indicates the
presence of non-mixed populations from accreted objects. The average
metallicity out to the largest radii probed remains relatively high
when compared to the values of the MW halo; metallicity gradients are
also detected in half of the considered galaxies.
Surprisingly, and in contrast to the results presented by
\cite{mouhcine05b}, the spiral galaxies in this sample do not show a
strong correlation between the halo metallicity and the total mass of the
galaxies, highlighting instead the stochasticity inherent to the halo
formation process through accretion events (e.g., \citealt{cooper10}). 
The advantage of the GHOSTS dataset over the one
from \cite{mouhcine05b} is that the GHOSTS fields are deeper, there
are several pointings per galaxy and they reach significantly larger
galactocentric distances, thus offering a more global view of the
haloes of the targets.

In an effort to increase the sample of nearby galaxies for which
stellar haloes are resolved and characterized, several groups have 
individually targeted Local Volume objects with either
ground-based or space-borne facilities: the low-mass spirals
NGC~2403 (\citealt{barker11}, with Subaru/SuprimeCam),
NGC~300 (\citealt{vlajic09}, with Gemini/GMOS), and
NGC~55 (\citealt{tanaka11}, with Subaru/SuprimeCam),
the ellipticals NGC~3379 (\citealt{harris07a}, with {\it HST})
and NGC~3377 (\citealt{harris07b}, with {\it HST}),
and the lenticular NGC~3115 (\citealt{peacock15}, with {\it HST}).
In most of these galaxies, a resolved faint halo 
(or at least an extended, faint and diffuse component) has been detected
and is characterized by populations more metal-poor than the central/disk
regions. Most of these haloes also show signs of substructure, pointing at
past accretion/merger events as predicted by a hierarchical galaxy
formation model.
Even galaxies as far as the central elliptical of the Virgo cluster, M87,
($\sim16$\,Mpc) are starting to be targeted with {\it HST}, although pushing 
its resolution capabilities to the technical limits (\citealt{bird10}).

While spectroscopically targeting individual RGB stars to obtain radial
velocity and metallicity information is still prohibitive beyond
the LG (see Sect.~\ref{dwarfs_halo}), some cutting-edge studies have pushed
the limits of spectroscopy for dwarf galaxies within $\sim1.5$\,Mpc 
(e.g., \citealt{kirby12}, and references therein).
At the same time, novel spectroscopic techniques are being developed
to take full advantage of the information locked into galaxy
haloes. One example is the use of co-added spectra of individual stars,
or stellar blends, to obtain radial velocities, metallicities and
possibly gradients in galaxies within $\sim4$\,Mpc, as robustly
demonstrated by \cite{toloba16}. The development of new analysis
methods and the advent of high-resolution spectrographs will soon allow for 
systematic spectroscopic investigations of nearby galaxy haloes which will
importantly complement the available photometric studies, similarly to
the studies of LG galaxies.

Besides the systematic studies presented here, which mostly involve deep space
observations, an increasing effort is being invested in producing
spatial density maps of outer haloes in some of the closest galaxies
with ground-based observations, akin to the panoramic view of M31
offered by PAndAS. In the following Section we describe some of these
efforts.

\subsection{Panoramic Views of Individual Galaxies\index{Panoramic halo surveys, PISCeS}} \label{panoramic}

Panoramic views of nearby galaxies can be obtained with the use of
remarkable ground-based wide-field imagers such as Subaru/SuprimeCam
and Hyper\-Suprime\-Cam and CFTH/MegaCam in the northern hemisphere,
and Magellan/Megacam, CTIO/DECam and VISTA/VIRCAM in the southern hemisphere. Clearly,
such CMDs cannot reach the depth of those obtained for M31; these
studies nevertheless represent cornerstones for our investigation of
global halo properties, and serve as precursor science cases for the
next generation of telescopes that will open new perspectives for this kind of
studies to be performed on a significantly larger sample of galaxies.
As mentioned in Sect.~\ref{dwarfs_halo}, the haloes of low-mass
galaxies are also starting to be systematically investigated, to gain
a more complete picture of galaxy formation at all mass scales.
Here we further describe the few examples of spatially extended imaging
obtained to date for some of the closest spiral and elliptical galaxies.

\subsubsection{NGC~891}

\begin{figure}[h]
\centering
\sidecaption
\includegraphics[width=7.cm]{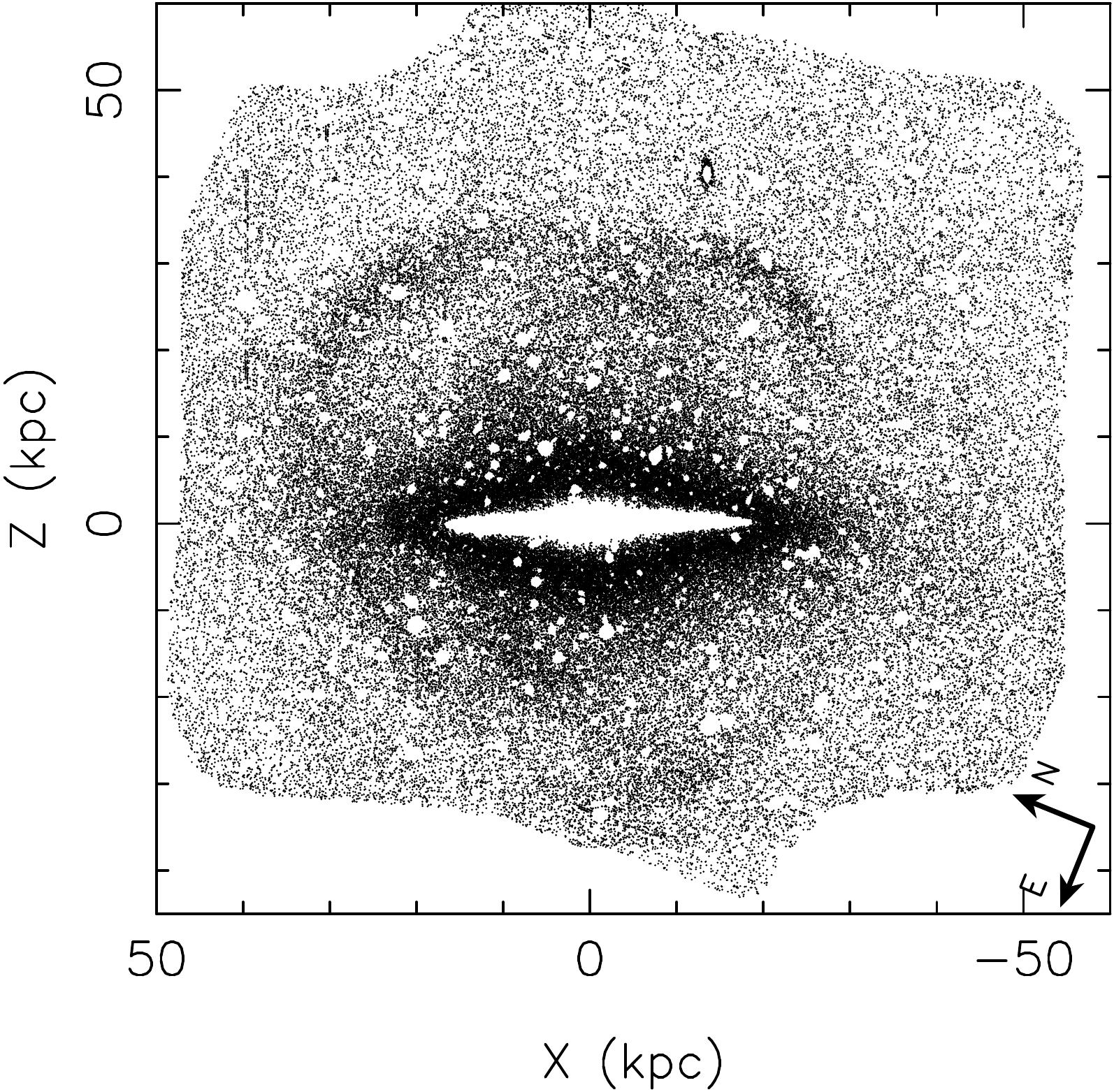}
\caption{Surface density map of RGB stars in the halo of NGC~891,
obtained with Subaru/SuprimeCam. The overdensities of old RGB stars
reveal a large complex of arcing streams that loops around the galaxy, tracing
the remnants of an ancient accretion. The second spectacular morphological
feature is the dark cocoon-like structure enveloping the high surface brightness
disk and bulge. Fig.~1 from \cite{mouhcine10}, reproduced by permission of the AAS}
\label{fig:mouhcine10} 
\end{figure}

Despite its relatively large distance ($\sim9$\,Mpc, \citealt{radburn11}), 
the ``MW-twin'' NCG~891 (\citealt{vanderkruit84}) is one of the first spirals to be 
individually investigated in resolved light. Its high inclination and 
absence of a prominent bulge make it an appealing target for halo studies. 

\citet{mouhcine07} exploit three {\it HST} pointings located approximately
10\,kpc above the disk of NGC~891 to investigate the properties of this
galaxy's halo. The broad observed RGB indicates a wide range of
metallicities in this population, with metal-rich peaks and 
extended metal-poor tails. The three fields also show a decreasing 
mean metallicity trend as a function of increasing distance 
along the major axis.
The mean metallicity of this sample of RGB stars ([Fe/H]$\sim-1$) 
falls on the halo metallicity-galaxy luminosity relation 
pointed out by \cite{mouhcine05b}: this, together with the gradient
mentioned before, is in contrast with the lower metallicities
and absence of a gradient for non-rotating stars in the inner haloes of 
the MW and M31 (\citealt{chapman06, kalirai06}). \citet{mouhcine07} thus suggest 
that not all massive galaxies' outskirts
are dominated by metal-poor, pressure-supported stellar populations
(because of the inclination and absence of a bulge, the studied 
RGB sample is thought to be representative of the true halo
population). A possible explanation is suggested with the presence
of two separate populations: a metal-rich one that is present in the most massive 
galaxies' outskirts, and one constituting the metal-poor, pressure-supported
halo, coming from the accretion of moderate-mass satellites.
For smaller-mass galaxies, the halo would instead be dominated
by debris of small satellites with lower metallicities.

Follow-up analysis on the same {\it HST} dataset has been carried out 
by \cite{ibata09} and \cite{rejkuba09}. After careful accounting for the 
internal reddening of the galaxy, a mild metallicity gradient
is confirmed in NGC~891's spheroidal component, which is surveyed out 
to $\sim20$\,kpc (assuming elliptical radii), and suggested to arise 
from the presence of a distinct outer halo, similarly to the MW (\citealt{ibata09}). 
Most importantly, and for the first time, this refined analysis reveals a substantial 
amount of substructure not only in the RGB spatial distribution 
but also as metallicity fluctuations in the halo of NGC~891. This evidence 
points at multiple small accretion events that have not fully blended 
into the smooth halo. 

Motivated by these studies, \citet{mouhcine10} provide the first attempt to derive a 
PAndAS-like map of a MW-analogue beyond the LG: their wide-field map of NGC~891's
halo is shown in Fig.~\ref{fig:mouhcine10}. The panoramic survey, performed
contiguously with Subaru/SuprimeCam, covers an impressive $\sim90\times90$\,kpc$^2$ 
in the halo of NGC~891 with the $V$ and $i$ filters, reaching $\sim2$\,mag
below the TRGB. Among the abundant substructures uncovered by the RGB map around
NGC~891, a system of arcs/streams reaches out some $\sim50$\,kpc into the halo,
including the first giant stream detected beyond the LG with ground-based
imaging. The latter's shape does not rule out a single accretion event
origin, but a possible progenitor cannot be identified as a surviving stellar
overdensity. These structures appear to be old, given the absence of corresponding
overdensities in the luminous AGB (i.e., intermediate-age populations) maps.
Another surprising feature highlighted by the RGB map is a flattened, super-thick
envelope surrounding the disk and bulge of NGC~891, which does not seem to
constitute a simple extension of its thick disk but is instead believed to 
generate from the tidal disruption of satellites given its non-smooth
nature (\citealt{ibata09}).

\subsubsection{M81} \label{m81}

\begin{figure}
\centering
\sidecaption
\includegraphics[width=7.cm]{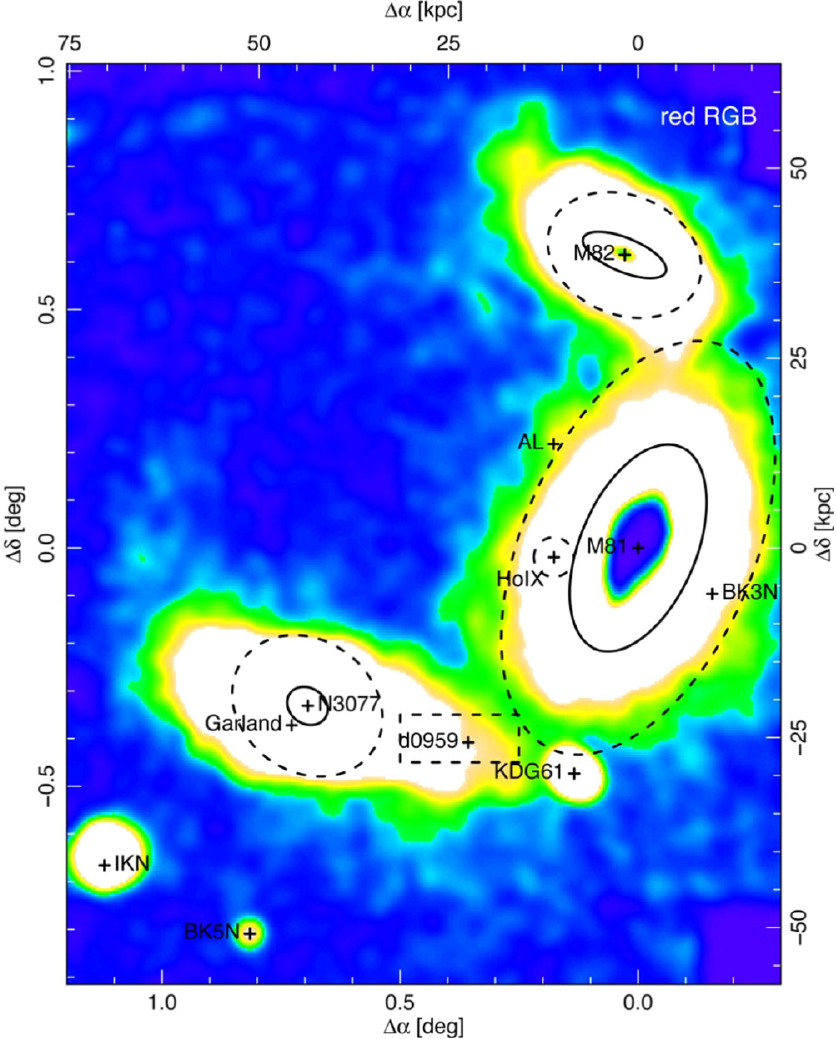}
\caption{Isodensity contour map of red RGB stars in the M81 group, as
  observed by Subaru/HyperSuprimeCam. Structures up to $20\sigma$
  above the background level are visible; the cross marks represent
  the centres of known M81 group members, while solid lines are
  ${R}_{25}$ of galaxies. The high degree of substructure underlines
  the ongoing tidal interactions in this group; note in particular the
  S-shape of the outer regions in NGC~3077 and M82. 
  Fig.~4 from \cite{okamoto15}, reproduced by
  permission of the AAS}
\label{fig:okamoto}  
\end{figure}

Located at a distance of 3.6\,Mpc (\citealt{radburn11}) and with a dynamical mass inside 
20\,kpc of $\sim10^{11}\,M_\odot$, M81 is one of the closest MW-analogues, 
and has thus been among the first targets for extended 
halo studies beyond the LG. The earliest H{\sc i} imaging
of the galaxy group dominated by this spiral unambiguously shows a spectacular
amount of substructure, most prominently a bridge of gas
between M81 and its brightest companions NGC~3077 and M82, located
at a projected distance of $\sim60$\,kpc
(\citealt{vanderhulst79, yun94}). 

Given the high level of interaction and H{\sc i} substructure 
in a group that can be considered as a LG-analogue, it is
natural to pursue the investigation of this complex environment
even further.
The intergalactic gas clouds embedding this environment are traced
by young stellar systems identified in resolved stellar
studies (\citealt{durrell04, davidge08, demello08}).
Some of them are classified as tidal
dwarf galaxies, such as Holmberg IX and the Garland 
(\citealt{makarova02, kara04, sabbi08, weisz08}),
characterized by a predominance of young stellar
populations. This type of galaxy has no counterpart in our own LG, 
and it is believed to be DM-free (see, e.g., \citealt{duc00}).

The first detailed look into the resolved populations in the outskirts of M81
is through the eye of {\it HST}: the predominantly old halo RGB
stars show a broad range of metallicities and a radial
gradient (\citealt{tikhonov05, mouhcine05c}). The radial
stellar counts (along several different directions) also 
reveal a break at a radius of $\sim25$\,kpc, which is 
interpreted as the transition point between thick 
disk and halo (\citealt{tikhonov05}).
In a similar fashion, the ground-based wide-field imager 
Subaru/SuprimeCam has been
used to uncover a faint and extended component beyond
M81's disk with a flat surface brightness profile
extending out to $\sim0.5$\,deg (or $\sim30$\,kpc) to the north
of M81 (\citealt{barker09}). This low surface brightness feature 
($\sim28$\,mag\,arcsec$^{-2}$) traced by the brightest RGB star 
counts appears bluer than the disk,
suggesting a metallicity lower than that of M81's main body,
but its true nature remains unclear. The authors suggest
this component to have intermediate properties between 
the MW's halo and its thick disk, but the limited
surveyed area ($0.3$\,deg$^2$) precludes any robust conclusions.

As part of a campaign to obtain panoramic views of nearby galaxy
haloes, \cite{mouhcine09} present a $0.9\times 0.9$\,deg$^2$
view of M81's surroundings obtained with the CFHT/MegaCam
imager. The images resolve individual RGB stars down to
$\sim2$\,mag below the TRGB, but this study focusses on
the younger, bright populations such as massive main sequence stars 
and red supergiants, which reveal further young systems 
tracing the H{\sc i} tidal distribution between M81 and its companions.
These systems are younger than the estimated dynamical age 
of the large-scale interaction and do not have an old population counterpart,
suggesting that they are not simply being detached from the 
main body of the primary galaxies but are instead formed
within the H{\sc i} clouds.

\cite{durrell10} recently conducted a deeper, albeit spatially
limited, {\it HST} study of a field at a galactocentric distance
of $\sim20$\,kpc. This field reveals an [M/H]$\sim-1.15$
population with an approximate old age of $\sim9$\,Gyr.
This field thus contains the most metal-poor stars 
found in M81's halo to that date,
which led the authors to the conclusion that they 
were dealing with an authentic halo component.
This study is extended by \cite{monachesi13} with the
{\it HST} GHOSTS dataset (see Sect.~\ref{systematic}): they construct
a colour profile out to a radius of $\sim50$\,kpc, and this dataset
does not show a significant gradient. The mean photometric
metallicity derived is [Fe/H]$\sim-1.2$, similarly to 
\citet{durrell10}. This result is found
to be in good agreement with simulations and the authors suggest
that the halo of M81 could have been assembled through an 
early accretion of satellites with comparable mass
(e.g., \citealt{cooper10, font06}).

As a further step in the investigation of M81's halo,
the \cite{barker09} and \cite{mouhcine09} ground-based
imaging of M81 is being improved by means of 
the Subaru/HyperSuprimeCam. The first $\sim2\times2$\,deg$^2$
($\sim100\times115$\,kpc$^2$) resolved stellar maps from 
different subpopulations
(upper main sequence, red supergiants, RGB and AGB stars)
are presented in \cite{okamoto15} and constitute a preview 
of an even wider-field effort to map the extended halo of
this group. These first maps (see Fig.~\ref{fig:okamoto}) 
confirm a high degree of substructure, most interestingly: the youngest
populations nicely trace the H{\sc i} gas content, confirming
previous small FoV studies; the
RGB distributions are smoother and significantly more extended 
than the young component, and show stream-like overlaps between
the dominant group galaxies, e.g., M82's stars clearly being stripped
by M81; a redder RGB distribution is detected for 
M81 and NGC~3077 compared to M82, indicating a lower
metallicity in the latter; in addition, M82 and NGC~3077's
outer regions present S-shaped morphologies, a smoking gun
of the tidal interaction with M81 and typical of 
interacting dwarf galaxies with larger companions
(e.g., \citealt{penarrubia09}).

Not less importantly, the widest-field survey to date
($\sim65$\,deg$^2$) of the M81 group has been performed by
\cite{chiboucas09} with CFHT/MegaCam, although with only one filter. 
The main goal of 
this survey was to identify new, faint dwarf galaxies and 
investigate the satellite LF in a highly interacting 
group environment as compared to the LG. This is the first
survey to systematically search for faint dwarfs beyond the
LG. Resolved spatial 
overdensities consistent with candidate dwarfs have been
followed up with two-band {\it HST}/ACS and {\it HST}/WFPC2 observations.
Fourteen of the 22 candidates turned out
to be real satellites of M81 based on their CMDs and TRGB distances, 
extending the previously known galaxy LF in this group by three 
orders of magnitude down to $M_r\sim-9.0$ (\citealt{chiboucas13}), 
with an additional possibly ultra-faint member at $M_r\sim-7.0$.
The measured slope of the LF in the M81 group appears to
be flatter than cosmological predictions ($\alpha\sim-1.27$,
in contrast to the theoretical value of $\alpha\sim-1.8$),
similar to what has been found for the MW and M31 satellites.

\subsubsection{NGC~253} \label{ngc253}

Another obvious MW-mass spiral target for halo studies 
is NCG~253 ($\sim3.5$\,Mpc, \citealt{radburn11}). Its role of brightest
object within the loose Sculptor filament of galaxies makes it
ideally suited to investigate the effects of 
external environment on the assembly of haloes.
As already apparent from old photographic plates,
NGC~253's outskirts show faint perturbation signs,
such as an extended shelf to the south of its disk 
(\citealt{malin97}), pointing at a possible accretion event.
This spiral galaxy, despite its relative isolation, is experiencing a
recent starburst and a pronounced nuclear outflow: the latter is
believed to host local star formation extending as high as
$\sim15$\,kpc above the disk in the minor axis direction
(see \citealt{comeron01}, and references therein).

The resolved near-infrared study of \cite{davidge10} allowed them to
detect bright AGB stars, but not RGB stars, extending out to
$\sim13$\,kpc from the disk plane in the south direction: 
these are interpreted as being
expelled from the disk into the halo as consequence of a recent interaction.
Subsequently, \cite{bailin11} exploited a combination of {\it HST} data from the GHOSTS
survey and ground-based Magellan/IMACS imaging, the former being
deeper while the latter have a more extended FoV (out to $\sim30$\,kpc
in the halo NGC~253 in the south direction). The authors are able to estimate NGC~253's halo
mass as $\sim2\times10^9\,M_\odot$, or 6\% of the galaxy's total
stellar mass: this value is broadly consistent with those derived from the MW
and M31 but higher, reminiscent of the halo-to-halo scatter seen in
simulations. A power law is fit to the RGB radial profile which
is found to be slightly steeper than that of the two LG spirals, and appears to be
flattened in the same direction as the disk component. This is the one
of the few studies to date to quantitatively measure such parameters for a halo
beyond the LG, and it sets the stage for the possibilities opened by
similar studies of other nearby galaxies.
The RGB density maps derived in \cite{bailin11} from IMACS imaging 
confirm the early detection of a shelf structure, and uncover
several additional kpc-scale substructures in the halo of this spiral.

A more recent wide-field study of NGC~253 is presented by
\cite{greggio14}, who exploit the near-infrared VISTA/VIRCAM imager to
study the RGB and AGB stellar content of this galaxy out to
$\sim40-50$\,kpc, covering also the northern portion which was not
included in previous studies. This portion, in particular, reveals an
RGB substructure symmetric (and likely connected) to the one in the
south. A prominent arc ($\sim20$\,kpc in length) to the north-west of
the disk is detected and
estimated to arise from a progenitor with a stellar mass of roughly
$\sim7\times10^6\,M_\odot$. The RGB radial density profile shows a
break at a radius of $\sim25$\,kpc, indicative of the transition from
disk to halo. The elongated halo component already
discussed in \cite{bailin11} is confirmed here, but is considered to
be an inner halo: an outer, more spherical and homogeneous component extends at least
out to the galactocentric distances covered by this
survey. Intriguingly, the AGB density map reveals that 25\% of this
intermediate-age (i.e., up to a few Gyr old) population is 
spread out to $\sim30$\,kpc above the disk: this component cannot
easily be explained with either an in-situ or an accreted origin.

NGC~253 is also one of the two targets of the Panoramic Imaging
Survey of Centaurus and Sculptor (PISCeS), recently initiated with the
wide-field imager Magellan/Megacam. This ambitious survey aims at
obtaining RGB stellar maps of this galaxy and of the elliptical
Centaurus~A (Cen~A; see next Section) out to galactocentric radii of
$\sim150$\,kpc, similarly to the PAndAS survey of M31. Early results
from this survey include the discovery of two new faint
satellites of NGC~253, one of which is clearly elongated and in the process of
being disrupted by its host (\citealt{sand14, toloba16b}).

\subsubsection{NGC~5128 (Centaurus~A)} \label{cena}

\begin{figure}[h]
\centering
\sidecaption
\includegraphics[width=11cm]{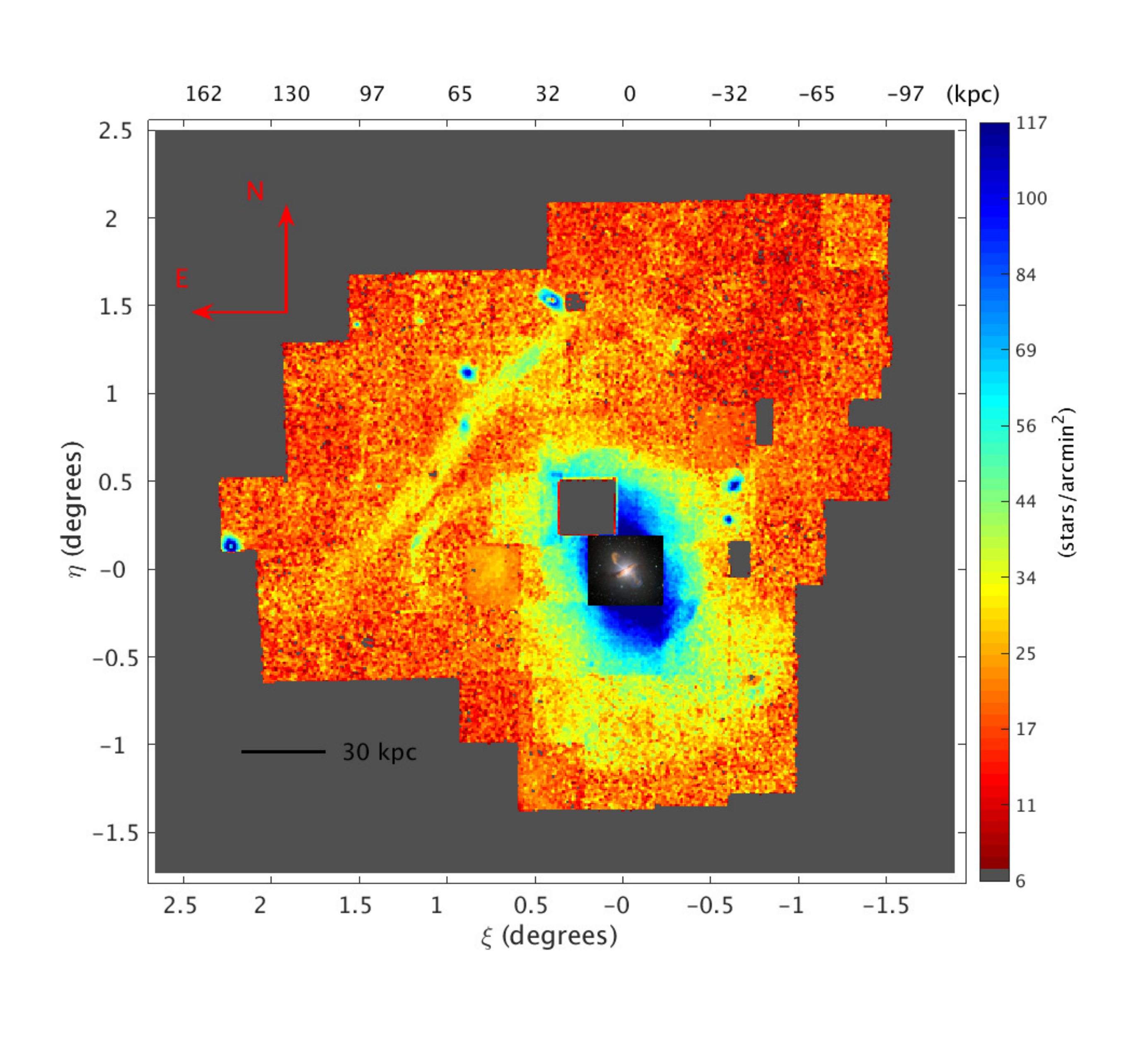}
\caption{Surface density map of RGB stars in the halo of Cen~A,
obtained with Magellan/Megacam as part of the PISCeS survey.
The map extends out to a radius of 150\,kpc in the north
and east directions (physical and density scales are reported). 
Several tidal features are easily recognized, including a
stunning disrupting dwarf with tails 2\,deg long in the outer halo, an
extended sparse cloud to the south of the galaxy, as
well as arcs and plumes around the inner regions, tracing both
ongoing and past accretion events. 
Fig.~3 from \cite{crnojevic16}, reproduced by permission of the AAS}
\label{fig:crnojevic16}   
\end{figure}

It is important to target galaxies of different morphologies
and environments to thoroughly investigate the assembly of haloes.
The closest ($\sim3.8$\,Mpc; \citealt{harrisg09}) 
elliptical galaxy is Centaurus~A (Cen~A; technically speaking, 
Maffei~1 is slightly closer but it lies behind the Galactic disk and 
is thus heavily reddened, see \citealt{wu14}). Cen~A is the dominant galaxy
of a rich and dense group, which also has a second subgroup component 
centred on the spiral M83 (e.g., \citealt{kara07}).

Despite having often been referred to as a peculiar galaxy, due to 
its pronounced radio activity, its central dust lanes, and a perturbed
morphology, the luminosity of Cen~A is quite typical of field
elliptical galaxies: a recent ($<1$\,Gyr) merger event 
is believed to be the culprit for its peculiar features
(see \citealt{israel98}, and references therein). Besides
this main merger event, \cite{peng02} uncover a system
of faint shells and an arc within $\sim25$\,kpc of Cen~A's centre 
from integrated light observations; the arc is believed to have
been produced by the infall of a low-mass, star forming galaxy
around $\sim300$\,Myr ago.

This elliptical galaxy has been the subject of a systematic
study conducted with {\it HST}/ACS and {\it HST}/WFPC2 throughout the past 
couple of decades: a number of pointings at increasingly large
galactocentric radii (from a few out to $\sim150$\,kpc) 
have been used to investigate the properties and gradients of Cen~A's 
halo populations (\citealt{rejkuba14}, and references therein).
The considered pointings out to 40\,kpc reveal metal-rich
populations ([Fe/H]$>-1.0$), not dissimilar to what
has been observed for the haloes of spiral galaxies.
The deepest CMD to date of this elliptical is presented by
\cite{rejkuba11} for the {\it HST} field at 40\,kpc: this study
concludes that the vast majority of Cen~A's halo population is old
($\sim12$\,Gyr), with a younger ($\sim2-4$\,Gyr) component 
accounting for $\sim20\%$ of the total population. 

The first wide-field study of Cen~A was performed with the 
ground-based VLT/VIMOS imager, reaching out to $\sim85$\,kpc
along both minor and major axes (\citealt{crnojevic13}). Cen~A's
halo population seems to extend all the way out to this large radius.
This study confirms the relatively high metallicity for halo populations
found by the {\it HST} studies, although with a considerable presence
of metal-poor stars at all radii; the authors also highlight the absence of a strong
metallicity gradient from a $\sim30$\,kpc radius out to the most
distant regions probed. This study suggests that the outer regions
of Cen~A's halo show an increase in ellipticity as a function of radius,
which could, however, be interpreted as the presence of substructure
contaminating the observed fields. 
A subsequent study exploits additional {\it HST} pointings out to a remarkably large
radius of $\sim150$\,kpc: the edge of Cen~A's halo is not reached even
by this study (\citealt{rejkuba14}). This dataset, analysed together with the previous
{\it HST} pointings, confirms that a very mild metallicity gradient
is present, with median metallicities remaining high 
out to the largest distances probed. \cite{rejkuba14},
however, also detect a significant pointing-to-pointing variation
in both the RGB star counts and the median metallicity, which is 
likely indicative of non-mixed accreted populations.

Recently, the PISCeS survey (see previous Section) has sketched
a PAndAS-like picture of Cen~A's halo out to $\sim150$\,kpc:
the RGB stellar density map derived from a mosaic of Magellan/Megacam
images is presented in Fig.~\ref{fig:crnojevic16}.
This map, very much like the ones obtained for M31 and NGC~891, 
uncovers a plethora of faint substructures, both in the inner
regions of the target galaxy and in its outskirts. The
morphological variety of these features is reminiscent of that
observed in PAndAS, with shells, plumes, an extended cloud and 
long tidal streams. In particular, one of the newly discovered
dwarf satellites of Cen~A is clearly in the process of being
disrupted, with $\sim2$\,deg long tails: taking into account the
stellar content of these tails, this galaxy's pre-disruption
luminosity could have been similar to that of Sagittarius in the LG.
This survey also led to the discovery of nine (confirmed) dwarf 
satellites down to $M_V\sim-7$. Their properties are consistent
with those of faint LG satellites, but some of them lie at
the faint/diffuse end of the LG luminosity/surface brightness/radius
distribution: this indicates that we might be looking at 
previously unexplored physical regimes for these faintest
satellites, which opens new exciting perspectives for future
studies.


\section{Summary and Future Prospects\index{Future facilities}} \label{conclusions}

In a $\rm \Lambda$CDM hierarchical model, all galaxies are predicted to
have experienced mergers, of which many should be recognizable as
debris/streams that make up for a large fraction of their haloes. Haloes
and their substructures thus provide a unique glimpse into the
assembly history of galaxies,
and can inform the models at the smallest galactic scales, where they
still fall short in reproducing observations.
The time is now ripe for in-depth systematic studies of the resolved
stellar populations in galaxy haloes, which will dramatically increase
our understanding of galaxy evolution over the next decade.

The challenges for this type of studies are of a different nature: for
our own Galaxy, state-of-the-art results on its halo shape, profile and mass
inevitably suffer from assumptions on underlying density models and 
extrapolations of the available data to radii larger than observed.
The major current limitation of MW halo studies lies in
observational biases due to small field-of-view samples, which 
preclude the identification of possible substructure contamination.
Future surveys hold the promise to advance the
knowledge of our Galaxy by obtaining significantly larger samples
of tracers, especially in areas so far not covered. Most notably, 
the astrometric {\it Gaia} mission (which will provide unprecedented 
six-dimensional phase space information for two billion stars out to the
inner MW halo) and the Large Synoptic Survey Telescope (LSST; 
designed to provide a southern sky counterpart to SDSS, and reaching
$\sim4$ magnitudes fainter than its predecessor for a total sample of 
tens of billions of stars), are going to revolutionize our view of
the MW. At the same time, the current and future generation of high-resolution
spectrographs will follow up these surveys from the ground, providing
comprehensive kinematic and chemical information to assess the origin
of halo stars and characterize their birthplaces (see also Figueras, this volume).

The pioneering studies of an increasing number of haloes beyond the LG,
and across a range of masses, will soon be extended by the next
generation of ground-based extremely large telescopes (E-ELT,
GMT, TMT), as well as space-borne missions ({\it JWST}, {\it Euclid}, {\it WFIRST}). The
PAndAS survey of M31 has extensively demonstrated that only the
synergy of wide-field ground-based observations, deep (but spatially
limited) observations from space, and spectroscopy can return a truly global
understanding of haloes made up of a complex mixture of in-situ and
accreted populations. The aforementioned facilities will open new perspectives
with wide-field optical and infrared imagers in concert with
high-resolution spectrographs, which will allow us to systematically
survey hundreds of galaxies within tens of Mpc in the next decade or
two. For example, with the E-ELT/MICADO and {\it JWST}/NIRcam imagers (the
former having higher
resolving power and the latter a wider field-of-view), we should
resolve stars down to the HB within $\sim10$\,Mpc, thus identifying
and characterizing the SFHs of streams and faint satellites; derive radial
profiles, MDFs and stellar population gradients in haloes within
20\,Mpc from the uppermost few magnitudes of the RGB; and trace 
the halo shape and possible overdensities down to $\mu_V\sim33$\,mag\,arcsec$^{-2}$ from the uppermost $\sim0.5$\,mag of the RGB out to
50\,Mpc (\citealt{greggio16}). 

These observational constraints will be crucial to inform
increasingly sophisticated theoretical models, and ultimately answer
intriguing open questions (as well as possibly unexpected ones that
will likely be raised by these observations themselves), such as:

\begin{itemize}
\item{Do all galaxies have haloes?}
\item{What are the relative fractions of in-situ versus accreted
    populations in galaxy haloes, and how does this depend on
    galactocentric distance, galaxy morphology, and environment?}
\item{What are the properties of the objects currently being accreted, i.e., mass, chemical content, SFH,
    orbital properties, and how do they relate to
    those of the present day low-mass satellites?}
\item{Do low-mass galaxies possess haloes/satellites of their own, and
    what is their fate and contribution upon infall onto a massive galaxy?}
\item{How extended really are the haloes of massive galaxies?}
\item{What is the shape and mass of the DM haloes underlying galaxies?}
\item{What is the relation between the outer halo and the bulge/disk
    of a galaxy?}
\item{What is the role of internal versus external processes in
    shaping a galaxy's properties, especially at the low-mass end of
    the galaxy LF?}
\item{What is the relation between the present-day haloes/satellites
    and their unresolved, high-redshift counterparts?}
\end{itemize}

The era of resolved populations in galaxy haloes has just begun, and it
holds the promise to be a golden one.


\begin{acknowledgement}
I would like to thank the organizers for a lively and stimulating
conference. I am indebted to S. Pasetto for his advice and support
throughout the preparation of this contribution. I acknowledge the
hospitality of the Carnegie Observatories during the completion of
this work.
\end{acknowledgement}

\bibliographystyle{spbasic}

\bibliography{biblio}

\begin{thebibliography}{198}
\providecommand{\natexlab}[1]{#1}
\providecommand{\url}[1]{{#1}}
\providecommand{\urlprefix}{URL }
\expandafter\ifx\csname urlstyle\endcsname\relax
  \providecommand{\doi}[1]{DOI~\discretionary{}{}{}#1}\else
  \providecommand{\doi}{DOI~\discretionary{}{}{}\begingroup
  \urlstyle{rm}\Url}\fi
\providecommand{\eprint}[2][]{\url{#2}}

\bibitem[{{Abadi} et~al(2006){Abadi}, {Navarro}, and {Steinmetz}}]{abadi06}
{Abadi} MG, {Navarro} JF, {Steinmetz} M (2006) {Stars beyond galaxies: the
  origin of extended luminous haloes around galaxies}. \mnras 365:747--758,
  \doi{10.1111/j.1365-2966.2005.09789.x}, \eprint{astro-ph/0506659}

\bibitem[{{Adams} et~al(2012){Adams}, {Zaritsky}, {Sand}, {Graham}, {Bildfell},
  {Hoekstra}, and {Pritchet}}]{adams12}
{Adams} SM, {Zaritsky} D, {Sand} DJ, {Graham} ML, {Bildfell} C, {Hoekstra} H,
  {Pritchet} C (2012) {The Environmental Dependence of the Incidence of
  Galactic Tidal Features}. \aj 144:128, \doi{10.1088/0004-6256/144/5/128},
  \eprint{1208.4843}

\bibitem[{{Atkinson} et~al(2013){Atkinson}, {Abraham}, and
  {Ferguson}}]{atkinson13}
{Atkinson} AM, {Abraham} RG, {Ferguson} AMN (2013) {Faint Tidal Features in
  Galaxies within the Canada-France-Hawaii Telescope Legacy Survey Wide
  Fields}. \apj 765:28, \doi{10.1088/0004-637X/765/1/28}, \eprint{1301.4275}

\bibitem[{{Bailin} et~al(2011){Bailin}, {Bell}, {Chappell}, {Radburn-Smith},
  and {de Jong}}]{bailin11}
{Bailin} J, {Bell} EF, {Chappell} SN, {Radburn-Smith} DJ, {de Jong} RS (2011)
  {The Resolved Stellar Halo of NGC 253}. \apj 736:24,
  \doi{10.1088/0004-637X/736/1/24}, \eprint{1105.0005}

\bibitem[{{Barker} et~al(2009){Barker}, {Ferguson}, {Irwin}, {Arimoto}, and
  {Jablonka}}]{barker09}
{Barker} MK, {Ferguson} AMN, {Irwin} M, {Arimoto} N, {Jablonka} P (2009)
  {Resolving the Stellar Outskirts of M81: Evidence for a Faint, Extended
  Structural Component}. \aj 138:1469--1484,
  \doi{10.1088/0004-6256/138/5/1469}, \eprint{0909.1430}

\bibitem[{{Barker} et~al(2012){Barker}, {Ferguson}, {Irwin}, {Arimoto}, and
  {Jablonka}}]{barker11}
{Barker} MK, {Ferguson} AMN, {Irwin} MJ, {Arimoto} N, {Jablonka} P (2012)
  {Quantifying the faint structure of galaxies: the late-type spiral NGC 2403}.
  \mnras 419:1489--1506, \doi{10.1111/j.1365-2966.2011.19814.x},
  \eprint{1109.2625}

\bibitem[{{Bechtol} et~al(2015){Bechtol}, {Drlica-Wagner}, {Balbinot},
  {Pieres}, {Simon}, {Yanny}, {Santiago}, {Wechsler}, {Frieman}, {Walker},
  {Williams}, {Rozo}, {Rykoff}, {Queiroz}, {Luque}, {Benoit-L{\'e}vy},
  {Tucker}, {Sevilla}, {Gruendl}, {da Costa}, {Fausti Neto}, {Maia}, {Abbott},
  {Allam}, {Armstrong}, {Bauer}, {Bernstein}, {Bernstein}, {Bertin}, {Brooks},
  {Buckley-Geer}, {Burke}, {Carnero Rosell}, {Castander}, {Covarrubias},
  {D'Andrea}, {DePoy}, {Desai}, {Diehl}, {Eifler}, {Estrada}, {Evrard},
  {Fernandez}, {Finley}, {Flaugher}, {Gaztanaga}, {Gerdes}, {Girardi},
  {Gladders}, {Gruen}, {Gutierrez}, {Hao}, {Honscheid}, {Jain}, {James},
  {Kent}, {Kron}, {Kuehn}, {Kuropatkin}, {Lahav}, {Li}, {Lin}, {Makler},
  {March}, {Marshall}, {Martini}, {Merritt}, {Miller}, {Miquel}, {Mohr},
  {Neilsen}, {Nichol}, {Nord}, {Ogando}, {Peoples}, {Petravick}, {Plazas},
  {Romer}, {Roodman}, {Sako}, {Sanchez}, {Scarpine}, {Schubnell}, {Smith},
  {Soares-Santos}, {Sobreira}, {Suchyta}, {Swanson}, {Tarle}, {Thaler},
  {Thomas}, {Wester}, {Zuntz}, and {DES Collaboration}}]{bechtol15}
{Bechtol} K, {Drlica-Wagner} A, {Balbinot} E, {Pieres} A, {Simon} JD, {Yanny}
  B, {Santiago} B, {Wechsler} RH, {Frieman} J, {Walker} AR, {Williams} P,
  {Rozo} E, {Rykoff} ES, {Queiroz} A, {Luque} E, {Benoit-L{\'e}vy} A, {Tucker}
  D, {Sevilla} I, {Gruendl} RA, {da Costa} LN, {Fausti Neto} A, {Maia} MAG,
  {Abbott} T, {Allam} S, {Armstrong} R, {Bauer} AH, {Bernstein} GM, {Bernstein}
  RA, {Bertin} E, {Brooks} D, {Buckley-Geer} E, {Burke} DL, {Carnero Rosell} A,
  {Castander} FJ, {Covarrubias} R, {D'Andrea} CB, {DePoy} DL, {Desai} S,
  {Diehl} HT, {Eifler} TF, {Estrada} J, {Evrard} AE, {Fernandez} E, {Finley}
  DA, {Flaugher} B, {Gaztanaga} E, {Gerdes} D, {Girardi} L, {Gladders} M,
  {Gruen} D, {Gutierrez} G, {Hao} J, {Honscheid} K, {Jain} B, {James} D, {Kent}
  S, {Kron} R, {Kuehn} K, {Kuropatkin} N, {Lahav} O, {Li} TS, {Lin} H, {Makler}
  M, {March} M, {Marshall} J, {Martini} P, {Merritt} KW, {Miller} C, {Miquel}
  R, {Mohr} J, {Neilsen} E, {Nichol} R, {Nord} B, {Ogando} R, {Peoples} J,
  {Petravick} D, {Plazas} AA, {Romer} AK, {Roodman} A, {Sako} M, {Sanchez} E,
  {Scarpine} V, {Schubnell} M, {Smith} RC, {Soares-Santos} M, {Sobreira} F,
  {Suchyta} E, {Swanson} MEC, {Tarle} G, {Thaler} J, {Thomas} D, {Wester} W,
  {Zuntz} J, {DES Collaboration} (2015) {Eight New Milky Way Companions
  Discovered in First-year Dark Energy Survey Data}. \apj 807:50,
  \doi{10.1088/0004-637X/807/1/50}, \eprint{1503.02584}

\bibitem[{{Bell} et~al(2008){Bell}, {Zucker}, {Belokurov}, {Sharma},
  {Johnston}, {Bullock}, {Hogg}, {Jahnke}, {de Jong}, {Beers}, {Evans},
  {Grebel}, {Ivezi{\'c}}, {Koposov}, {Rix}, {Schneider}, {Steinmetz}, and
  {Zolotov}}]{bell08}
{Bell} EF, {Zucker} DB, {Belokurov} V, {Sharma} S, {Johnston} KV, {Bullock} JS,
  {Hogg} DW, {Jahnke} K, {de Jong} JTA, {Beers} TC, {Evans} NW, {Grebel} EK,
  {Ivezi{\'c}} {\v Z}, {Koposov} SE, {Rix} HW, {Schneider} DP, {Steinmetz} M,
  {Zolotov} A (2008) {The Accretion Origin of the Milky Way's Stellar Halo}.
  \apj 680:295-311, \doi{10.1086/588032}, \eprint{0706.0004}

\bibitem[{{Belokurov}(2013)}]{belokurov13}
{Belokurov} V (2013) {Galactic Archaeology: The dwarfs that survived and
  perished}. \nar 57:100--121, \doi{10.1016/j.newar.2013.07.001},
  \eprint{1307.0041}

\bibitem[{{Belokurov} et~al(2006){Belokurov}, {Zucker}, {Evans}, {Gilmore},
  {Vidrih}, {Bramich}, {Newberg}, {Wyse}, {Irwin}, {Fellhauer}, {Hewett},
  {Walton}, {Wilkinson}, {Cole}, {Yanny}, {Rockosi}, {Beers}, {Bell},
  {Brinkmann}, {Ivezi{\'c}}, and {Lupton}}]{belokurov06a}
{Belokurov} V, {Zucker} DB, {Evans} NW, {Gilmore} G, {Vidrih} S, {Bramich} DM,
  {Newberg} HJ, {Wyse} RFG, {Irwin} MJ, {Fellhauer} M, {Hewett} PC, {Walton}
  NA, {Wilkinson} MI, {Cole} N, {Yanny} B, {Rockosi} CM, {Beers} TC, {Bell} EF,
  {Brinkmann} J, {Ivezi{\'c}} {\v Z}, {Lupton} R (2006) {The Field of Streams:
  Sagittarius and Its Siblings}. \apjl 642:L137--L140, \doi{10.1086/504797},
  \eprint{astro-ph/0605025}

\bibitem[{{Bernard} et~al(2015){Bernard}, {Ferguson}, {Richardson}, {Irwin},
  {Barker}, {Hidalgo}, {Aparicio}, {Chapman}, {Ibata}, {Lewis}, {McConnachie},
  and {Tanvir}}]{bernard15}
{Bernard} EJ, {Ferguson} AMN, {Richardson} JC, {Irwin} MJ, {Barker} MK,
  {Hidalgo} SL, {Aparicio} A, {Chapman} SC, {Ibata} RA, {Lewis} GF,
  {McConnachie} AW, {Tanvir} NR (2015) {The nature and origin of substructure
  in the outskirts of M31 - II. Detailed star formation histories}. \mnras
  446:2789--2801, \doi{10.1093/mnras/stu2309}, \eprint{1406.2247}

\bibitem[{{Bernard} et~al(2016){Bernard}, {Ferguson}, {Schlafly}, {Martin},
  {Rix}, {Bell}, {Finkbeiner}, {Goldman}, {Mart{\'{\i}}nez-Delgado}, {Sesar},
  {Wyse}, {Burgett}, {Chambers}, {Draper}, {Hodapp}, {Kaiser}, {Kudritzki},
  {Magnier}, {Metcalfe}, {Wainscoat}, and {Waters}}]{bernard16}
{Bernard} EJ, {Ferguson} AMN, {Schlafly} EF, {Martin} NF, {Rix} HW, {Bell} EF,
  {Finkbeiner} DP, {Goldman} B, {Mart{\'{\i}}nez-Delgado} D, {Sesar} B, {Wyse}
  RFG, {Burgett} WS, {Chambers} KC, {Draper} PW, {Hodapp} KW, {Kaiser} N,
  {Kudritzki} RP, {Magnier} EA, {Metcalfe} N, {Wainscoat} RJ, {Waters} C (2016)
  {A Synoptic Map of Halo Substructures from the Pan-STARRS1 3{$\pi$} Survey}.
  \mnras 463:1759--1768, \doi{10.1093/mnras/stw2134}, \eprint{1607.06088}

\bibitem[{{Bird} et~al(2010){Bird}, {Harris}, {Blakeslee}, and
  {Flynn}}]{bird10}
{Bird} S, {Harris} WE, {Blakeslee} JP, {Flynn} C (2010) {The inner halo of M
  87: a first direct view of the red-giant population}. \aap 524:A71,
  \doi{10.1051/0004-6361/201014876}, \eprint{1009.3202}

\bibitem[{{Bland-Hawthorn} and {Gerhard}(2016)}]{bland16}
{Bland-Hawthorn} J, {Gerhard} O (2016) {The Galaxy in Context: Structural,
  Kinematic, and Integrated Properties}. \araa 54:529--596,
  \doi{10.1146/annurev-astro-081915-023441}, \eprint{1602.07702}

\bibitem[{{Boylan-Kolchin} et~al(2011){Boylan-Kolchin}, {Bullock}, and
  {Kaplinghat}}]{boylan11}
{Boylan-Kolchin} M, {Bullock} JS, {Kaplinghat} M (2011) {Too big to fail? The
  puzzling darkness of massive Milky Way subhaloes}. \mnras 415:L40--L44,
  \doi{10.1111/j.1745-3933.2011.01074.x}, \eprint{1103.0007}

\bibitem[{{Brooks} et~al(2013){Brooks}, {Kuhlen}, {Zolotov}, and
  {Hooper}}]{brooks13}
{Brooks} AM, {Kuhlen} M, {Zolotov} A, {Hooper} D (2013) {A Baryonic Solution to
  the Missing Satellites Problem}. \apj 765:22,
  \doi{10.1088/0004-637X/765/1/22}, \eprint{1209.5394}

\bibitem[{{Brown} et~al(2006){Brown}, {Smith}, {Ferguson}, {Rich},
  {Guhathakurta}, {Renzini}, {Sweigart}, and {Kimble}}]{brown06}
{Brown} TM, {Smith} E, {Ferguson} HC, {Rich} RM, {Guhathakurta} P, {Renzini} A,
  {Sweigart} AV, {Kimble} RA (2006) {The Detailed Star Formation History in the
  Spheroid, Outer Disk, and Tidal Stream of the Andromeda Galaxy}. \apj
  652:323--353, \doi{10.1086/508015}, \eprint{astro-ph/0607637}

\bibitem[{{Bullock} and {Johnston}(2005)}]{bullock05}
{Bullock} JS, {Johnston} KV (2005) {Tracing Galaxy Formation with Stellar
  Halos. I. Methods}. \apj 635:931--949, \doi{10.1086/497422},
  \eprint{arXiv:astro-ph/0506467}

\bibitem[{{Carlin} et~al(2016){Carlin}, {Sand}, {Price}, {Willman},
  {Karunakaran}, {Spekkens}, {Bell}, {Brodie}, {Crnojevi{\'c}}, {Forbes},
  {Hargis}, {Kirby}, {Lupton}, {Peter}, {Romanowsky}, and {Strader}}]{carlin16}
{Carlin} JL, {Sand} DJ, {Price} P, {Willman} B, {Karunakaran} A, {Spekkens} K,
  {Bell} EF, {Brodie} JP, {Crnojevi{\'c}} D, {Forbes} DA, {Hargis} J, {Kirby}
  E, {Lupton} R, {Peter} AHG, {Romanowsky} AJ, {Strader} J (2016) {First
  Results from the MADCASH Survey: A Faint Dwarf Galaxy Companion to the
  Low-mass Spiral Galaxy NGC 2403 at 3.2 Mpc}. \apjl 828:L5,
  \doi{10.3847/2041-8205/828/1/L5}, \eprint{1608.02591}

\bibitem[{{Carollo} et~al(2007){Carollo}, {Beers}, {Lee}, {Chiba}, {Norris},
  {Wilhelm}, {Sivarani}, {Marsteller}, {Munn}, {Bailer-Jones}, {Fiorentin}, and
  {York}}]{carollo07}
{Carollo} D, {Beers} TC, {Lee} YS, {Chiba} M, {Norris} JE, {Wilhelm} R,
  {Sivarani} T, {Marsteller} B, {Munn} JA, {Bailer-Jones} CAL, {Fiorentin} PR,
  {York} DG (2007) {Two stellar components in the halo of the Milky Way}. \nat
  450:1020--1025, \doi{10.1038/nature06460}, \eprint{0706.3005}

\bibitem[{{Chapman} et~al(2006){Chapman}, {Ibata}, {Lewis}, {Ferguson},
  {Irwin}, {McConnachie}, and {Tanvir}}]{chapman06}
{Chapman} SC, {Ibata} R, {Lewis} GF, {Ferguson} AMN, {Irwin} M, {McConnachie}
  A, {Tanvir} N (2006) {A Kinematically Selected, Metal-poor Stellar Halo in
  the Outskirts of M31}. \apj 653:255--266, \doi{10.1086/508599},
  \eprint{astro-ph/0602604}

\bibitem[{{Chiba} and {Beers}(2000)}]{chiba00}
{Chiba} M, {Beers} TC (2000) {Kinematics of Metal-poor Stars in the Galaxy.
  III. Formation of the Stellar Halo and Thick Disk as Revealed from a Large
  Sample of Nonkinematically Selected Stars}. \aj 119:2843--2865,
  \doi{10.1086/301409}, \eprint{astro-ph/0003087}

\bibitem[{{Chiboucas} et~al(2009){Chiboucas}, {Karachentsev}, and
  {Tully}}]{chiboucas09}
{Chiboucas} K, {Karachentsev} ID, {Tully} RB (2009) {Discovery of New Dwarf
  Galaxies in the M81 Group}. \aj 137:3009--3037,
  \doi{10.1088/0004-6256/137/2/3009}, \eprint{0805.1250}

\bibitem[{{Chiboucas} et~al(2013){Chiboucas}, {Jacobs}, {Tully}, and
  {Karachentsev}}]{chiboucas13}
{Chiboucas} K, {Jacobs} BA, {Tully} RB, {Karachentsev} ID (2013) {Confirmation
  of Faint Dwarf Galaxies in the M81 Group}. \aj 146:126,
  \doi{10.1088/0004-6256/146/5/126}, \eprint{1309.4130}

\bibitem[{{Cockcroft} et~al(2013){Cockcroft}, {McConnachie}, {Harris}, {Ibata},
  {Irwin}, {Ferguson}, {Fardal}, {Babul}, {Chapman}, {Lewis}, {Martin}, and
  {Puzia}}]{cockcroft13}
{Cockcroft} R, {McConnachie} AW, {Harris} WE, {Ibata} R, {Irwin} MJ, {Ferguson}
  AMN, {Fardal} MA, {Babul} A, {Chapman} SC, {Lewis} GF, {Martin} NF, {Puzia}
  TH (2013) {Unearthing foundations of a cosmic cathedral: searching the stars
  for M33's halo}. \mnras 428:1248--1262, \doi{10.1093/mnras/sts112},
  \eprint{1210.4114}

\bibitem[{{Collins} et~al(2014){Collins}, {Chapman}, {Rich}, {Ibata}, {Martin},
  {Irwin}, {Bate}, {Lewis}, {Pe{\~n}arrubia}, {Arimoto}, {Casey}, {Ferguson},
  {Koch}, {McConnachie}, and {Tanvir}}]{collins14}
{Collins} MLM, {Chapman} SC, {Rich} RM, {Ibata} RA, {Martin} NF, {Irwin} MJ,
  {Bate} NF, {Lewis} GF, {Pe{\~n}arrubia} J, {Arimoto} N, {Casey} CM,
  {Ferguson} AMN, {Koch} A, {McConnachie} AW, {Tanvir} N (2014) {The Masses of
  Local Group Dwarf Spheroidal Galaxies: The Death of the Universal Mass
  Profile}. \apj 783:7, \doi{10.1088/0004-637X/783/1/7}, \eprint{1309.3053}

\bibitem[{{Comer{\'o}n} et~al(2001){Comer{\'o}n}, {Torra}, {M{\'e}ndez}, and
  {G{\'o}mez}}]{comeron01}
{Comer{\'o}n} F, {Torra} J, {M{\'e}ndez} RA, {G{\'o}mez} AE (2001) {Possible
  star formation in the halo of NGC 253}. \aap 366:796--810,
  \doi{10.1051/0004-6361:20000336}

\bibitem[{{Cooper} et~al(2010){Cooper}, {Cole}, {Frenk}, {White}, {Helly},
  {Benson}, {De Lucia}, {Helmi}, and {et al.}}]{cooper10}
{Cooper} AP, {Cole} S, {Frenk} CS, {White} SDM, {Helly} J, {Benson} AJ, {De
  Lucia} G, {Helmi} A, {et al} (2010) {Galactic stellar haloes in the CDM
  model}. \mnras 406:744--766, \doi{10.1111/j.1365-2966.2010.16740.x},
  \eprint{0910.3211}

\bibitem[{{Cooper} et~al(2015){Cooper}, {Parry}, {Lowing}, {Cole}, and
  {Frenk}}]{cooper15}
{Cooper} AP, {Parry} OH, {Lowing} B, {Cole} S, {Frenk} C (2015) {Formation of
  in situ stellar haloes in Milky Way-mass galaxies}. \mnras 454:3185--3199,
  \doi{10.1093/mnras/stv2057}, \eprint{1501.04630}

\bibitem[{{Crnojevi{\'c}} et~al(2010){Crnojevi{\'c}}, {Grebel}, and
  {Koch}}]{crnojevic10}
{Crnojevi{\'c}} D, {Grebel} EK, {Koch} A (2010) {A close look at the Centaurus
  A group of galaxies. I. Metallicity distribution functions and population
  gradients in early-type dwarfs}. \aap 516:A85,
  \doi{10.1051/0004-6361/200913429}, \eprint{1002.0341}

\bibitem[{{Crnojevi{\'c}} et~al(2013){Crnojevi{\'c}}, {Ferguson}, {Irwin},
  {Bernard}, {Arimoto}, {Jablonka}, and {Kobayashi}}]{crnojevic13}
{Crnojevi{\'c}} D, {Ferguson} AMN, {Irwin} MJ, {Bernard} EJ, {Arimoto} N,
  {Jablonka} P, {Kobayashi} C (2013) {The outer halo of the nearest giant
  elliptical: a VLT/VIMOS survey of the resolved stellar populations in
  Centaurus A to 85 kpc}. \mnras 432:832--847, \doi{10.1093/mnras/stt494},
  \eprint{1303.4736}

\bibitem[{{Crnojevi{\'c}} et~al(2014){Crnojevi{\'c}}, {Ferguson}, {Irwin},
  {McConnachie}, {Bernard}, {Fardal}, {Ibata}, {Lewis}, {Martin}, {Navarro},
  {No{\"e}l}, and {Pasetto}}]{crnojevic14a}
{Crnojevi{\'c}} D, {Ferguson} AMN, {Irwin} MJ, {McConnachie} AW, {Bernard} EJ,
  {Fardal} MA, {Ibata} RA, {Lewis} GF, {Martin} NF, {Navarro} JF, {No{\"e}l}
  NED, {Pasetto} S (2014) {A PAndAS view of M31 dwarf elliptical satellites:
  NGC 147 and NGC 185}. \mnras 445:3862--3877, \doi{10.1093/mnras/stu2003},
  \eprint{1409.7065}

\bibitem[{{Crnojevi{\'c}} et~al(2016){Crnojevi{\'c}}, {Sand}, {Spekkens},
  {Caldwell}, {Guhathakurta}, {McLeod}, {Seth}, {Simon}, {Strader}, and
  {Toloba}}]{crnojevic16}
{Crnojevi{\'c}} D, {Sand} DJ, {Spekkens} K, {Caldwell} N, {Guhathakurta} P,
  {McLeod} B, {Seth} A, {Simon} JD, {Strader} J, {Toloba} E (2016) {The
  Extended Halo of Centaurus A: Uncovering Satellites, Streams, and
  Substructures}. \apj 823:19, \doi{10.3847/0004-637X/823/1/19},
  \eprint{1512.05366}

\bibitem[{{Davidge}(2008)}]{davidge08}
{Davidge} TJ (2008) {An Arc of Young Stars in the Halo of M82}. \apjl 678:L85,
  \doi{10.1086/588551}, \eprint{0803.3613}

\bibitem[{{Davidge}(2010)}]{davidge10}
{Davidge} TJ (2010) {Shaken, Not Stirred: The Disrupted Disk of the Starburst
  Galaxy NGC 253}. \apj 725:1342--1365, \doi{10.1088/0004-637X/725/1/1342},
  \eprint{1011.3006}

\bibitem[{{Davidge} and {Jones}(1989)}]{davidge89}
{Davidge} TJ, {Jones} JH (1989) {The evolved stellar content of Holmberg IX}.
  \aj 97:1607--1613, \doi{10.1086/115102}

\bibitem[{{de Jong} et~al(2007){de Jong}, {Seth}, {Radburn-Smith}, {Bell},
  {Brown}, {Bullock}, {Courteau}, {Dalcanton}, {Ferguson}, {Goudfrooij},
  {Holfeltz}, {Holwerda}, {Purcell}, {Sick}, and {Zucker}}]{dejong07}
{de Jong} RS, {Seth} AC, {Radburn-Smith} DJ, {Bell} EF, {Brown} TM, {Bullock}
  JS, {Courteau} S, {Dalcanton} JJ, {Ferguson} HC, {Goudfrooij} P, {Holfeltz}
  S, {Holwerda} BW, {Purcell} C, {Sick} J, {Zucker} DB (2007) {Stellar
  Populations across the NGC 4244 Truncated Galactic Disk}. \apjl 667:L49--L52,
  \doi{10.1086/522035}, \eprint{0708.0826}

\bibitem[{{de Mello} et~al(2008){de Mello}, {Smith}, {Sabbi}, {Gallagher},
  {Mountain}, and {Harbeck}}]{demello08}
{de Mello} DF, {Smith} LJ, {Sabbi} E, {Gallagher} JS, {Mountain} M, {Harbeck}
  DR (2008) {Star Formation in the H I Bridge Between M81 and M82}. \aj
  135:548--554, \doi{10.1088/0004-6256/135/2/548}, \eprint{0711.2685}

\bibitem[{{Deason} et~al(2011){Deason}, {Belokurov}, and {Evans}}]{deason11}
{Deason} AJ, {Belokurov} V, {Evans} NW (2011) {The Milky Way stellar halo out
  to 40 kpc: squashed, broken but smooth}. \mnras 416:2903--2915,
  \doi{10.1111/j.1365-2966.2011.19237.x}, \eprint{1104.3220}

\bibitem[{{Deason} et~al(2013){Deason}, {Belokurov}, {Evans}, and
  {Johnston}}]{deason13}
{Deason} AJ, {Belokurov} V, {Evans} NW, {Johnston} KV (2013) {Broken and
  Unbroken: The Milky Way and M31 Stellar Halos}. \apj 763:113,
  \doi{10.1088/0004-637X/763/2/113}, \eprint{1210.4929}

\bibitem[{{D'Onghia} and {Lake}(2008)}]{donghia08}
{D'Onghia} E, {Lake} G (2008) {Small Dwarf Galaxies within Larger Dwarfs: Why
  Some Are Luminous while Most Go Dark}. \apjl 686:L61, \doi{10.1086/592995},
  \eprint{0802.0001}

\bibitem[{{Dooley} et~al(2016){Dooley}, {Peter}, {Yang}, {Willman}, {Griffen},
  and {Frebel}}]{dooley16}
{Dooley} GA, {Peter} AHG, {Yang} T, {Willman} B, {Griffen} BF, {Frebel} A
  (2016) {An observer's guide to the (Local Group) dwarf galaxies: predictions
  for their own dwarf satellite populations}. ArXiv e-prints
  \eprint{1610.00708}

\bibitem[{{Dorman} et~al(2013){Dorman}, {Widrow}, {Guhathakurta}, {Seth},
  {Foreman-Mackey}, {Bell}, {Dalcanton}, {Gilbert}, {Skillman}, and
  {Williams}}]{dorman13}
{Dorman} CE, {Widrow} LM, {Guhathakurta} P, {Seth} AC, {Foreman-Mackey} D,
  {Bell} EF, {Dalcanton} JJ, {Gilbert} KM, {Skillman} ED, {Williams} BF (2013)
  {A New Approach to Detailed Structural Decomposition from the SPLASH and PHAT
  Surveys: Kicked-up Disk Stars in the Andromeda Galaxy?} \apj 779:103,
  \doi{10.1088/0004-637X/779/2/103}, \eprint{1310.4179}

\bibitem[{{Drlica-Wagner} et~al(2015){Drlica-Wagner}, {Bechtol}, {Rykoff},
  {Luque}, {Queiroz}, {Mao}, {Wechsler}, {Simon}, {Santiago}, {Yanny},
  {Balbinot}, {Dodelson}, {Fausti Neto}, {James}, {Li}, {Maia}, {Marshall},
  {Pieres}, {Stringer}, {Walker}, {Abbott}, {Abdalla}, {Allam},
  {Benoit-L{\'e}vy}, {Bernstein}, {Bertin}, {Brooks}, {Buckley-Geer}, {Burke},
  {Carnero Rosell}, {Carrasco Kind}, {Carretero}, {Crocce}, {da Costa},
  {Desai}, {Diehl}, {Dietrich}, {Doel}, {Eifler}, {Evrard}, {Finley},
  {Flaugher}, {Fosalba}, {Frieman}, {Gaztanaga}, {Gerdes}, {Gruen}, {Gruendl},
  {Gutierrez}, {Honscheid}, {Kuehn}, {Kuropatkin}, {Lahav}, {Martini},
  {Miquel}, {Nord}, {Ogando}, {Plazas}, {Reil}, {Roodman}, {Sako}, {Sanchez},
  {Scarpine}, {Schubnell}, {Sevilla-Noarbe}, {Smith}, {Soares-Santos},
  {Sobreira}, {Suchyta}, {Swanson}, {Tarle}, {Tucker}, {Vikram}, {Wester},
  {Zhang}, {Zuntz}, and {DES Collaboration}}]{drlica16}
{Drlica-Wagner} A, {Bechtol} K, {Rykoff} ES, {Luque} E, {Queiroz} A, {Mao} YY,
  {Wechsler} RH, {Simon} JD, {Santiago} B, {Yanny} B, {Balbinot} E, {Dodelson}
  S, {Fausti Neto} A, {James} DJ, {Li} TS, {Maia} MAG, {Marshall} JL, {Pieres}
  A, {Stringer} K, {Walker} AR, {Abbott} TMC, {Abdalla} FB, {Allam} S,
  {Benoit-L{\'e}vy} A, {Bernstein} GM, {Bertin} E, {Brooks} D, {Buckley-Geer}
  E, {Burke} DL, {Carnero Rosell} A, {Carrasco Kind} M, {Carretero} J, {Crocce}
  M, {da Costa} LN, {Desai} S, {Diehl} HT, {Dietrich} JP, {Doel} P, {Eifler}
  TF, {Evrard} AE, {Finley} DA, {Flaugher} B, {Fosalba} P, {Frieman} J,
  {Gaztanaga} E, {Gerdes} DW, {Gruen} D, {Gruendl} RA, {Gutierrez} G,
  {Honscheid} K, {Kuehn} K, {Kuropatkin} N, {Lahav} O, {Martini} P, {Miquel} R,
  {Nord} B, {Ogando} R, {Plazas} AA, {Reil} K, {Roodman} A, {Sako} M, {Sanchez}
  E, {Scarpine} V, {Schubnell} M, {Sevilla-Noarbe} I, {Smith} RC,
  {Soares-Santos} M, {Sobreira} F, {Suchyta} E, {Swanson} MEC, {Tarle} G,
  {Tucker} D, {Vikram} V, {Wester} W, {Zhang} Y, {Zuntz} J, {DES Collaboration}
  (2015) {Eight Ultra-faint Galaxy Candidates Discovered in Year Two of the
  Dark Energy Survey}. \apj 813:109, \doi{10.1088/0004-637X/813/2/109},
  \eprint{1508.03622}

\bibitem[{{Duc} et~al(2000){Duc}, {Brinks}, {Springel}, {Pichardo},
  {Weilbacher}, and {Mirabel}}]{duc00}
{Duc} P, {Brinks} E, {Springel} V, {Pichardo} B, {Weilbacher} P, {Mirabel} IF
  (2000) {Formation of a Tidal Dwarf Galaxy in the Interacting System Arp 245
  (NGC 2992/93)}. \aj 120:1238--1264, \doi{10.1086/301516},
  \eprint{arXiv:astro-ph/0006038}

\bibitem[{{Duffau} et~al(2006){Duffau}, {Zinn}, {Vivas}, {Carraro},
  {M{\'e}ndez}, {Winnick}, and {Gallart}}]{duffau06}
{Duffau} S, {Zinn} R, {Vivas} AK, {Carraro} G, {M{\'e}ndez} RA, {Winnick} R,
  {Gallart} C (2006) {Spectroscopy of QUEST RR Lyrae Variables: The New Virgo
  Stellar Stream}. \apjl 636:L97--L100, \doi{10.1086/500130},
  \eprint{astro-ph/0510589}

\bibitem[{{Durrell} et~al(2001){Durrell}, {Harris}, and {Pritchet}}]{durrell01}
{Durrell} PR, {Harris} WE, {Pritchet} CJ (2001) {Photometry and the Metallicity
  Distribution of the Outer Halo of M31}. \aj 121:2557--2571,
  \doi{10.1086/320403}, \eprint{arXiv:astro-ph/0101436}

\bibitem[{{Durrell} et~al(2004){Durrell}, {Decesar}, {Ciardullo},
  {Hurley-Keller}, and {Feldmeier}}]{durrell04}
{Durrell} PR, {Decesar} ME, {Ciardullo} R, {Hurley-Keller} D, {Feldmeier} JJ
  (2004) {A CFH12K Survey of Red Giant Stars in the M81 Group}. In: {Duc} PA,
  {Braine} J, {Brinks} E (eds) Recycling Intergalactic and Interstellar Matter,
  IAU Symposium, vol 217, p~90, \eprint{astro-ph/0311130}

\bibitem[{{Durrell} et~al(2010){Durrell}, {Sarajedini}, and
  {Chandar}}]{durrell10}
{Durrell} PR, {Sarajedini} A, {Chandar} R (2010) {Deep HST/ACS Photometry of
  the M81 Halo}. \apj 718:1118--1127, \doi{10.1088/0004-637X/718/2/1118},
  \eprint{1006.2036}

\bibitem[{{Eggen} et~al(1962){Eggen}, {Lynden-Bell}, and {Sandage}}]{eggen62}
{Eggen} OJ, {Lynden-Bell} D, {Sandage} AR (1962) {Evidence from the motions of
  old stars that the Galaxy collapsed.} \apj 136:748, \doi{10.1086/147433}

\bibitem[{{Elson}(1997)}]{elson97}
{Elson} RAW (1997) {Red giants in the halo of the S0 galaxy NGC 3115: a
  distance and a bimodal metallicity distribution}. \mnras 286:771--776,
  \doi{10.1093/mnras/286.3.771}, \eprint{astro-ph/9612037}

\bibitem[{{Fardal} et~al(2013){Fardal}, {Weinberg}, {Babul}, {Irwin},
  {Guhathakurta}, {Gilbert}, {Ferguson}, {Ibata}, {Lewis}, {Tanvir}, and
  {Huxor}}]{fardal13}
{Fardal} MA, {Weinberg} MD, {Babul} A, {Irwin} MJ, {Guhathakurta} P, {Gilbert}
  KM, {Ferguson} AMN, {Ibata} RA, {Lewis} GF, {Tanvir} NR, {Huxor} AP (2013)
  {Inferring the Andromeda Galaxy's mass from its giant southern stream with
  Bayesian simulation sampling}. \mnras 434:2779--2802,
  \doi{10.1093/mnras/stt1121}, \eprint{1307.3219}

\bibitem[{{Ferguson} and {Mackey}(2016)}]{ferguson16}
{Ferguson} AMN, {Mackey} AD (2016) {Substructure and Tidal Streams in the
  Andromeda Galaxy and its Satellites}. In: {Newberg} HJ, {Carlin} JL (eds)
  Astrophysics and Space Science Library, Astrophysics and Space Science
  Library, vol 420, p 191, \doi{10.1007/978-3-319-19336-6\_8},
  \eprint{1603.01993}

\bibitem[{{Ferguson} et~al(2002){Ferguson}, {Irwin}, {Ibata}, {Lewis}, and
  {Tanvir}}]{ferguson02}
{Ferguson} AMN, {Irwin} MJ, {Ibata} RA, {Lewis} GF, {Tanvir} NR (2002)
  {Evidence for Stellar Substructure in the Halo and Outer Disk of M31}. \aj
  124:1452--1463, \doi{10.1086/342019}, \eprint{arXiv:astro-ph/0205530}

\bibitem[{{Ferguson} et~al(2005){Ferguson}, {Johnson}, {Faria}, {Irwin},
  {Ibata}, {Johnston}, {Lewis}, and {Tanvir}}]{ferguson05}
{Ferguson} AMN, {Johnson} RA, {Faria} DC, {Irwin} MJ, {Ibata} RA, {Johnston}
  KV, {Lewis} GF, {Tanvir} NR (2005) {The Stellar Populations of the M31 Halo
  Substructure}. \apjl 622:L109--L112, \doi{10.1086/429371},
  \eprint{astro-ph/0501511}

\bibitem[{{Fernando} et~al(2016){Fernando}, {Arias}, {Guglielmo}, {Lewis},
  {Ibata}, and {Power}}]{fernando16}
{Fernando} N, {Arias} V, {Guglielmo} M, {Lewis} GF, {Ibata} RA, {Power} C
  (2016) {On the Stability of Satellite Planes I: Effects of Mass, Velocity,
  Halo Shape and Alignment}. ArXiv e-prints \eprint{1610.05393}

\bibitem[{{Font} et~al(2006){Font}, {Johnston}, {Bullock}, and
  {Robertson}}]{font06}
{Font} AS, {Johnston} KV, {Bullock} JS, {Robertson} BE (2006) {Chemical
  Abundance Distributions of Galactic Halos and Their Satellite Systems in a
  {$\Lambda$}CDM Universe}. \apj 638:585--595, \doi{10.1086/498970},
  \eprint{astro-ph/0507114}

\bibitem[{{Font} et~al(2011){Font}, {McCarthy}, {Crain}, {Theuns}, {Schaye},
  {Wiersma}, and {Dalla Vecchia}}]{font11}
{Font} AS, {McCarthy} IG, {Crain} RA, {Theuns} T, {Schaye} J, {Wiersma} RPC,
  {Dalla Vecchia} C (2011) {Cosmological simulations of the formation of the
  stellar haloes around disc galaxies}. \mnras 416:2802--2820,
  \doi{10.1111/j.1365-2966.2011.19227.x}, \eprint{1102.2526}

\bibitem[{{Frebel} and {Norris}(2015)}]{frebel15}
{Frebel} A, {Norris} JE (2015) {Near-Field Cosmology with Extremely Metal-Poor
  Stars}. \araa 53:631--688, \doi{10.1146/annurev-astro-082214-122423},
  \eprint{1501.06921}

\bibitem[{{Freeman} and {Bland-Hawthorn}(2002)}]{freeman02}
{Freeman} K, {Bland-Hawthorn} J (2002) {The New Galaxy: Signatures of Its
  Formation}. \araa 40:487--537, \doi{10.1146/annurev.astro.40.060401.093840},
  \eprint{astro-ph/0208106}

\bibitem[{{Gallart} et~al(2004){Gallart}, {Stetson}, {Hardy}, {Pont}, and
  {Zinn}}]{gallart04}
{Gallart} C, {Stetson} PB, {Hardy} E, {Pont} F, {Zinn} R (2004) {Surface
  Brightness and Stellar Populations at the Outer Edge of the Large Magellanic
  Cloud: No Stellar Halo Yet}. \apjl 614:L109--L112, \doi{10.1086/425866},
  \eprint{astro-ph/0409023}

\bibitem[{{Gallart} et~al(2005){Gallart}, {Zoccali}, and
  {Aparicio}}]{gallart05}
{Gallart} C, {Zoccali} M, {Aparicio} A (2005) {The Adequacy of Stellar
  Evolution Models for the Interpretation of the Color-Magnitude Diagrams of
  Resolved Stellar Populations}. \araa 43:387--434,
  \doi{10.1146/annurev.astro.43.072103.150608}

\bibitem[{{Garrison-Kimmel} et~al(2014){Garrison-Kimmel}, {Boylan-Kolchin},
  {Bullock}, and {Kirby}}]{garrison14}
{Garrison-Kimmel} S, {Boylan-Kolchin} M, {Bullock} JS, {Kirby} EN (2014) {Too
  big to fail in the Local Group}. \mnras 444:222--236,
  \doi{10.1093/mnras/stu1477}, \eprint{1404.5313}

\bibitem[{{Georgiev} et~al(1992){Georgiev}, {Bilkina}, and
  {Tikhonov}}]{georgiev92}
{Georgiev} TB, {Bilkina} BI, {Tikhonov} NA (1992) {The distribution of blue and
  red stars around the M81 galaxy}. \aaps 96:569--581

\bibitem[{{Gilbert} et~al(2009){Gilbert}, {Guhathakurta}, {Kollipara},
  {Beaton}, {Geha}, {Kalirai}, {Kirby}, {Majewski}, and
  {Patterson}}]{gilbert09}
{Gilbert} KM, {Guhathakurta} P, {Kollipara} P, {Beaton} RL, {Geha} MC,
  {Kalirai} JS, {Kirby} EN, {Majewski} SR, {Patterson} RJ (2009) {The Splash
  Survey: A Spectroscopic Portrait of Andromeda's Giant Southern Stream}. \apj
  705:1275--1297, \doi{10.1088/0004-637X/705/2/1275}, \eprint{0909.4540}

\bibitem[{{Gilbert} et~al(2012){Gilbert}, {Guhathakurta}, {Beaton}, {Bullock},
  {Geha}, {Kalirai}, {Kirby}, {Majewski}, {Ostheimer}, {Patterson}, {Tollerud},
  {Tanaka}, and {Chiba}}]{gilbert12}
{Gilbert} KM, {Guhathakurta} P, {Beaton} RL, {Bullock} J, {Geha} MC, {Kalirai}
  JS, {Kirby} EN, {Majewski} SR, {Ostheimer} JC, {Patterson} RJ, {Tollerud} EJ,
  {Tanaka} M, {Chiba} M (2012) {Global Properties of M31's Stellar Halo from
  the SPLASH Survey. I. Surface Brightness Profile}. \apj 760:76,
  \doi{10.1088/0004-637X/760/1/76}, \eprint{1210.3362}

\bibitem[{{Gilbert} et~al(2014){Gilbert}, {Kalirai}, {Guhathakurta}, {Beaton},
  {Geha}, {Kirby}, {Majewski}, {Patterson}, {Tollerud}, {Bullock}, {Tanaka},
  and {Chiba}}]{gilbert14}
{Gilbert} KM, {Kalirai} JS, {Guhathakurta} P, {Beaton} RL, {Geha} MC, {Kirby}
  EN, {Majewski} SR, {Patterson} RJ, {Tollerud} EJ, {Bullock} JS, {Tanaka} M,
  {Chiba} M (2014) {Global Properties of M31's Stellar Halo from the SPLASH
  Survey. II. Metallicity Profile}. \apj 796:76,
  \doi{10.1088/0004-637X/796/2/76}, \eprint{1409.3843}

\bibitem[{{Grebel} et~al(2000){Grebel}, {Seitzer}, {Dolphin}, {Geisler},
  {Guhathakurta}, {Hodge}, {Karachentseva}, and {Sarajedini}}]{grebel00}
{Grebel} EK, {Seitzer} P, {Dolphin} A, {Geisler} D, {Guhathakurta} P, {Hodge}
  P, {Karachentseva} I, {Sarajedini} A (2000) {A Dwarf Galaxy Survey in the
  Local Volume}. In: {D~Alloin, K~Olsen, \& G~Galaz} (ed) Stars, Gas and Dust
  in Galaxies: Exploring the Links, Astronomical Society of the Pacific
  Conference Series, vol 221, p 147

\bibitem[{{Greggio} et~al(2014){Greggio}, {Rejkuba}, {Gonzalez}, {Arnaboldi},
  {Iodice}, {Irwin}, {Neeser}, and {Emerson}}]{greggio14}
{Greggio} L, {Rejkuba} M, {Gonzalez} OA, {Arnaboldi} M, {Iodice} E, {Irwin} M,
  {Neeser} MJ, {Emerson} J (2014) {A panoramic VISTA of the stellar halo of NGC
  253}. \aap 562:A73, \doi{10.1051/0004-6361/201322759}, \eprint{1401.1665}

\bibitem[{{Greggio} et~al(2016){Greggio}, {Falomo}, and {Uslenghi}}]{greggio16}
{Greggio} L, {Falomo} R, {Uslenghi} M (2016) {Studying stellar halos with
  future facilities}. In: {Bragaglia} A, {Arnaboldi} M, {Rejkuba} M, {Romano} D
  (eds) The General Assembly of Galaxy Halos: Structure, Origin and Evolution,
  IAU Symposium, vol 317, pp 209--214, \doi{10.1017/S1743921315007024},
  \eprint{1510.03181}

\bibitem[{{Grillmair}(2006)}]{grillmair06}
{Grillmair} CJ (2006) {Detection of a 60${\deg}$-long Dwarf Galaxy Debris
  Stream}. \apjl 645:L37--L40, \doi{10.1086/505863}, \eprint{astro-ph/0605396}

\bibitem[{{Guhathakurta} et~al(2006){Guhathakurta}, {Rich}, {Reitzel},
  {Cooper}, {Gilbert}, {Majewski}, {Ostheimer}, {Geha}, {Johnston}, and
  {Patterson}}]{guha06}
{Guhathakurta} P, {Rich} RM, {Reitzel} DB, {Cooper} MC, {Gilbert} KM,
  {Majewski} SR, {Ostheimer} JC, {Geha} MC, {Johnston} KV, {Patterson} RJ
  (2006) {Dynamics and Stellar Content of the Giant Southern Stream in M31. I.
  Keck Spectroscopy of Red Giant Stars}. \aj 131:2497--2513,
  \doi{10.1086/499562}, \eprint{astro-ph/0406145}

\bibitem[{{Hammer} et~al(2007){Hammer}, {Puech}, {Chemin}, {Flores}, and
  {Lehnert}}]{hammer07}
{Hammer} F, {Puech} M, {Chemin} L, {Flores} H, {Lehnert} MD (2007) {The Milky
  Way, an Exceptionally Quiet Galaxy: Implications for the Formation of Spiral
  Galaxies}. \apj 662:322--334, \doi{10.1086/516727}, \eprint{astro-ph/0702585}

\bibitem[{{Harris} et~al(2010){Harris}, {Rejkuba}, and {Harris}}]{harrisg09}
{Harris} GLH, {Rejkuba} M, {Harris} WE (2010) {The Distance to NGC 5128
  (Centaurus A)}. \pasa 27:457--462, \doi{10.1071/AS09061}, \eprint{0911.3180}

\bibitem[{{Harris} et~al(2007{\natexlab{a}}){Harris}, {Harris}, {Layden}, and
  {Stetson}}]{harris07b}
{Harris} WE, {Harris} GLH, {Layden} AC, {Stetson} PB (2007{\natexlab{a}})
  {Hubble Space Telescope Photometry for the Halo Stars in the Leo Elliptical
  NGC 3377}. \aj 134:43--55, \doi{10.1086/518233}, \eprint{0706.1997}

\bibitem[{{Harris} et~al(2007{\natexlab{b}}){Harris}, {Harris}, {Layden}, and
  {Wehner}}]{harris07a}
{Harris} WE, {Harris} GLH, {Layden} AC, {Wehner} EMH (2007{\natexlab{b}}) {The
  Leo Elliptical NGC 3379: A Metal-Poor Halo Emerges}. \apj 666:903--918,
  \doi{10.1086/520799}, \eprint{0706.1995}

\bibitem[{{Hidalgo} et~al(2013){Hidalgo}, {Monelli}, {Aparicio}, {Gallart},
  {Skillman}, {Cassisi}, {Bernard}, {Mayer}, {Stetson}, {Cole}, and
  {Dolphin}}]{hidalgo13}
{Hidalgo} SL, {Monelli} M, {Aparicio} A, {Gallart} C, {Skillman} ED, {Cassisi}
  S, {Bernard} EJ, {Mayer} L, {Stetson} P, {Cole} A, {Dolphin} A (2013) {The
  ACS LCID Project. IX. Imprints of the Early Universe in the Radial Variation
  of the Star Formation History of Dwarf Galaxies}. \apj 778:103,
  \doi{10.1088/0004-637X/778/2/103}, \eprint{1309.6130}

\bibitem[{{Higgs} et~al(2016){Higgs}, {McConnachie}, {Irwin}, {Bate}, {Lewis},
  {Walker}, {C{\^o}t{\'e}}, {Venn}, and {Battaglia}}]{higgs16}
{Higgs} CR, {McConnachie} AW, {Irwin} M, {Bate} NF, {Lewis} GF, {Walker} MG,
  {C{\^o}t{\'e}} P, {Venn} K, {Battaglia} G (2016) {Solo dwarfs I: survey
  introduction and first results for the Sagittarius dwarf irregular galaxy}.
  \mnras 458:1678--1695, \doi{10.1093/mnras/stw257}, \eprint{1602.01881}

\bibitem[{{Humphreys} et~al(1986){Humphreys}, {Aaronson}, {Lebofsky},
  {McAlary}, {Strom}, and {Capps}}]{humphreys86}
{Humphreys} RM, {Aaronson} M, {Lebofsky} M, {McAlary} CW, {Strom} SE, {Capps}
  RW (1986) {The luminosities of M supergiants and the distances to M101, NGC
  2403, and M81}. \aj 91:808--821, \doi{10.1086/114061}

\bibitem[{{Huxor} et~al(2014){Huxor}, {Mackey}, {Ferguson}, {Irwin}, {Martin},
  {Tanvir}, {Veljanoski}, {McConnachie}, {Fishlock}, {Ibata}, and
  {Lewis}}]{huxor14}
{Huxor} AP, {Mackey} AD, {Ferguson} AMN, {Irwin} MJ, {Martin} NF, {Tanvir} NR,
  {Veljanoski} J, {McConnachie} A, {Fishlock} CK, {Ibata} R, {Lewis} GF (2014)
  {The outer halo globular cluster system of M31 - I. The final PAndAS
  catalogue}. \mnras 442:2165--2187, \doi{10.1093/mnras/stu771},
  \eprint{1404.5807}

\bibitem[{{Ibata} et~al(2001){Ibata}, {Irwin}, {Lewis}, {Ferguson}, and
  {Tanvir}}]{ibata01}
{Ibata} R, {Irwin} M, {Lewis} G, {Ferguson} AMN, {Tanvir} N (2001) {A giant
  stream of metal-rich stars in the halo of the galaxy M31}. \nat 412:49--52,
  \eprint{astro-ph/0107090}

\bibitem[{{Ibata} et~al(2007){Ibata}, {Martin}, {Irwin}, {Chapman}, {Ferguson},
  {Lewis}, and {McConnachie}}]{ibata07}
{Ibata} R, {Martin} NF, {Irwin} M, {Chapman} S, {Ferguson} AMN, {Lewis} GF,
  {McConnachie} AW (2007) {The Haunted Halos of Andromeda and Triangulum: A
  Panorama of Galaxy Formation in Action}. \apj 671:1591--1623,
  \doi{10.1086/522574}, \eprint{0704.1318}

\bibitem[{{Ibata} et~al(2009){Ibata}, {Mouhcine}, and {Rejkuba}}]{ibata09}
{Ibata} R, {Mouhcine} M, {Rejkuba} M (2009) {An HST/ACS investigation of the
  spatial and chemical structure and sub-structure of NGC 891, a Milky Way
  analogue}. \mnras 395:126--143, \doi{10.1111/j.1365-2966.2009.14536.x},
  \eprint{0903.4209}

\bibitem[{{Ibata} et~al(1994){Ibata}, {Gilmore}, and {Irwin}}]{ibata94}
{Ibata} RA, {Gilmore} G, {Irwin} MJ (1994) {A dwarf satellite galaxy in
  Sagittarius}. \nat 370:194--196, \doi{10.1038/370194a0}

\bibitem[{{Ibata} et~al(2013){Ibata}, {Lewis}, {Conn}, {Irwin}, {McConnachie},
  {Chapman}, {Collins}, {Fardal}, {Ferguson}, {Ibata}, {Mackey}, {Martin},
  {Navarro}, {Rich}, {Valls-Gabaud}, and {Widrow}}]{ibata13}
{Ibata} RA, {Lewis} GF, {Conn} AR, {Irwin} MJ, {McConnachie} AW, {Chapman} SC,
  {Collins} ML, {Fardal} M, {Ferguson} AMN, {Ibata} NG, {Mackey} AD, {Martin}
  NF, {Navarro} J, {Rich} RM, {Valls-Gabaud} D, {Widrow} LM (2013) {A vast,
  thin plane of corotating dwarf galaxies orbiting the Andromeda galaxy}. \nat
  493:62--65, \doi{10.1038/nature11717}, \eprint{1301.0446}

\bibitem[{{Ibata} et~al(2014){Ibata}, {Lewis}, {McConnachie}, {Martin},
  {Irwin}, {Ferguson}, {Babul}, {Bernard}, {Chapman}, {Collins}, {Fardal},
  {Mackey}, {Navarro}, {Pe{\~n}arrubia}, {Rich}, {Tanvir}, and
  {Widrow}}]{ibata14}
{Ibata} RA, {Lewis} GF, {McConnachie} AW, {Martin} NF, {Irwin} MJ, {Ferguson}
  AMN, {Babul} A, {Bernard} EJ, {Chapman} SC, {Collins} M, {Fardal} M, {Mackey}
  AD, {Navarro} J, {Pe{\~n}arrubia} J, {Rich} RM, {Tanvir} N, {Widrow} L (2014)
  {The Large-scale Structure of the Halo of the Andromeda Galaxy. I. Global
  Stellar Density, Morphology and Metallicity Properties}. \apj 780:128,
  \doi{10.1088/0004-637X/780/2/128}, \eprint{1311.5888}

\bibitem[{{Irwin} et~al(2005){Irwin}, {Ferguson}, {Ibata}, {Lewis}, and
  {Tanvir}}]{irwin05}
{Irwin} MJ, {Ferguson} AMN, {Ibata} RA, {Lewis} GF, {Tanvir} NR (2005) {A
  Minor-Axis Surface Brightness Profile for M31}. \apjl 628:L105--L108,
  \doi{10.1086/432718}, \eprint{astro-ph/0505077}

\bibitem[{{Israel}(1998)}]{israel98}
{Israel} FP (1998) {Centaurus A - NGC 5128}. A\&AR 8:237--278,
  \doi{10.1007/s001590050011}, \eprint{arXiv:astro-ph/9811051}

\bibitem[{{Ivezi{\'c}} et~al(2000){Ivezi{\'c}}, {Goldston}, {Finlator},
  {Knapp}, {Yanny}, {McKay}, {Amrose}, {Krisciunas}, {Willman}, {Anderson},
  {Schaber}, {Erb}, {Logan}, {Stubbs}, {Chen}, {Neilsen}, {Uomoto}, {Pier},
  {Fan}, {Gunn}, {Lupton}, {Rockosi}, {Schlegel}, {Strauss}, {Annis},
  {Brinkmann}, {Csabai}, {Doi}, {Fukugita}, {Hennessy}, {Hindsley}, {Margon},
  {Munn}, {Newberg}, {Schneider}, {Smith}, {Szokoly}, {Thakar}, {Vogeley},
  {Waddell}, {Yasuda}, {York}, and {SDSS Collaboration}}]{ivezic00}
{Ivezi{\'c}} {\v Z}, {Goldston} J, {Finlator} K, {Knapp} GR, {Yanny} B, {McKay}
  TA, {Amrose} S, {Krisciunas} K, {Willman} B, {Anderson} S, {Schaber} C, {Erb}
  D, {Logan} C, {Stubbs} C, {Chen} B, {Neilsen} E, {Uomoto} A, {Pier} JR, {Fan}
  X, {Gunn} JE, {Lupton} RH, {Rockosi} CM, {Schlegel} D, {Strauss} MA, {Annis}
  J, {Brinkmann} J, {Csabai} I, {Doi} M, {Fukugita} M, {Hennessy} GS,
  {Hindsley} RB, {Margon} B, {Munn} JA, {Newberg} HJ, {Schneider} DP, {Smith}
  JA, {Szokoly} GP, {Thakar} AR, {Vogeley} MS, {Waddell} P, {Yasuda} N, {York}
  DG, {SDSS Collaboration} (2000) {Candidate RR Lyrae Stars Found in Sloan
  Digital Sky Survey Commissioning Data}. \aj 120:963--977,
  \doi{10.1086/301455}, \eprint{astro-ph/0004130}

\bibitem[{{Jablonka} et~al(2010){Jablonka}, {Tafelmeyer}, {Courbin}, and
  {Ferguson}}]{jablonka10}
{Jablonka} P, {Tafelmeyer} M, {Courbin} F, {Ferguson} AMN (2010) {Direct
  detection of galaxy stellar halos: NGC 3957 as a test case}. \aap 513:A78,
  \doi{10.1051/0004-6361/200913320}, \eprint{1001.3067}

\bibitem[{{Janowiecki} et~al(2010){Janowiecki}, {Mihos}, {Harding},
  {Feldmeier}, {Rudick}, and {Morrison}}]{janowiecki10}
{Janowiecki} S, {Mihos} JC, {Harding} P, {Feldmeier} JJ, {Rudick} C, {Morrison}
  H (2010) {Diffuse Tidal Structures in the Halos of Virgo Ellipticals}. \apj
  715:972--985, \doi{10.1088/0004-637X/715/2/972}, \eprint{1004.1473}

\bibitem[{{Johnston} et~al(2008){Johnston}, {Bullock}, {Sharma}, {Font},
  {Robertson}, and {Leitner}}]{johnston08}
{Johnston} KV, {Bullock} JS, {Sharma} S, {Font} A, {Robertson} BE, {Leitner} SN
  (2008) {Tracing Galaxy Formation with Stellar Halos. II. Relating
  Substructure in Phase and Abundance Space to Accretion Histories}. \apj
  689:936-957, \doi{10.1086/592228}, \eprint{0807.3911}

\bibitem[{{Juri{\'c}} et~al(2008){Juri{\'c}}, {Ivezi{\'c}}, {Brooks}, {Lupton},
  {Schlegel}, {Finkbeiner}, {Padmanabhan}, {Bond}, {Sesar}, {Rockosi}, {Knapp},
  {Gunn}, {Sumi}, {Schneider}, {Barentine}, {Brewington}, {Brinkmann},
  {Fukugita}, {Harvanek}, {Kleinman}, {Krzesinski}, {Long}, {Neilsen}, {Nitta},
  {Snedden}, and {York}}]{juric08}
{Juri{\'c}} M, {Ivezi{\'c}} {\v Z}, {Brooks} A, {Lupton} RH, {Schlegel} D,
  {Finkbeiner} D, {Padmanabhan} N, {Bond} N, {Sesar} B, {Rockosi} CM, {Knapp}
  GR, {Gunn} JE, {Sumi} T, {Schneider} DP, {Barentine} JC, {Brewington} HJ,
  {Brinkmann} J, {Fukugita} M, {Harvanek} M, {Kleinman} SJ, {Krzesinski} J,
  {Long} D, {Neilsen} EH Jr, {Nitta} A, {Snedden} SA, {York} DG (2008) {The
  Milky Way Tomography with SDSS. I. Stellar Number Density Distribution}. \apj
  673:864-914, \doi{10.1086/523619}, \eprint{astro-ph/0510520}

\bibitem[{{Kalirai} et~al(2006){Kalirai}, {Gilbert}, {Guhathakurta},
  {Majewski}, {Ostheimer}, {Rich}, {Cooper}, {Reitzel}, and
  {Patterson}}]{kalirai06}
{Kalirai} JS, {Gilbert} KM, {Guhathakurta} P, {Majewski} SR, {Ostheimer} JC,
  {Rich} RM, {Cooper} MC, {Reitzel} DB, {Patterson} RJ (2006) {The Metal-poor
  Halo of the Andromeda Spiral Galaxy (M31)1,}. \apj 648:389--404,
  \doi{10.1086/505697}, \eprint{astro-ph/0605170}

\bibitem[{{Karachentsev} et~al(2004){Karachentsev}, {Karachentseva},
  {Huchtmeier}, and {Makarov}}]{kara04}
{Karachentsev} ID, {Karachentseva} VE, {Huchtmeier} WK, {Makarov} DI (2004) {A
  Catalog of Neighboring Galaxies}. \aj 127:2031--2068, \doi{10.1086/382905}

\bibitem[{{Karachentsev} et~al(2007){Karachentsev}, {Tully}, {Dolphin},
  {Sharina}, {Makarova}, {Makarov}, {Sakai}, {Shaya}, and {et al.}}]{kara07}
{Karachentsev} ID, {Tully} RB, {Dolphin} A, {Sharina} M, {Makarova} L,
  {Makarov} D, {Sakai} S, {Shaya} EJ, {et al} (2007) {The Hubble Flow around
  the Centaurus A/M83 Galaxy Complex}. \aj 133:504--517, \doi{10.1086/510125},
  \eprint{arXiv:astro-ph/0603091}

\bibitem[{{Kim} et~al(2015){Kim}, {Jerjen}, {Mackey}, {Da Costa}, and
  {Milone}}]{kim15a}
{Kim} D, {Jerjen} H, {Mackey} D, {Da Costa} GS, {Milone} AP (2015) {A Hero's
  Dark Horse: Discovery of an Ultra-faint Milky Way Satellite in Pegasus}.
  \apjl 804:L44, \doi{10.1088/2041-8205/804/2/L44}, \eprint{1503.08268}

\bibitem[{{Kirby} et~al(2012){Kirby}, {Cohen}, and {Bellazzini}}]{kirby12}
{Kirby} EN, {Cohen} JG, {Bellazzini} M (2012) {The Dynamics and Metallicity
  Distribution of the Distant Dwarf Galaxy VV124}. \apj 751:46,
  \doi{10.1088/0004-637X/751/1/46}, \eprint{1203.4561}

\bibitem[{{Klypin} et~al(1999){Klypin}, {Kravtsov}, {Valenzuela}, and
  {Prada}}]{klypin99}
{Klypin} A, {Kravtsov} AV, {Valenzuela} O, {Prada} F (1999) {Where Are the
  Missing Galactic Satellites?} \apj 522:82--92, \doi{10.1086/307643},
  \eprint{arXiv:astro-ph/9901240}

\bibitem[{{Koch} et~al(2006){Koch}, {Grebel}, {Wyse}, {Kleyna}, {Wilkinson},
  {Harbeck}, {Gilmore}, and {Evans}}]{koch06}
{Koch} A, {Grebel} EK, {Wyse} RFG, {Kleyna} JT, {Wilkinson} MI, {Harbeck} DR,
  {Gilmore} GF, {Evans} NW (2006) {Complexity on Small Scales: The Metallicity
  Distribution of the Carina Dwarf Spheroidal Galaxy}. \aj 131:895--911,
  \doi{10.1086/499490}, \eprint{arXiv:astro-ph/0511087}

\bibitem[{{Koposov} et~al(2015){Koposov}, {Belokurov}, {Torrealba}, and
  {Evans}}]{koposov15}
{Koposov} SE, {Belokurov} V, {Torrealba} G, {Evans} NW (2015) {Beasts of the
  Southern Wild : Discovery of nine Ultra Faint satellites in the vicinity of
  the Magellanic Clouds.} \apj 805:130, \doi{10.1088/0004-637X/805/2/130},
  \eprint{1503.02079}

\bibitem[{{Lewis} et~al(2013){Lewis}, {Braun}, {McConnachie}, {Irwin}, {Ibata},
  {Chapman}, {Ferguson}, {Martin}, {Fardal}, {Dubinski}, {Widrow}, {Mackey},
  {Babul}, {Tanvir}, and {Rich}}]{lewis13}
{Lewis} GF, {Braun} R, {McConnachie} AW, {Irwin} MJ, {Ibata} RA, {Chapman} SC,
  {Ferguson} AMN, {Martin} NF, {Fardal} M, {Dubinski} J, {Widrow} L, {Mackey}
  AD, {Babul} A, {Tanvir} NR, {Rich} M (2013) {PAndAS in the Mist: The Stellar
  and Gaseous Mass within the Halos of M31 and M33}. \apj 763:4,
  \doi{10.1088/0004-637X/763/1/4}, \eprint{1211.4059}

\bibitem[{{Lu} et~al(2014){Lu}, {Mo}, {Lu}, {Katz}, {Weinberg}, {van den
  Bosch}, and {Yang}}]{lu14}
{Lu} Z, {Mo} HJ, {Lu} Y, {Katz} N, {Weinberg} MD, {van den Bosch} FC, {Yang} X
  (2014) {An empirical model for the star formation history in dark matter
  haloes}. \mnras 439:1294--1312, \doi{10.1093/mnras/stu016},
  \eprint{1306.0650}

\bibitem[{{Mackey} et~al(2010){Mackey}, {Huxor}, {Ferguson}, {Irwin}, {Tanvir},
  {McConnachie}, {Ibata}, {Chapman}, and {Lewis}}]{mackey10}
{Mackey} AD, {Huxor} AP, {Ferguson} AMN, {Irwin} MJ, {Tanvir} NR, {McConnachie}
  AW, {Ibata} RA, {Chapman} SC, {Lewis} GF (2010) {Evidence for an Accretion
  Origin for the Outer Halo Globular Cluster System of M31}. \apjl
  717:L11--L16, \doi{10.1088/2041-8205/717/1/L11}, \eprint{1005.3812}

\bibitem[{{Majewski}(1999)}]{majewski99a}
{Majewski} SR (1999) {The Role of Accretion in the Formation of the Halo:
  Observational View}. In: {Gibson} BK, {Axelrod} RS, {Putman} ME (eds) The
  Third Stromlo Symposium: The Galactic Halo, Astronomical Society of the
  Pacific Conference Series, vol 165, p~76

\bibitem[{{Majewski} et~al(2003){Majewski}, {Skrutskie}, {Weinberg}, and
  {Ostheimer}}]{majewski03}
{Majewski} SR, {Skrutskie} MF, {Weinberg} MD, {Ostheimer} JC (2003) {A Two
  Micron All Sky Survey View of the Sagittarius Dwarf Galaxy. I. Morphology of
  the Sagittarius Core and Tidal Arms}. \apj 599:1082--1115,
  \doi{10.1086/379504}, \eprint{astro-ph/0304198}

\bibitem[{{Makarova} et~al(2002){Makarova}, {Grebel}, {Karachentsev},
  {Dolphin}, {Karachentseva}, {Sharina}, {Geisler}, {Guhathakurta}, {Hodge},
  {Sarajedini}, and {Seitzer}}]{makarova02}
{Makarova} LN, {Grebel} EK, {Karachentsev} ID, {Dolphin} AE, {Karachentseva}
  VE, {Sharina} ME, {Geisler} D, {Guhathakurta} P, {Hodge} PW, {Sarajedini} A,
  {Seitzer} P (2002) {Tidal dwarfs in the M81 group: The second generation?}
  \aap 396:473--487, \doi{10.1051/0004-6361:20021426}

\bibitem[{{Malin} and {Hadley}(1997)}]{malin97}
{Malin} D, {Hadley} B (1997) {HI in Shell Galaxies and Other Merger Remnants}.
  \pasa 14:52--58, \doi{10.1071/AS97052}

\bibitem[{{Malin} et~al(1983){Malin}, {Quinn}, and {Graham}}]{malin83}
{Malin} DF, {Quinn} PJ, {Graham} JA (1983) {Shell structure in NGC 5128}. \apjl
  272:L5--L7, \doi{10.1086/184106}

\bibitem[{{Martell} and {Grebel}(2010)}]{martell10}
{Martell} SL, {Grebel} EK (2010) {Light-element abundance variations in the
  Milky Way halo}. \aap 519:A14, \doi{10.1051/0004-6361/201014135},
  \eprint{1005.4070}

\bibitem[{{Martin} et~al(2015){Martin}, {Nidever}, {Besla}, {Olsen}, {Walker},
  {Vivas}, {Gruendl}, {Kaleida}, {Mu{\~n}oz}, {Blum}, {Saha}, {Conn}, {Bell},
  {Chu}, {Cioni}, {de Boer}, {Gallart}, {Jin}, {Kunder}, {Majewski},
  {Martinez-Delgado}, {Monachesi}, {Monelli}, {Monteagudo}, {No{\"e}l},
  {Olszewski}, {Stringfellow}, {van der Marel}, and {Zaritsky}}]{martin15}
{Martin} NF, {Nidever} DL, {Besla} G, {Olsen} K, {Walker} AR, {Vivas} AK,
  {Gruendl} RA, {Kaleida} CC, {Mu{\~n}oz} RR, {Blum} RD, {Saha} A, {Conn} BC,
  {Bell} EF, {Chu} YH, {Cioni} MRL, {de Boer} TJL, {Gallart} C, {Jin} S,
  {Kunder} A, {Majewski} SR, {Martinez-Delgado} D, {Monachesi} A, {Monelli} M,
  {Monteagudo} L, {No{\"e}l} NED, {Olszewski} EW, {Stringfellow} GS, {van der
  Marel} RP, {Zaritsky} D (2015) {Hydra II: A Faint and Compact Milky Way Dwarf
  Galaxy Found in the Survey of the Magellanic Stellar History}. \apjl 804:L5,
  \doi{10.1088/2041-8205/804/1/L5}, \eprint{1503.06216}

\bibitem[{{Martin} et~al(2016){Martin}, {Ibata}, {Lewis}, {McConnachie},
  {Babul}, {Bate}, {Bernard}, {Chapman}, {Collins}, {Conn}, {Crnojevi{\'c}},
  {Fardal}, {Ferguson}, {Irwin}, {Mackey}, {McMonigal}, {Navarro}, and
  {Rich}}]{martin16}
{Martin} NF, {Ibata} RA, {Lewis} GF, {McConnachie} A, {Babul} A, {Bate} NF,
  {Bernard} E, {Chapman} SC, {Collins} MML, {Conn} AR, {Crnojevi{\'c}} D,
  {Fardal} MA, {Ferguson} AMN, {Irwin} M, {Mackey} AD, {McMonigal} B, {Navarro}
  JF, {Rich} RM (2016) {The PAndAS view of the Andromeda satellite system - II.
  Detailed properties of 23 M31 dwarf spheroidal galaxies}. ArXiv e-prints
  \eprint{1610.01158}

\bibitem[{{Mart{\'{\i}}nez-Delgado} et~al(2010){Mart{\'{\i}}nez-Delgado},
  {Gabany}, {Crawford}, {Zibetti}, {Majewski}, {Rix}, {Fliri}, and {et
  al.}}]{martinez10}
{Mart{\'{\i}}nez-Delgado} D, {Gabany} RJ, {Crawford} K, {Zibetti} S, {Majewski}
  SR, {Rix} H, {Fliri} J, {et al} (2010) {Stellar Tidal Streams in Spiral
  Galaxies of the Local Volume: A Pilot Survey with Modest Aperture
  Telescopes}. \aj 140:962--967, \doi{10.1088/0004-6256/140/4/962},
  \eprint{1003.4860}

\bibitem[{{Mateo}(1998)}]{mateo98}
{Mateo} ML (1998) {Dwarf Galaxies of the Local Group}. \araa 36:435--506,
  \doi{10.1146/annurev.astro.36.1.435}, \eprint{arXiv:astro-ph/9810070}

\bibitem[{{McConnachie} et~al(2003){McConnachie}, {Irwin}, {Ibata}, {Ferguson},
  {Lewis}, and {Tanvir}}]{mcconnachie03}
{McConnachie} AW, {Irwin} MJ, {Ibata} RA, {Ferguson} AMN, {Lewis} GF, {Tanvir}
  N (2003) {The three-dimensional structure of the giant stellar stream in
  Andromeda}. \mnras 343:1335--1340, \doi{10.1046/j.1365-8711.2003.06785.x},
  \eprint{astro-ph/0305160}

\bibitem[{{McConnachie} et~al(2009){McConnachie}, {Irwin}, {Ibata}, {Dubinski},
  {Widrow}, {Martin}, {C{\^o}t{\'e}}, {Dotter}, and {et al.}}]{mcconnachie09}
{McConnachie} AW, {Irwin} MJ, {Ibata} RA, {Dubinski} J, {Widrow} LM, {Martin}
  NF, {C{\^o}t{\'e}} P, {Dotter} AL, {et al} (2009) {The remnants of galaxy
  formation from a panoramic survey of the region around M31}. \nat 461:66--69,
  \doi{10.1038/nature08327}, \eprint{0909.0398}

\bibitem[{{McMonigal} et~al(2014){McMonigal}, {Bate}, {Lewis}, {Irwin},
  {Battaglia}, {Ibata}, {Martin}, {McConnachie}, {Guglielmo}, and
  {Conn}}]{mcmonigal14}
{McMonigal} B, {Bate} NF, {Lewis} GF, {Irwin} MJ, {Battaglia} G, {Ibata} RA,
  {Martin} NF, {McConnachie} AW, {Guglielmo} M, {Conn} AR (2014) {Sailing under
  the Magellanic Clouds: a DECam view of the Carina dwarf}. \mnras
  444:3139--3149, \doi{10.1093/mnras/stu1659}, \eprint{1408.2907}

\bibitem[{{McMonigal} et~al(2016){McMonigal}, {Lewis}, {Brewer}, {Irwin},
  {Martin}, {McConnachie}, {Ibata}, {Ferguson}, {Mackey}, and
  {Chapman}}]{mcmonigal16}
{McMonigal} B, {Lewis} GF, {Brewer} BJ, {Irwin} MJ, {Martin} NF, {McConnachie}
  AW, {Ibata} RA, {Ferguson} AMN, {Mackey} AD, {Chapman} SC (2016) {The elusive
  stellar halo of the Triangulum galaxy}. \mnras 461:4374--4388,
  \doi{10.1093/mnras/stw1657}, \eprint{1607.02190}

\bibitem[{{Monachesi} et~al(2013){Monachesi}, {Bell}, {Radburn-Smith},
  {Vlaji{\'c}}, {de Jong}, {Bailin}, {Dalcanton}, {Holwerda}, and
  {Streich}}]{monachesi13}
{Monachesi} A, {Bell} EF, {Radburn-Smith} DJ, {Vlaji{\'c}} M, {de Jong} RS,
  {Bailin} J, {Dalcanton} JJ, {Holwerda} BW, {Streich} D (2013) {Testing Galaxy
  Formation Models with the GHOSTS Survey: The Color Profile of M81's Stellar
  Halo}. \apj 766:106, \doi{10.1088/0004-637X/766/2/106}, \eprint{1302.2626}

\bibitem[{{Monachesi} et~al(2016){Monachesi}, {Bell}, {Radburn-Smith},
  {Bailin}, {de Jong}, {Holwerda}, {Streich}, and {Silverstein}}]{monachesi16}
{Monachesi} A, {Bell} EF, {Radburn-Smith} DJ, {Bailin} J, {de Jong} RS,
  {Holwerda} B, {Streich} D, {Silverstein} G (2016) {The GHOSTS survey - II.
  The diversity of halo colour and metallicity profiles of massive disc
  galaxies}. \mnras 457:1419--1446, \doi{10.1093/mnras/stv2987},
  \eprint{1507.06657}

\bibitem[{{Moore} et~al(1999){Moore}, {Ghigna}, {Governato}, {Lake}, {Quinn},
  {Stadel}, and {Tozzi}}]{moore99}
{Moore} B, {Ghigna} S, {Governato} F, {Lake} G, {Quinn} T, {Stadel} J, {Tozzi}
  P (1999) {Dark Matter Substructure within Galactic Halos}. \apjl
  524:L19--L22, \doi{10.1086/312287}, \eprint{arXiv:astro-ph/9907411}

\bibitem[{{Morrison}(1993)}]{morrison93}
{Morrison} HL (1993) {The local density of halo giants}. \aj 106:578--590,
  \doi{10.1086/116662}

\bibitem[{{Morrison} et~al(1994){Morrison}, {Boroson}, and
  {Harding}}]{morrison94}
{Morrison} HL, {Boroson} TA, {Harding} P (1994) {Stellar populations in edge-on
  galaxies from deep CCD surface photometry, 1: NGC 5907}. \aj 108:1191--1208,
  \doi{10.1086/117148}

\bibitem[{{Mouhcine}(2006)}]{mouhcine06}
{Mouhcine} M (2006) {The Outskirts of Spiral Galaxies: Evidence for Multiple
  Stellar Populations}. \apj 652:277--282, \doi{10.1086/504104},
  \eprint{astro-ph/0603191}

\bibitem[{{Mouhcine} and {Ibata}(2009)}]{mouhcine09}
{Mouhcine} M, {Ibata} R (2009) {A panoramic view of M81: new stellar systems in
  the debris field}. \mnras 399:737--743,
  \doi{10.1111/j.1365-2966.2009.15135.x}

\bibitem[{{Mouhcine} et~al(2005{\natexlab{a}}){Mouhcine}, {Ferguson}, {Rich},
  {Brown}, and {Smith}}]{mouhcine05a}
{Mouhcine} M, {Ferguson} HC, {Rich} RM, {Brown} TM, {Smith} TE
  (2005{\natexlab{a}}) {Halos of Spiral Galaxies. I. The Tip of the Red Giant
  Branch as a Distance Indicator}. \apj 633:810--820, \doi{10.1086/468177},
  \eprint{astro-ph/0510253}

\bibitem[{{Mouhcine} et~al(2005{\natexlab{b}}){Mouhcine}, {Ferguson}, {Rich},
  {Brown}, and {Smith}}]{mouhcine05b}
{Mouhcine} M, {Ferguson} HC, {Rich} RM, {Brown} TM, {Smith} TE
  (2005{\natexlab{b}}) {Halos of Spiral Galaxies. II. Halo
  Metallicity-Luminosity Relation}. \apj 633:821--827, \doi{10.1086/468178},
  \eprint{astro-ph/0510254}

\bibitem[{{Mouhcine} et~al(2005{\natexlab{c}}){Mouhcine}, {Rich}, {Ferguson},
  {Brown}, and {Smith}}]{mouhcine05c}
{Mouhcine} M, {Rich} RM, {Ferguson} HC, {Brown} TM, {Smith} TE
  (2005{\natexlab{c}}) {Halos of Spiral Galaxies. III. Metallicity
  Distributions}. \apj 633:828--843, \doi{10.1086/468179},
  \eprint{arXiv:astro-ph/0510255}

\bibitem[{{Mouhcine} et~al(2007){Mouhcine}, {Rejkuba}, and
  {Ibata}}]{mouhcine07}
{Mouhcine} M, {Rejkuba} M, {Ibata} R (2007) {The stellar halo of the edge-on
  galaxy NGC 891}. \mnras 381:873--880, \doi{10.1111/j.1365-2966.2007.12291.x}

\bibitem[{{Mouhcine} et~al(2010){Mouhcine}, {Ibata}, and
  {Rejkuba}}]{mouhcine10}
{Mouhcine} M, {Ibata} R, {Rejkuba} M (2010) {A Panoramic View of the Milky Way
  Analog NGC 891}. \apjl 714:L12--L15, \doi{10.1088/2041-8205/714/1/L12},
  \eprint{1002.0461}

\bibitem[{{Mould} and {Kristian}(1986)}]{mould86}
{Mould} J, {Kristian} J (1986) {The stellar population in the halos of M31 and
  M33}. \apj 305:591--599, \doi{10.1086/164273}

\bibitem[{{Newberg} et~al(2002){Newberg}, {Yanny}, {Rockosi}, {Grebel}, {Rix},
  {Brinkmann}, {Csabai}, {Hennessy}, {Hindsley}, {Ibata}, {Ivezi{\'c}}, {Lamb},
  {Nash}, {Odenkirchen}, {Rave}, {Schneider}, {Smith}, {Stolte}, and
  {York}}]{newberg02}
{Newberg} HJ, {Yanny} B, {Rockosi} C, {Grebel} EK, {Rix} HW, {Brinkmann} J,
  {Csabai} I, {Hennessy} G, {Hindsley} RB, {Ibata} R, {Ivezi{\'c}} Z, {Lamb} D,
  {Nash} ET, {Odenkirchen} M, {Rave} HA, {Schneider} DP, {Smith} JA, {Stolte}
  A, {York} DG (2002) {The Ghost of Sagittarius and Lumps in the Halo of the
  Milky Way}. \apj 569:245--274, \doi{10.1086/338983},
  \eprint{astro-ph/0111095}

\bibitem[{{Norris} and {Ryan}(1991)}]{norris91}
{Norris} JE, {Ryan} SG (1991) {Population studies. XI - The extended disk, halo
  configuration}. \apj 380:403--418, \doi{10.1086/170599}

\bibitem[{{Odenkirchen} et~al(2001){Odenkirchen}, {Grebel}, {Rockosi},
  {Dehnen}, {Ibata}, {Rix}, {Stolte}, {Wolf}, {Anderson}, {Bahcall},
  {Brinkmann}, {Csabai}, {Hennessy}, {Hindsley}, {Ivezi{\'c}}, {Lupton},
  {Munn}, {Pier}, {Stoughton}, and {York}}]{odenkirchen01}
{Odenkirchen} M, {Grebel} EK, {Rockosi} CM, {Dehnen} W, {Ibata} R, {Rix} HW,
  {Stolte} A, {Wolf} C, {Anderson} JE Jr, {Bahcall} NA, {Brinkmann} J, {Csabai}
  I, {Hennessy} G, {Hindsley} RB, {Ivezi{\'c}} {\v Z}, {Lupton} RH, {Munn} JA,
  {Pier} JR, {Stoughton} C, {York} DG (2001) {Detection of Massive Tidal Tails
  around the Globular Cluster Palomar 5 with Sloan Digital Sky Survey
  Commissioning Data}. \apjl 548:L165--L169, \doi{10.1086/319095},
  \eprint{astro-ph/0012311}

\bibitem[{{Okamoto} et~al(2015){Okamoto}, {Arimoto}, {Ferguson}, {Bernard},
  {Irwin}, {Yamada}, and {Utsumi}}]{okamoto15}
{Okamoto} S, {Arimoto} N, {Ferguson} AMN, {Bernard} EJ, {Irwin} MJ, {Yamada} Y,
  {Utsumi} Y (2015) {A Hyper Suprime-Cam View of the Interacting Galaxies of
  the M81 Group}. \apjl 809:L1, \doi{10.1088/2041-8205/809/1/L1},
  \eprint{1507.04889}

\bibitem[{{Pawlowski} et~al(2014){Pawlowski}, {Famaey}, {Jerjen}, {Merritt},
  {Kroupa}, {Dabringhausen}, {L{\"u}ghausen}, {Forbes}, {Hensler}, {Hammer},
  {Puech}, {Fouquet}, {Flores}, and {Yang}}]{pawlowski14}
{Pawlowski} MS, {Famaey} B, {Jerjen} H, {Merritt} D, {Kroupa} P,
  {Dabringhausen} J, {L{\"u}ghausen} F, {Forbes} DA, {Hensler} G, {Hammer} F,
  {Puech} M, {Fouquet} S, {Flores} H, {Yang} Y (2014) {Co-orbiting satellite
  galaxy structures are still in conflict with the distribution of primordial
  dwarf galaxies}. \mnras 442:2362--2380, \doi{10.1093/mnras/stu1005},
  \eprint{1406.1799}

\bibitem[{{Pe{\~n}arrubia} et~al(2009){Pe{\~n}arrubia}, {Navarro},
  {McConnachie}, and {Martin}}]{penarrubia09}
{Pe{\~n}arrubia} J, {Navarro} JF, {McConnachie} AW, {Martin} NF (2009) {The
  Signature of Galactic Tides in Local Group Dwarf Spheroidals}. \apj
  698:222--232, \doi{10.1088/0004-637X/698/1/222}, \eprint{0811.1579}

\bibitem[{{Peacock} et~al(2015){Peacock}, {Strader}, {Romanowsky}, and
  {Brodie}}]{peacock15}
{Peacock} MB, {Strader} J, {Romanowsky} AJ, {Brodie} JP (2015) {Detection of a
  Distinct Metal-poor Stellar Halo in the Early-type Galaxy NGC 3115}. \apj
  800:13, \doi{10.1088/0004-637X/800/1/13}, \eprint{1412.2752}

\bibitem[{{Peng} et~al(2002){Peng}, {Ford}, {Freeman}, and {White}}]{peng02}
{Peng} EW, {Ford} HC, {Freeman} KC, {White} RL (2002) {A Young Blue Tidal
  Stream in NGC 5128}. \aj 124:3144--3156, \doi{10.1086/344308},
  \eprint{arXiv:astro-ph/0208422}

\bibitem[{{Peng} et~al(2006){Peng}, {Jord{\'a}n}, {C{\^o}t{\'e}}, {Blakeslee},
  {Ferrarese}, {Mei}, {West}, {Merritt}, {Milosavljevi{\'c}}, and
  {Tonry}}]{peng06}
{Peng} EW, {Jord{\'a}n} A, {C{\^o}t{\'e}} P, {Blakeslee} JP, {Ferrarese} L,
  {Mei} S, {West} MJ, {Merritt} D, {Milosavljevi{\'c}} M, {Tonry} JL (2006)
  {The ACS Virgo Cluster Survey. IX. The Color Distributions of Globular
  Cluster Systems in Early-Type Galaxies}. \apj 639:95--119,
  \doi{10.1086/498210}, \eprint{astro-ph/0509654}

\bibitem[{{Pillepich} et~al(2014){Pillepich}, {Vogelsberger}, {Deason},
  {Rodriguez-Gomez}, {Genel}, {Nelson}, {Torrey}, {Sales}, {Marinacci},
  {Springel}, {Sijacki}, and {Hernquist}}]{pillepich14}
{Pillepich} A, {Vogelsberger} M, {Deason} A, {Rodriguez-Gomez} V, {Genel} S,
  {Nelson} D, {Torrey} P, {Sales} LV, {Marinacci} F, {Springel} V, {Sijacki} D,
  {Hernquist} L (2014) {Halo mass and assembly history exposed in the faint
  outskirts: the stellar and dark matter haloes of Illustris galaxies}. \mnras
  444:237--249, \doi{10.1093/mnras/stu1408}, \eprint{1406.1174}

\bibitem[{{Pohlen} et~al(2004){Pohlen}, {Mart{\'{\i}}nez-Delgado}, {Majewski},
  {Palma}, {Prada}, and {Balcells}}]{pohlen04}
{Pohlen} M, {Mart{\'{\i}}nez-Delgado} D, {Majewski} S, {Palma} C, {Prada} F,
  {Balcells} M (2004) {Tidal Streams around External Galaxies}. In: {Prada} F,
  {Martinez Delgado} D, {Mahoney} TJ (eds) Satellites and Tidal Streams,
  Astronomical Society of the Pacific Conference Series, vol 327, p 288,
  \eprint{astro-ph/0308142}

\bibitem[{{Radburn-Smith} et~al(2011){Radburn-Smith}, {de Jong}, {Seth},
  {Bailin}, {Bell}, {Brown}, {Bullock}, {Courteau}, {Dalcanton}, {Ferguson},
  {Goudfrooij}, {Holfeltz}, {Holwerda}, {Purcell}, {Sick}, {Streich}, {Vlajic},
  and {Zucker}}]{radburn11}
{Radburn-Smith} DJ, {de Jong} RS, {Seth} AC, {Bailin} J, {Bell} EF, {Brown} TM,
  {Bullock} JS, {Courteau} S, {Dalcanton} JJ, {Ferguson} HC, {Goudfrooij} P,
  {Holfeltz} S, {Holwerda} BW, {Purcell} C, {Sick} J, {Streich} D, {Vlajic} M,
  {Zucker} DB (2011) {The GHOSTS Survey. I. Hubble Space Telescope Advanced
  Camera for Surveys Data}. \apjs 195:18, \doi{10.1088/0067-0049/195/2/18}

\bibitem[{{Read}(2014)}]{read14}
{Read} JI (2014) {The local dark matter density}. Journal of Physics G Nuclear
  Physics 41(6):063101, \doi{10.1088/0954-3899/41/6/063101}, \eprint{1404.1938}

\bibitem[{{Rejkuba} et~al(2009){Rejkuba}, {Mouhcine}, and {Ibata}}]{rejkuba09}
{Rejkuba} M, {Mouhcine} M, {Ibata} R (2009) {The stellar population content of
  the thick disc and halo of the Milky Way analogue NGC 891}. \mnras
  396:1231--1246, \doi{10.1111/j.1365-2966.2009.14821.x}

\bibitem[{{Rejkuba} et~al(2011){Rejkuba}, {Harris}, {Greggio}, and
  {Harris}}]{rejkuba11}
{Rejkuba} M, {Harris} WE, {Greggio} L, {Harris} GLH (2011) {How old are the
  stars in the halo of NGC 5128 (Centaurus A)?} \aap 526:A123,
  \doi{10.1051/0004-6361/201015640}, \eprint{1011.4309}

\bibitem[{{Rejkuba} et~al(2014){Rejkuba}, {Harris}, {Greggio}, {Harris},
  {Jerjen}, and {Gonzalez}}]{rejkuba14}
{Rejkuba} M, {Harris} WE, {Greggio} L, {Harris} GLH, {Jerjen} H, {Gonzalez} OA
  (2014) {Tracing the Outer Halo in a Giant Elliptical to 25 R $_{eff}$}. \apjl
  791:L2, \doi{10.1088/2041-8205/791/1/L2}, \eprint{1406.4627}

\bibitem[{{Renda} et~al(2005){Renda}, {Gibson}, {Mouhcine}, {Ibata}, {Kawata},
  {Flynn}, and {Brook}}]{renda05}
{Renda} A, {Gibson} BK, {Mouhcine} M, {Ibata} RA, {Kawata} D, {Flynn} C,
  {Brook} CB (2005) {The stellar halo metallicity-luminosity relationship for
  spiral galaxies}. \mnras 363:L16--L20,
  \doi{10.1111/j.1745-3933.2005.00075.x}, \eprint{astro-ph/0507281}

\bibitem[{{Richardson} et~al(2008){Richardson}, {Ferguson}, {Johnson}, {Irwin},
  {Tanvir}, {Faria}, {Ibata}, {Johnston}, {Lewis}, {McConnachie}, and
  {Chapman}}]{richardson08}
{Richardson} JC, {Ferguson} AMN, {Johnson} RA, {Irwin} MJ, {Tanvir} NR, {Faria}
  DC, {Ibata} RA, {Johnston} KV, {Lewis} GF, {McConnachie} AW, {Chapman} SC
  (2008) {The Nature and Origin of Substructure in the Outskirts of M31. I.
  Surveying the Stellar Content with the Hubble Space Telescope Advanced Camera
  for Surveys}. \aj 135:1998--2012, \doi{10.1088/0004-6256/135/6/1998},
  \eprint{0803.2614}

\bibitem[{{Richardson} et~al(2011){Richardson}, {Irwin}, {McConnachie},
  {Martin}, {Dotter}, {Ferguson}, {Ibata}, {Chapman}, {Lewis}, {Tanvir}, and
  {Rich}}]{richardson11}
{Richardson} JC, {Irwin} MJ, {McConnachie} AW, {Martin} NF, {Dotter} AL,
  {Ferguson} AMN, {Ibata} RA, {Chapman} SC, {Lewis} GF, {Tanvir} NR, {Rich} RM
  (2011) {PAndAS' Progeny: Extending the M31 Dwarf Galaxy Cabal}. \apj 732:76,
  \doi{10.1088/0004-637X/732/2/76}, \eprint{1102.2902}

\bibitem[{{Ryan} and {Norris}(1991)}]{ryan91}
{Ryan} SG, {Norris} JE (1991) {Subdwarf Studies. III. The Halo Metallicity
  Distribution}. \aj 101:1865--1878, \doi{10.1086/115812}

\bibitem[{{Sabbi} et~al(2008){Sabbi}, {Gallagher}, {Smith}, {de Mello}, and
  {Mountain}}]{sabbi08}
{Sabbi} E, {Gallagher} JS, {Smith} LJ, {de Mello} DF, {Mountain} M (2008)
  {Holmberg IX: The Nearest Young Galaxy}. \apjl 676:L113,
  \doi{10.1086/587548}, \eprint{0802.4446}

\bibitem[{{Sackett} et~al(1994){Sackett}, {Morrisoni}, {Harding}, and
  {Boroson}}]{sackett94}
{Sackett} PD, {Morrisoni} HL, {Harding} P, {Boroson} TA (1994) {A faint
  luminous halo that may trace the dark matter around spiral galaxy NGC5907}.
  \nat 370:441--443, \doi{10.1038/370441a0}, \eprint{astro-ph/9407068}

\bibitem[{{Sadoun} et~al(2014){Sadoun}, {Mohayaee}, and {Colin}}]{sadoun14}
{Sadoun} R, {Mohayaee} R, {Colin} J (2014) {A single-merger scenario for the
  formation of the giant stream and the warp of M31}. \mnras 442:160--175,
  \doi{10.1093/mnras/stu850}, \eprint{1307.5044}

\bibitem[{{Sakai} and {Madore}(1999)}]{sakai99}
{Sakai} S, {Madore} BF (1999) {Detection of the Red Giant Branch Stars in M82
  Using the Hubble Space Telescope}. \apj 526:599--606, \doi{10.1086/308032},
  \eprint{astro-ph/9906484}

\bibitem[{{Sales} et~al(2016){Sales}, {Navarro}, {Kallivayalil}, and
  {Frenk}}]{sales16}
{Sales} LV, {Navarro} JF, {Kallivayalil} N, {Frenk} CS (2016) {Identifying true
  satellites of the Magellanic Clouds}. ArXiv e-prints \eprint{1605.03574}

\bibitem[{{Sand} et~al(2012){Sand}, {Strader}, {Willman}, {Zaritsky}, {McLeod},
  {Caldwell}, {Seth}, and {Olszewski}}]{sand12}
{Sand} DJ, {Strader} J, {Willman} B, {Zaritsky} D, {McLeod} B, {Caldwell} N,
  {Seth} A, {Olszewski} E (2012) {Tidal Signatures in the Faintest Milky Way
  Satellites: The Detailed Properties of Leo V, Pisces II, and Canes Venatici
  II}. \apj 756:79, \doi{10.1088/0004-637X/756/1/79}, \eprint{1111.6608}

\bibitem[{{Sand} et~al(2014){Sand}, {Crnojevi{\'c}}, {Strader}, {Toloba},
  {Simon}, {Caldwell}, {Guhathakurta}, {McLeod}, and {Seth}}]{sand14}
{Sand} DJ, {Crnojevi{\'c}} D, {Strader} J, {Toloba} E, {Simon} JD, {Caldwell}
  N, {Guhathakurta} P, {McLeod} B, {Seth} AC (2014) {Discovery of a New Faint
  Dwarf Galaxy Associated with NGC 253}. \apjl 793:L7,
  \doi{10.1088/2041-8205/793/1/L7}, \eprint{1406.6687}

\bibitem[{{Sand} et~al(2015){Sand}, {Spekkens}, {Crnojevi{\'c}}, {Hargis},
  {Willman}, {Strader}, and {Grillmair}}]{sand15}
{Sand} DJ, {Spekkens} K, {Crnojevi{\'c}} D, {Hargis} JR, {Willman} B, {Strader}
  J, {Grillmair} CJ (2015) {Antlia B: A Faint Dwarf Galaxy Member of the NGC
  3109 Association}. \apjl 812:L13, \doi{10.1088/2041-8205/812/1/L13},
  \eprint{1508.01800}

\bibitem[{{Sawala} et~al(2016){Sawala}, {Frenk}, {Fattahi}, {Navarro}, {Bower},
  {Crain}, {Dalla Vecchia}, {Furlong}, {Helly}, {Jenkins}, {Oman}, {Schaller},
  {Schaye}, {Theuns}, {Trayford}, and {White}}]{sawala16}
{Sawala} T, {Frenk} CS, {Fattahi} A, {Navarro} JF, {Bower} RG, {Crain} RA,
  {Dalla Vecchia} C, {Furlong} M, {Helly} JC, {Jenkins} A, {Oman} KA,
  {Schaller} M, {Schaye} J, {Theuns} T, {Trayford} J, {White} SDM (2016) {The
  APOSTLE simulations: solutions to the Local Group's cosmic puzzles}. \mnras
  457:1931--1943, \doi{10.1093/mnras/stw145}, \eprint{1511.01098}

\bibitem[{{Schiavon} et~al(2016){Schiavon}, {Zamora}, {Carrera}, {Lucatello},
  {Robin}, {Ness}, {Martell}, {Smith}, {Garc{\'{\i}}a-Hern{\'a}ndez},
  {Manchado}, {Sch{\"o}nrich}, {Bastian}, {Chiappini}, {Shetrone}, {Mackereth},
  {Williams}, {M{\'e}sz{\'a}ros}, {Allende Prieto}, {Anders}, {Bizyaev},
  {Beers}, {Chojnowski}, {Cunha}, {Epstein}, {Frinchaboy}, {Garc{\'{\i}}a
  P{\'e}rez}, {Hearty}, {Holtzman}, {Johnson}, {Kinemuchi}, {Majewski}, {Muna},
  {Nidever}, {Nguyen}, {O'Connell}, {Oravetz}, {Pan}, {Pinsonneault},
  {Schneider}, {Schultheis}, {Simmons}, {Skrutskie}, {Sobeck}, {Wilson}, and
  {Zasowski}}]{schiavon16}
{Schiavon} RP, {Zamora} O, {Carrera} R, {Lucatello} S, {Robin} AC, {Ness} M,
  {Martell} SL, {Smith} VV, {Garc{\'{\i}}a-Hern{\'a}ndez} DA, {Manchado} A,
  {Sch{\"o}nrich} R, {Bastian} N, {Chiappini} C, {Shetrone} M, {Mackereth} JT,
  {Williams} RA, {M{\'e}sz{\'a}ros} S, {Allende Prieto} C, {Anders} F,
  {Bizyaev} D, {Beers} TC, {Chojnowski} SD, {Cunha} K, {Epstein} C,
  {Frinchaboy} PM, {Garc{\'{\i}}a P{\'e}rez} AE, {Hearty} FR, {Holtzman} JA,
  {Johnson} JA, {Kinemuchi} K, {Majewski} SR, {Muna} D, {Nidever} DL, {Nguyen}
  DC, {O'Connell} RW, {Oravetz} D, {Pan} K, {Pinsonneault} M, {Schneider} DP,
  {Schultheis} M, {Simmons} A, {Skrutskie} MF, {Sobeck} J, {Wilson} JC,
  {Zasowski} G (2016) {Chemical tagging with APOGEE: Discovery of a large
  population of N-rich stars in the inner Galaxy}. \mnras
  \doi{10.1093/mnras/stw2162}, \eprint{1606.05651}

\bibitem[{{Sch{\"o}nrich} et~al(2014){Sch{\"o}nrich}, {Asplund}, and
  {Casagrande}}]{schoenrich14}
{Sch{\"o}nrich} R, {Asplund} M, {Casagrande} L (2014) {Does SEGUE/SDSS indicate
  a dual Galactic halo?} \apj 786:7, \doi{10.1088/0004-637X/786/1/7},
  \eprint{1403.0937}

\bibitem[{{Searle} and {Zinn}(1978)}]{searle78}
{Searle} L, {Zinn} R (1978) {Compositions of halo clusters and the formation of
  the galactic halo}. \apj 225:357--379, \doi{10.1086/156499}

\bibitem[{{Sesar} et~al(2013){Sesar}, {Ivezi{\'c}}, {Stuart}, {Morgan},
  {Becker}, {Sharma}, {Palaversa}, {Juri{\'c}}, {Wozniak}, and
  {Oluseyi}}]{sesar13}
{Sesar} B, {Ivezi{\'c}} {\v Z}, {Stuart} JS, {Morgan} DM, {Becker} AC, {Sharma}
  S, {Palaversa} L, {Juri{\'c}} M, {Wozniak} P, {Oluseyi} H (2013) {Exploring
  the Variable Sky with LINEAR. II. Halo Structure and Substructure Traced by
  RR Lyrae Stars to 30 kpc}. \aj 146:21, \doi{10.1088/0004-6256/146/2/21},
  \eprint{1305.2160}

\bibitem[{{Simon} and {Geha}(2007)}]{simon07}
{Simon} JD, {Geha} M (2007) {The Kinematics of the Ultra-faint Milky Way
  Satellites: Solving the Missing Satellite Problem}. \apj 670:313--331,
  \doi{10.1086/521816}, \eprint{0706.0516}

\bibitem[{{Soria} et~al(1996){Soria}, {Mould}, {Watson}, {Gallagher},
  {Ballester}, {Burrows}, {Casertano}, {Clarke}, and {et al.}}]{soria96}
{Soria} R, {Mould} JR, {Watson} AM, {Gallagher} JS III, {Ballester} GE,
  {Burrows} CJ, {Casertano} S, {Clarke} JT, {et al} (1996) {Detection of the
  Tip of the Red Giant Branch in NGC 5128}. \apj 465:79, \doi{10.1086/177403}

\bibitem[{{Springel} et~al(2006){Springel}, {Frenk}, and {White}}]{springel06}
{Springel} V, {Frenk} CS, {White} SDM (2006) {The large-scale structure of the
  Universe}. \nat 440:1137--1144, \doi{10.1038/nature04805},
  \eprint{astro-ph/0604561}

\bibitem[{{Stinson} et~al(2009){Stinson}, {Dalcanton}, {Quinn}, {Gogarten},
  {Kaufmann}, and {Wadsley}}]{stinson09}
{Stinson} GS, {Dalcanton} JJ, {Quinn} T, {Gogarten} SM, {Kaufmann} T, {Wadsley}
  J (2009) {Feedback and the formation of dwarf galaxy stellar haloes}. \mnras
  395:1455--1466, \doi{10.1111/j.1365-2966.2009.14555.x}, \eprint{0902.0775}

\bibitem[{{Tal} et~al(2009){Tal}, {van Dokkum}, {Nelan}, and
  {Bezanson}}]{tal09}
{Tal} T, {van Dokkum} PG, {Nelan} J, {Bezanson} R (2009) {The Frequency of
  Tidal Features Associated with Nearby Luminous Elliptical Galaxies From a
  Statistically Complete Sample}. \aj 138:1417--1427,
  \doi{10.1088/0004-6256/138/5/1417}, \eprint{0908.1382}

\bibitem[{{Tanaka} et~al(2010){Tanaka}, {Chiba}, {Komiyama}, {Guhathakurta},
  {Kalirai}, and {Iye}}]{tanaka10}
{Tanaka} M, {Chiba} M, {Komiyama} Y, {Guhathakurta} P, {Kalirai} JS, {Iye} M
  (2010) {Structure and Population of the Andromeda Stellar Halo from a
  Subaru/Suprime-Cam Survey}. \apj 708:1168--1203,
  \doi{10.1088/0004-637X/708/2/1168}, \eprint{0908.0245}

\bibitem[{{Tanaka} et~al(2011){Tanaka}, {Chiba}, {Komiyama}, {Guhathakurta},
  and {Kalirai}}]{tanaka11}
{Tanaka} M, {Chiba} M, {Komiyama} Y, {Guhathakurta} P, {Kalirai} JS (2011)
  {Structure and Population of the NGC 55 Stellar Halo from A
  Subaru/Suprime-Cam Survey}. \apj 738:150, \doi{10.1088/0004-637X/738/2/150},
  \eprint{1107.0911}

\bibitem[{{Tikhonov} et~al(2003){Tikhonov}, {Galazutdinova}, and
  {Aparicio}}]{tikhonov03}
{Tikhonov} NA, {Galazutdinova} OA, {Aparicio} A (2003) {Stellar content of NGC
  404 - The nearest S0 Galaxy}. \aap 401:863--872,
  \doi{10.1051/0004-6361:20021819}

\bibitem[{{Tikhonov} et~al(2005){Tikhonov}, {Galazutdinova}, and
  {Drozdovsky}}]{tikhonov05}
{Tikhonov} NA, {Galazutdinova} OA, {Drozdovsky} IO (2005) {Thick disks and
  halos of spiral galaxies M 81, NGC 55 and NGC 300}. \aap 431:127--142,
  \doi{10.1051/0004-6361:20047042}, \eprint{astro-ph/0407389}

\bibitem[{{Tissera} and {Scannapieco}(2014)}]{tissera14}
{Tissera} PB, {Scannapieco} C (2014) {Low-metallicity stellar halo populations
  as tracers of dark matter haloes}. \mnras 445:L21--L25,
  \doi{10.1093/mnrasl/slu114}, \eprint{1407.5800}

\bibitem[{{Tissera} et~al(2013){Tissera}, {Scannapieco}, {Beers}, and
  {Carollo}}]{tissera13}
{Tissera} PB, {Scannapieco} C, {Beers} TC, {Carollo} D (2013) {Stellar haloes
  of simulated Milky-Way-like galaxies: chemical and kinematic properties}.
  \mnras 432:3391--3400, \doi{10.1093/mnras/stt691}, \eprint{1301.1301}

\bibitem[{{Toloba} et~al(2016{\natexlab{a}}){Toloba}, {Sand}, {Guhathakurta},
  {Chiboucas}, {Crnojevi{\'c}}, and {Simon}}]{toloba16}
{Toloba} E, {Sand} D, {Guhathakurta} P, {Chiboucas} K, {Crnojevi{\'c}} D,
  {Simon} JD (2016{\natexlab{a}}) {Spectroscopic Confirmation of the Dwarf
  Spheroidal Galaxy d0994+71 as a Member of the M81 Group of Galaxies}. \apjl
  830:L21, \doi{10.3847/2041-8205/830/1/L21}, \eprint{1610.03856}

\bibitem[{{Toloba} et~al(2016{\natexlab{b}}){Toloba}, {Sand}, {Spekkens},
  {Crnojevi{\'c}}, {Simon}, {Guhathakurta}, {Strader}, {Caldwell}, {McLeod},
  and {Seth}}]{toloba16b}
{Toloba} E, {Sand} DJ, {Spekkens} K, {Crnojevi{\'c}} D, {Simon} JD,
  {Guhathakurta} P, {Strader} J, {Caldwell} N, {McLeod} B, {Seth} AC
  (2016{\natexlab{b}}) {A Tidally Disrupting Dwarf Galaxy in the Halo of NGC
  253}. \apjl 816:L5, \doi{10.3847/2041-8205/816/1/L5}, \eprint{1512.03816}

\bibitem[{{Tolstoy} et~al(2009){Tolstoy}, {Hill}, and {Tosi}}]{tolstoy09}
{Tolstoy} E, {Hill} V, {Tosi} M (2009) {Star-Formation Histories, Abundances,
  and Kinematics of Dwarf Galaxies in the Local Group}. \araa 47:371--425,
  \doi{10.1146/annurev-astro-082708-101650}, \eprint{0904.4505}

\bibitem[{{Torrealba} et~al(2016){Torrealba}, {Koposov}, {Belokurov}, and
  {Irwin}}]{torrealba16}
{Torrealba} G, {Koposov} SE, {Belokurov} V, {Irwin} M (2016) {The feeble giant.
  Discovery of a large and diffuse Milky Way dwarf galaxy in the constellation
  of Crater}. \mnras 459:2370--2378, \doi{10.1093/mnras/stw733},
  \eprint{1601.07178}

\bibitem[{{van der Hulst}(1979)}]{vanderhulst79}
{van der Hulst} JM (1979) {The structure and kinematics of the neutral hydrogen
  bridge between M81 and NGC 3077}. \aap 75:97--111

\bibitem[{{van der Kruit}(1984)}]{vanderkruit84}
{van der Kruit} PC (1984) {A comparison of our Galaxy to NGC 891 and NGC 4565
  and the structure of their spheroids}. \aap 140:470--475

\bibitem[{{van Dokkum}(2005)}]{vandokkum05}
{van Dokkum} PG (2005) {The Recent and Continuing Assembly of Field Elliptical
  Galaxies by Red Mergers}. \aj 130:2647--2665, \doi{10.1086/497593},
  \eprint{astro-ph/0506661}

\bibitem[{{VandenBerg} et~al(2006){VandenBerg}, {Bergbusch}, and
  {Dowler}}]{vandenberg06}
{VandenBerg} DA, {Bergbusch} PA, {Dowler} PD (2006) {The Victoria-Regina
  Stellar Models: Evolutionary Tracks and Isochrones for a Wide Range in Mass
  and Metallicity that Allow for Empirically Constrained Amounts of Convective
  Core Overshooting}. \apjs 162:375--387, \doi{10.1086/498451},
  \eprint{arXiv:astro-ph/0510784}

\bibitem[{{Veljanoski} et~al(2014){Veljanoski}, {Mackey}, {Ferguson}, {Huxor},
  {C{\^o}t{\'e}}, {Irwin}, {Tanvir}, {Pe{\~n}arrubia}, {Bernard}, {Fardal},
  {Martin}, {McConnachie}, {Lewis}, {Chapman}, {Ibata}, and
  {Babul}}]{veljanoski14}
{Veljanoski} J, {Mackey} AD, {Ferguson} AMN, {Huxor} AP, {C{\^o}t{\'e}} P,
  {Irwin} MJ, {Tanvir} NR, {Pe{\~n}arrubia} J, {Bernard} EJ, {Fardal} M,
  {Martin} NF, {McConnachie} A, {Lewis} GF, {Chapman} SC, {Ibata} RA, {Babul} A
  (2014) {The outer halo globular cluster system of M31 - II. Kinematics}.
  \mnras 442:2929--2950, \doi{10.1093/mnras/stu1055}, \eprint{1406.0186}

\bibitem[{{Vlaji{\'c}} et~al(2009){Vlaji{\'c}}, {Bland-Hawthorn}, and
  {Freeman}}]{vlajic09}
{Vlaji{\'c}} M, {Bland-Hawthorn} J, {Freeman} KC (2009) {The Abundance Gradient
  in the Extremely Faint Outer Disk of NGC 300}. \apj 697:361--372,
  \doi{10.1088/0004-637X/697/1/361}, \eprint{0903.1855}

\bibitem[{{Walker} and {Pe{\~n}arrubia}(2011)}]{walker11}
{Walker} MG, {Pe{\~n}arrubia} J (2011) {A Method for Measuring (Slopes of) the
  Mass Profiles of Dwarf Spheroidal Galaxies}. \apj 742:20,
  \doi{10.1088/0004-637X/742/1/20}, \eprint{1108.2404}

\bibitem[{{Watkins} et~al(2009){Watkins}, {Evans}, {Belokurov}, {Smith},
  {Hewett}, {Bramich}, {Gilmore}, {Irwin}, {Vidrih}, {Wyrzykowski}, and
  {Zucker}}]{watkins09}
{Watkins} LL, {Evans} NW, {Belokurov} V, {Smith} MC, {Hewett} PC, {Bramich} DM,
  {Gilmore} GF, {Irwin} MJ, {Vidrih} S, {Wyrzykowski} {\L}, {Zucker} DB (2009)
  {Substructure revealed by RRLyraes in SDSS Stripe 82}. \mnras 398:1757--1770,
  \doi{10.1111/j.1365-2966.2009.15242.x}, \eprint{0906.0498}

\bibitem[{{Weisz} et~al(2008){Weisz}, {Skillman}, {Cannon}, {Dolphin},
  {Kennicutt}, {Lee}, and {Walter}}]{weisz08}
{Weisz} DR, {Skillman} ED, {Cannon} JM, {Dolphin} AE, {Kennicutt} RC Jr, {Lee}
  J, {Walter} F (2008) {The Recent Star Formation Histories of M81 Group Dwarf
  Irregular Galaxies}. \apj 689:160--183, \doi{10.1086/592323},
  \eprint{0809.5059}

\bibitem[{{Wetzel} et~al(2015){Wetzel}, {Deason}, and
  {Garrison-Kimmel}}]{wetzel15}
{Wetzel} AR, {Deason} AJ, {Garrison-Kimmel} S (2015) {Satellite Dwarf Galaxies
  in a Hierarchical Universe: Infall Histories, Group Preprocessing, and
  Reionization}. \apj 807:49, \doi{10.1088/0004-637X/807/1/49},
  \eprint{1501.01972}

\bibitem[{{Wetzel} et~al(2016){Wetzel}, {Hopkins}, {Kim},
  {Faucher-Gigu{\`e}re}, {Kere{\v s}}, and {Quataert}}]{wetzel16}
{Wetzel} AR, {Hopkins} PF, {Kim} Jh, {Faucher-Gigu{\`e}re} CA, {Kere{\v s}} D,
  {Quataert} E (2016) {Reconciling Dwarf Galaxies with {$\Lambda$}CDM
  Cosmology: Simulating a Realistic Population of Satellites around a Milky
  Way-mass Galaxy}. \apjl 827:L23, \doi{10.3847/2041-8205/827/2/L23},
  \eprint{1602.05957}

\bibitem[{{Wheeler} et~al(2015){Wheeler}, {O{\~n}orbe}, {Bullock},
  {Boylan-Kolchin}, {Elbert}, {Garrison-Kimmel}, {Hopkins}, and {Kere{\v
  s}}}]{wheeler15}
{Wheeler} C, {O{\~n}orbe} J, {Bullock} JS, {Boylan-Kolchin} M, {Elbert} OD,
  {Garrison-Kimmel} S, {Hopkins} PF, {Kere{\v s}} D (2015) {Sweating the small
  stuff: simulating dwarf galaxies, ultra-faint dwarf galaxies, and their own
  tiny satellites}. \mnras 453:1305--1316, \doi{10.1093/mnras/stv1691},
  \eprint{1504.02466}

\bibitem[{{White} and {Frenk}(1991)}]{white91}
{White} SDM, {Frenk} CS (1991) {Galaxy formation through hierarchical
  clustering}. \apj 379:52--79, \doi{10.1086/170483}

\bibitem[{{Willman}(2010)}]{willman10}
{Willman} B (2010) {In Pursuit of the Least Luminous Galaxies}. Advances in
  Astronomy 2010:285454, \doi{10.1155/2010/285454}, \eprint{0907.4758}

\bibitem[{{Wu} et~al(2014){Wu}, {Tully}, {Rizzi}, {Dolphin}, {Jacobs}, and
  {Karachentsev}}]{wu14}
{Wu} PF, {Tully} RB, {Rizzi} L, {Dolphin} AE, {Jacobs} BA, {Karachentsev} ID
  (2014) {Infrared Tip of the Red Giant Branch and Distances to the Maffei/IC
  342 Group}. \aj 148:7, \doi{10.1088/0004-6256/148/1/7}, \eprint{1404.2987}

\bibitem[{{Yanny} et~al(2000){Yanny}, {Newberg}, {Kent}, {Laurent-Muehleisen},
  {Pier}, {Richards}, {Stoughton}, {Anderson}, {Annis}, {Brinkmann}, {Chen},
  {Csabai}, {Doi}, {Fukugita}, {Hennessy}, {Ivezi{\'c}}, {Knapp}, {Lupton},
  {Munn}, {Nash}, {Rockosi}, {Schneider}, {Smith}, and {York}}]{yanny00}
{Yanny} B, {Newberg} HJ, {Kent} S, {Laurent-Muehleisen} SA, {Pier} JR,
  {Richards} GT, {Stoughton} C, {Anderson} JE Jr, {Annis} J, {Brinkmann} J,
  {Chen} B, {Csabai} I, {Doi} M, {Fukugita} M, {Hennessy} GS, {Ivezi{\'c}} {\v
  Z}, {Knapp} GR, {Lupton} R, {Munn} JA, {Nash} T, {Rockosi} CM, {Schneider}
  DP, {Smith} JA, {York} DG (2000) {Identification of A-colored Stars and
  Structure in the Halo of the Milky Way from Sloan Digital Sky Survey
  Commissioning Data}. \apj 540:825--841, \doi{10.1086/309386},
  \eprint{astro-ph/0004128}

\bibitem[{{Yun} et~al(1994){Yun}, {Ho}, and {Lo}}]{yun94}
{Yun} MS, {Ho} PTP, {Lo} KY (1994) {A high-resolution image of atomic hydrogen
  in the M81 group of galaxies}. \nat 372:530--532, \doi{10.1038/372530a0}

\bibitem[{{Zheng} et~al(1999){Zheng}, {Shang}, {Su}, {Burstein}, {Chen},
  {Deng}, {Byun}, {Chen}, {Chen}, {Deng}, {Fan}, {Fang}, {Hester}, {Jiang},
  {Li}, {Lin}, {Sun}, {Tsay}, {Windhorst}, {Wu}, {Xia}, {Xu}, {Xue}, {Yan},
  {Zheng}, {Zhou}, {Zhu}, {Zou}, and {Lu}}]{zheng99}
{Zheng} Z, {Shang} Z, {Su} H, {Burstein} D, {Chen} J, {Deng} Z, {Byun} YI,
  {Chen} R, {Chen} WP, {Deng} L, {Fan} X, {Fang} LZ, {Hester} JJ, {Jiang} Z,
  {Li} Y, {Lin} W, {Sun} WH, {Tsay} WS, {Windhorst} RA, {Wu} H, {Xia} X, {Xu}
  W, {Xue} S, {Yan} H, {Zheng} Z, {Zhou} X, {Zhu} J, {Zou} Z, {Lu} P (1999)
  {Deep Intermediate-Band Surface Photometry of NGC 5907}. \aj 117:2757--2780,
  \doi{10.1086/300866}

\bibitem[{{Zibetti} et~al(2004){Zibetti}, {White}, and {Brinkmann}}]{zibetti04}
{Zibetti} S, {White} SDM, {Brinkmann} J (2004) {Haloes around edge-on disc
  galaxies in the Sloan Digital Sky Survey}. \mnras 347:556--568,
  \doi{10.1111/j.1365-2966.2004.07235.x}, \eprint{astro-ph/0309623}

\end{thebibliography}


\end{document}